\pdfoutput=1
\documentclass[twocolumn,coverpage,10pt,prd,aps]{revtex4-1} 
\usepackage{atlasphysics} 
\usepackage{subfigure}
\usepackage{mathrsfs}
\usepackage{graphicx}
\usepackage{multirow}
\usepackage{rotating}
\usepackage{xspace}
\usepackage{upgreek}

\def\masslimitgeneric{832} 
\def\masslimitexpgeneric{731} 
\def\masslimitstop{379} 
\def\masslimitsbottom{344} 

\newcommand{\papertitle}{\bf \boldmath Search for long-lived stopped $R$-hadrons decaying out-of-time with pp collisions using the ATLAS detector}

\newcommand{\theabstract}{An updated search is performed for gluino, top squark, or bottom squark $R$-hadrons that have come to rest within the ATLAS calorimeter, and decay at some later time to hadronic jets and a neutralino, using 5.0 and 22.9~fb$^{-1}$ of $pp$ collisions at 7 and 8~TeV, respectively. Candidate decay events are triggered in selected empty bunch crossings of the LHC in order to remove $pp$ collision backgrounds. Selections based on jet shape and muon-system activity are applied to discriminate signal events from cosmic ray and beam-halo muon backgrounds. In the absence of an excess of events, improved limits are set on gluino, stop, and sbottom masses for different decays, lifetimes, and neutralino masses. With a neutralino of mass 100~GeV, the analysis excludes gluinos with mass below $\masslimitgeneric$~GeV (with an expected lower limit of $\masslimitexpgeneric$~GeV), for a gluino lifetime between 10~$\upmu$s and 1000~s in the generic $R$-hadron model with equal branching ratios for decays to $q\bar{q}\nino$ and $g\nino$. Under the same  assumptions for the neutralino mass and squark lifetime, top squarks and bottom squarks in the Regge $R$-hadron model are excluded with masses below $\masslimitstop$ and $\masslimitsbottom$~GeV, respectively.}

\usepackage{preprintcover}  
\PreprintCoverPaperTitle{\papertitle}  
\PreprintIdNumber{CERN-PH-EP-2013-161}  
\PreprintCoverAbstract{\theabstract}  
\PreprintJournalName{Physical Review D}  

\begin{document}
\title{\papertitle}
\author{The ATLAS Collaboration}

\begin{abstract}
\theabstract
\end{abstract}


\maketitle


\section{Introduction}
\label{sec:introduction}
Long-lived massive particles appear in many theories beyond the
Standard Model~\cite{Fairbairn:2006gg}. They are predicted in $R$-parity-conserving supersymmetry (SUSY)~\cite{Miyazawa:1966,Ramond:1971gb,Golfand:1971iw,Neveu:1971rx,Neveu:1971iv,Gervais:1971ji,Volkov:1973ix,Wess:1973kz,Wess:1974tw,
Fayet:1976et,Fayet:1977yc,Farrar:1978xj,Fayet:1979sa,Dimopoulos:1981zb}  models, such as split SUSY~\cite{SPLIT_SUSY,Aspects_SPLIT_SUSY} and gauge-mediated
SUSY breaking~\cite{Dine:1981gu,AlvarezGaume:1981wy,Nappi:1982hm,Dine:1993yw, Dine:1994vc,Dine:1995ag,Raby}, as well as other scenarios such as universal extra dimensions~\cite{UED} and leptoquark extensions~\cite{leptoquarks}. 
For instance, split SUSY addresses the hierarchy
problem via the same fine-tuning mechanism that solves the
cosmological constant problem; SUSY can be broken at a very high energy
scale, leading to heavy scalars, light fermions, and a light, finely
tuned, Higgs boson~\cite{SPLIT_SUSY}. Within this phenomenological
picture, squarks would thus be much heavier than gluinos, suppressing the
gluino decay.
If the lifetime of the gluino is long enough, it would hadronize into $R$-hadrons, color-singlet states
of $R$-mesons ($\tilde{g}q\bar{q}$), $R$-baryons ($\tilde{g}qqq$), and $R$-gluinoballs ($\tilde{g}g$).
Other models, notably $R$-parity-violating SUSY, could produce a long-lived
squark that would also form an $R$-hadron, e.g. $\tilde{t}\bar{q}$. The phenomenology of the top squark (stop) or the bottom squark (sbottom) is 
comparable to the gluino case but with a smaller production cross section~\cite{Johansen:2010ac,Evans:2012bf}.

$R$-hadron interactions in matter are highly uncertain, but some
features are well predicted. The gluino, stop, or sbottom can be regarded as a heavy,
non-interacting spectator, surrounded by a cloud of interacting
quarks. $R$-hadrons may change their properties through strong
interactions with the detector. Most $R$-mesons would turn into
$R$-baryons~\cite{Kraan:2004tz}, and they could also change their
electric charge through these interactions. At the Large Hadron Collider (LHC) at
CERN~\cite{LHC}, the $R$-hadrons would be produced in pairs and 
approximately back-to-back in the plane transverse to the beam direction. 
Some fraction of these $R$-hadrons would lose all
of their momentum, mainly from ionization energy loss, and come to
rest within the detector volume, only to decay to a neutralino (\nino) and hadronic jets at some later time.

A previous search for stopped gluino $R$-hadrons was performed by the
D0 Collaboration~\cite{dzero}, which excluded a signal for gluinos
with masses up to 250~GeV. That analysis, however, could use only 
the filled crossings in the Tevatron bunch scheme and suppressed
collision-related backgrounds by demanding that
there was no non-diffractive interaction in the events. Search techniques similar to those
described herein have also been employed by the CMS
Collaboration~\cite{cms,cms2} using 4~fb$^{-1}$ of 7 TeV data under the
assumptions that $m_{\tilde{g}}-m_{\nino}>$ 100~GeV and
BR($\tilde{g}\rightarrow g \nino$)$=100\%$. The
resulting limit, at 95$\%$ credibility level, is $m_{\tilde{g}}>$
640~GeV for gluino lifetimes from 10~$\upmu s$ to 1000~s. ATLAS
has up to now studied 31~pb$^{-1}$ of data from 2010~\cite{atlaslimits},
resulting in the limit $m_{\tilde{g}}>$ 341~GeV, under similar assumptions.

This analysis complements previous ATLAS searches for 
long-lived particles~\cite{bib:muon_agnostic,bib:muon_LLP}
that are less sensitive to particles with initial $\beta \ll 1$.
By relying primarily on calorimetric measurements, 
this analysis is also sensitive to events where $R$-hadron charge flipping
may make reconstruction in the inner tracker and the muon system impossible.
Detection of stopped $R$-hadrons could also lead to a measurement of their lifetime and decay properties. 
Moreover, the search is sensitive to any new physics scenario producing 
large out-of-time energy deposits in the calorimeter with minimal additional detector activity. 


\section{The ATLAS detector and event reconstruction}
\label{sec:atlas}

The ATLAS detector~\cite{ATLAS_performance} consists of an inner tracking system (ID) surrounded by a thin
superconducting solenoid, electromagnetic and hadronic calorimeters and a muon spectrometer (MS).
The ID consists of silicon pixel and microstrip detectors, surrounded by a transition radiation straw-tube tracker.
The calorimeter system is based on two active media for the electromagnetic
and hadronic calorimeters: liquid argon in the
inner barrel and end-cap/forward regions, and scintillator tiles 
(TileCal) in the outer barrel region for $|\eta|<1.7$~\footnote{
ATLAS uses a right-handed coordinate system 
with its origin at the nominal interaction point (IP) in the center of the detector and the $z$-axis along the beam 
pipe. The $x$-axis points from the IP to the center of the LHC ring, and the $y$-axis points upward. 
Cylindrical coordinates $(r,\phi)$ are used, $\phi$ being the azimuthal angle around the 
beam pipe. The pseudorapidity is defined in terms of the polar angle $\theta$ as $\eta=-\ln\tan(\theta/2)$.
}. 
The calorimeters are segmented into cells that have typical size 
0.1 by 0.1 in $\eta$--$\phi$ space in the TileCal section.
The MS, capable of reconstructing tracks within $|\eta|<2.7$, 
uses toroidal bending fields generated by three large superconducting magnet systems.
There are inner, middle, and outer muon detector stations, each consisting of several precision tracking layers.
Local muon track segments (abbreviated to simply ``muon segments'' from now on) are first found in each station, before being combined into extended muon tracks.

For this analysis, events are reconstructed using ``cosmic'' settings for the muon
system~\cite{cosmicmuon}, to find muon segments with high efficiency for muons
that are ``out-of-time'' with respect to the expected time for a
muon created from a $pp$ collision and traveling at near the speed of light.
Such out-of-time muons are present in the two most important
background sources. Cosmic ray muons are present at a random
time compared to the bunch-crossing time. Beam-halo muons are
in time with proton bunches but may appear early if they hit the
muon chamber before particles created from the bunch crossing. 
Using cosmic settings for
the muon reconstruction also loosens requirements on the segment direction 
and does not require the segment to point towards the interaction point. 

Jets are constructed from clusters of calorimeter energy deposits~\cite{Lampl:1099735} using the anti-$k_t$ jet algorithm~\cite{antikt}
with the radius parameter set to $R = 0.4$, which assumes the energetic particles
originate from the nominal interaction point.
This assumption, while not generally valid for this analysis, has been checked and still accurately quantifies
the energy released from the stopped $R$-hadron decays occurring
in the calorimeter, as shown by comparisons of test-beam studies of calorimeter energy response with simulation~\cite{Dotti2006106}.
Jet energy is quoted without correcting for the typical fraction of energy not deposited as ionization in the jet cone area, and the minimum jet transverse momentum (\pt) is 7~GeV. 
ATLAS jet reconstruction algorithms are described in more detail elsewhere~\cite{published_jetreco}. The missing transverse momentum (\met) is calculated from
the \pt\ of all reconstructed physics objects in the event, as well as all calorimeter energy clusters not associated with jets.


\section{LHC bunch structure and trigger strategy}
\label{sec:trigger}

The LHC accelerates two counter-rotating proton beams, each divided into 3564 slots for proton bunches separated by 25~ns.
When protons are injected into the LHC, not every bunch slot (BCID) is filled. During 2011 and 2012, alternate BCIDs within a ``bunch train'' were filled, leading to collisions every 50~ns, but there were also many gaps of various lengths between the bunch trains containing adjacent unfilled BCIDs.
Filled BCIDs typically had $>10^{11}$ protons.
Unfilled BCIDs could contain protons due to diffusion from filled BCIDs,
but typically $<10^{8}$ protons per BCID~\cite{Bartmann:2012zz,ATLAS-CONF-2010-102}.
The filled and unfilled BCIDs from the two beams can combine to make three different
``bunch crossing'' scenarios.  A \textit{paired} crossing consists of
a filled BCID from each beam colliding in ATLAS and is when $R$-hadrons would be produced. 
An \textit{unpaired}
crossing has a filled BCID from one beam and an unfilled BCID from the other.
Finally, in an \textit{empty} crossing the BCIDs from both beams are unfilled.

Standard ATLAS analyses use data collected from the
paired crossings, while this analysis searches for physics signatures of metastable $R$-hadrons produced in paired crossings and decaying during selected empty crossings.
This is accomplished with a set of dedicated low-threshold calorimeter
triggers that can fire only in the selected empty or unpaired crossings where the
background to this search is much lower. The type of each bunch crossing is defined at the start of each LHC ``fill'' using the ATLAS beam monitors~\cite{Ohm:2009pf}. Crossings at least six BCIDs after a filled crossing are selected, to reduce background in the muon system.

ATLAS has a three-level trigger system consisting of one hardware
and two software levels~\cite{atlas_trigger}.
Signal candidates are collected
using a hardware trigger requiring localized calorimeter activity, a so-called jet trigger, 
with a 30~GeV transverse energy threshold. 
This trigger could fire only during an empty crossing at
least 125~ns after the most recent paired crossing, such that the detector is mostly free of background from previous interactions.  
The highest-level software trigger then requires a jet with \pt$>50$~GeV, $|\eta|<1.3$, and \met$>50$~GeV. 
The software trigger is more robust against detector noise, keeping the final trigger rate to $<1$~Hz. After offline reconstruction, only 5\% of events with more than two muon segments are saved, and no events with more than 20 muon segments are saved, to reduce the data storage needs since events with muon segments are vetoed in the analysis.
A data sample enriched with beam-halo muons is also accepted with a lower-threshold jet trigger that fires in the
unpaired crossings, and a sample is collected using a trigger that accepts random events from the empty crossings to study background conditions.


\section{Data samples}
\label{sec:samples}

The data used are summarized in Table~\ref{tab:datasamples}, where
the corresponding {\it delivered} integrated luminosity and recorded live time in the selected empty BCIDs are
provided. The early periods of data taking in 2011 are selected as a ``cosmic background region'' to estimate the rate of
background events (mostly from cosmic ray muons, as discussed below). This is motivated by the low integrated
luminosity and small number of paired crossings during these initial periods. 
For a typical signal model that this analysis excludes, less than 3\% of events in the cosmic background region are expected to arise from signal processes.
As discussed in detail in
Sec.~\ref{sec:backgroundestimationcosmicmuon}, the cosmic ray muon background rate is constant, but the signal production rate scales with luminosity.
The later data of 2011 and all of 2012 are used as the ``search region'', where an excess of events from $R$-hadron decays is sought.
ATLAS data are taken in {\it runs}, which typically span one LHC fill, lasting approximately one day.

\begin{table}
\begin{center}
    \caption{The data analyzed in this work and the
             corresponding integrated delivered luminosity, center-of-mass energy, and live time of the ATLAS detector in the selected empty BCIDs during
             those periods.
    \label{tab:datasamples}}
    \begin{tabular}{l|c|c}\hline \hline
Data period & Delivered luminosity & Recorded empty \\
 & (fb$^{-1}$) @ energy (TeV)&  live time (hours) \\ 
      \hline
Cosmic  &  0.3 @ 7      & 125.8 \\ 
Search  &  5.0 @ 7  + 22.9 @ 8     &  389.3 \\ 
\hline
Total   & 5.3 @ 7  + 22.9 @ 8  & 515.1 \\ 
\hline \hline
    \end{tabular}
\end{center}
\end{table}


\section{Simulation of $R$-hadrons}
\label{sec:simulation}

Monte Carlo simulations are used primarily to determine the reconstruction 
efficiency and stopping fraction of the $R$-hadrons, and to study
associated systematic uncertainties on the quantities used in the selections.
The simulated samples have gluino or squark masses in the range 300--1000~GeV, 
to which the present analysis is sensitive.
The {\sc Pythia} program~\cite{Sjostrand:2006za}, version 6.427, with CTEQ6L1 parton distribution functions (PDF)~\cite{Pumplin:2002vw}, is used to simulate pair production of gluinos, stops, or sbottoms. 
The string hadronization model~\cite{Andersson:1983ia}, incorporating specialized hadronization 
routines~\cite{Fairbairn:2006gg} for $R$-hadrons, is used to produce final states containing two $R$-hadrons. 

To compensate for the fact that $R$-hadron scattering is not strongly constrained, 
the simulation of $R$-hadron interactions with matter is
handled by a special detector response simulation~\cite{Kraan:2004tz} 
using {\sc Geant4}~\cite{Agostinelli:2002hh,sim_infra} routines based on 
several scattering and spectrum models with different sets of assumptions: 
the {\it Generic}~\cite{Kraan:2004tz,Mackeprang:2006gx}, {\it Regge}~\cite{deBoer:2007ii,Mackeprang:2009ad}, and {\it Intermediate}~\cite{Farrar:2010ps} models. 
Each model makes different assumptions about the $R$-hadron nuclear interactions and
the spectrum of $R$-hadron states.
\begin{description}
\item[Generic] Limited constraints on allowed stable states permit the
occurrence of doubly-charged $R$-hadrons
and a wide variety of charge-exchange scenarios.
The nuclear scattering model is purely phase-space driven. 
This model is chosen as the nominal model for gluino $R$-hadrons.
\item[Regge] Only one (electrically neutral) baryonic state is allowed. 
The nuclear scattering model employs a triple-Regge formalism.
This model is chosen as the nominal model for stop and sbottom $R$-hadrons.
\item[Intermediate] The spectrum is more 
restricted than the generic model, while still featuring charged baryonic states. The scattering model used is that of the generic model.
\end{description}

\begin{figure}[htbp]
 \begin{center}
   \includegraphics[width=0.48\textwidth]{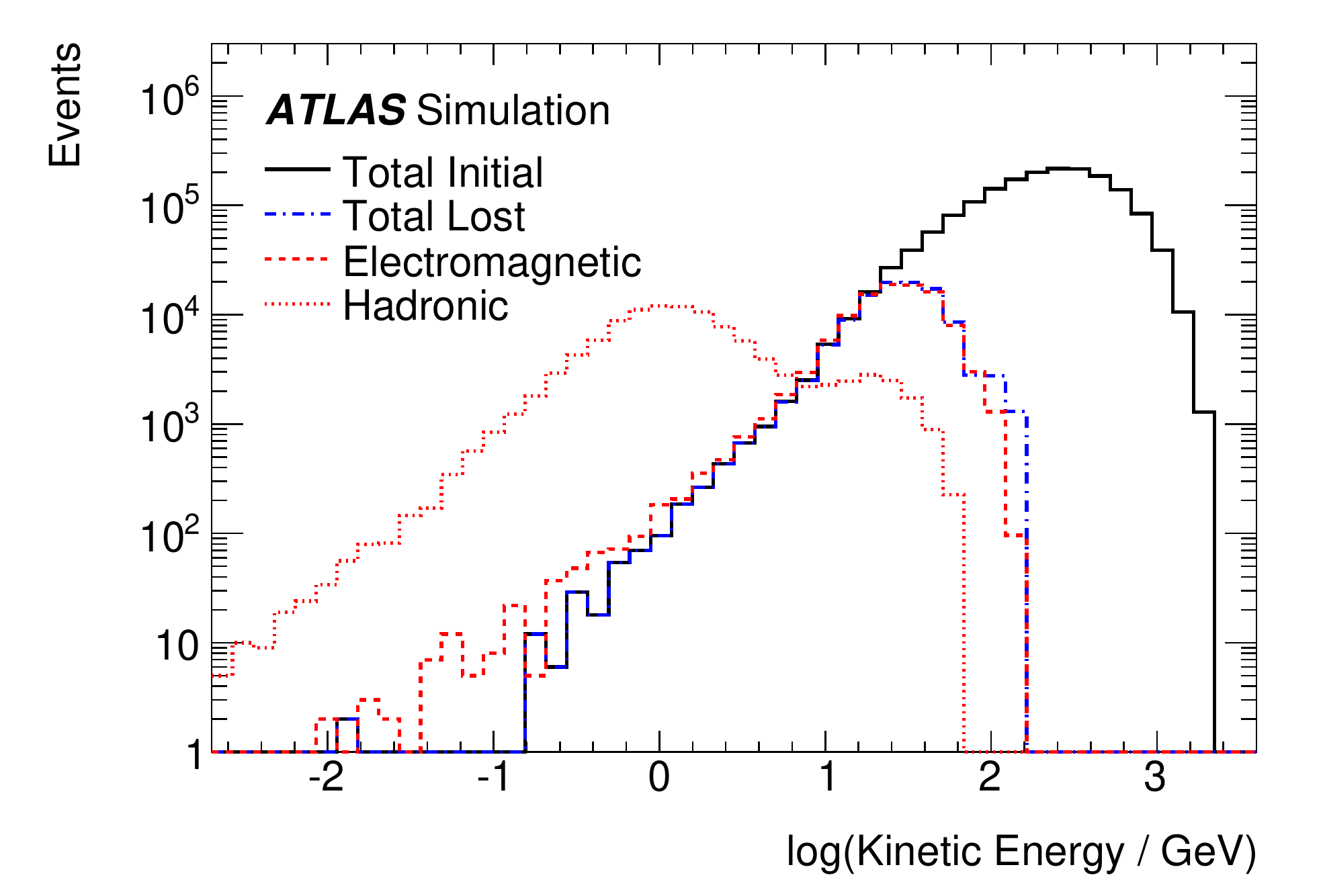}
 \caption{
The kinetic energies of simulated gluino $R$-hadrons with a mass of 800~GeV in the generic model are shown at initial production (black). The energy lost through hadronic interactions with the detector (red, dotted), electromagnetic ionization (red, dashed), and total (red, dashed-dotted) are also shown, for those $R$-hadrons that have stopped. \label{fig:ke}}
 \end{center}
\end{figure}

In the simulation, roughly equal numbers of singly-charged and neutral $R$-hadrons are generated. 
They undergo an average of 4--6 nuclear interactions with the detector, depending on the $R$-hadron model, during which their charge can change.
The $R$-hadrons are created on average with about 200~GeV of kinetic energy. Those created with less than about 20~GeV of kinetic energy tend to lose it all, mostly through ionization, and stop in the detector, as shown in Fig.~\ref{fig:ke}.
Those that stop in the detector are all charged when they stop, with roughly equal numbers of positive and negative singly-charged states. If the doubly-charged state is allowed (as in the generic model), about half of the stopped $R$-hadrons would be doubly positive charged.

If a simulated $R$-hadron comes to rest in the ATLAS detector volume, its location is recorded.
Such an $R$-hadron would bind to a heavy nucleus of an atom in the detector, once it slows down sufficiently, and remain in place indefinitely~\cite{Arvanitaki:2005nq}.
Table~\ref{tab:cutflow_signal_summary} shows the probability for a simulated signal event to have at least one $R$-hadron
stopped within the detector volume, for the models considered.  The stopping fraction shows no significant
dependence on the gluino, stop, or sbottom mass within the statistical uncertainty of the simulation.

\begin{table*}
\begin{center}
\begin{small}
    \caption{
        The selection efficiency after all selection criteria have been applied, its systematic uncertainty, and the stopping fraction, for
        all signal samples.
\label{tab:cutflow_signal_summary}}
\begin{tabular}{lccc|cc|cc|c}
 \hline\hline

 $R$-hadron & Gluino/squark & \multicolumn{2}{c|}{Mass (GeV)}& \multicolumn{2}{c|}{Selection efficiency} & \multicolumn{2}{c|}{Relative systematic uncert.} & Stopping \\
 model    & decay       &$\tilde{g}$ & $\nino$  & $E>$100~GeV & $E>$300~GeV    & $E>$100~GeV & $E>$300~GeV & fraction (\%) \\
\hline
Generic&$\tilde{g} \rightarrow g/q\bar{q}+\nino$&400 &100 & 14.1 $\%$ & 0.5   $\%$ & 15.9\% & --   & $12.2 \pm 0.1$ \\
Generic&$\tilde{g} \rightarrow g/q\bar{q}+\nino$&600 &100 & 15.0 $\%$ & 10.6  $\%$ & 15.7\% & 35.3\% \\
Generic&$\tilde{g} \rightarrow g/q\bar{q}+\nino$&800 &100 & 15.5 $\%$ & 13.9  $\%$ & 15.8\% & 16.2\% \\
Generic&$\tilde{g} \rightarrow g/q\bar{q}+\nino$&1000&100 & 14.8 $\%$ & 14.1 $\%$  & 15.1\% & 15.3\% \\
Generic&$\tilde{g} \rightarrow g/q\bar{q}+\nino$&400 &300 & 3.4  $\%$ & $<0.1\%$    & 60.1\% & --  \\
Generic&$\tilde{g} \rightarrow g/q\bar{q}+\nino$&600 &500 & 4.2  $\%$ & $<0.1\%$    & 48.7\% & --  \\
Generic&$\tilde{g} \rightarrow g/q\bar{q}+\nino$&800 &700 & 4.5  $\%$ & $<0.1\%$    & 35.6\% & --  \\
Generic&$\tilde{g} \rightarrow g/q\bar{q}+\nino$&1000&900 & 5.7  $\%$ & $<0.1\%$    & 33.7\% & --  \\
Generic&$\tilde{g} \rightarrow t\bar{t}+\nino$  &600 &100 & 9.9  $\%$ & 7.2   $\%$ & 18.5\% & 19.8\% \\
Generic&$\tilde{g} \rightarrow t\bar{t}+\nino$  &800 &100 & 10.1 $\%$ & 8.9   $\%$ & 17.7\% & 18.4\% \\
Generic&$\tilde{g} \rightarrow t\bar{t}+\nino$  &1000&100 & 9.5  $\%$ & 8.9   $\%$ & 16.3\% & 16.5\% \\
Generic&$\tilde{g} \rightarrow t\bar{t}+\nino$  &400 &20  & 8.7  $\%$ & 4.3   $\%$ & 18.8\% & 36.8\% \\
Generic&$\tilde{g} \rightarrow t\bar{t}+\nino$  &600 &220 & 9.8  $\%$ & 5.4   $\%$ & 17.0\% & 30.5\% \\
Generic&$\tilde{g} \rightarrow t\bar{t}+\nino$  &800 &420 & 8.3  $\%$ & 4.5   $\%$ & 17.4\% & 28.7\% \\
Generic&$\tilde{g} \rightarrow t\bar{t}+\nino$  &1000&620 & 8.7  $\%$ & 4.7   $\%$ & 17.4\% & 33.5\% \\
\hline
Intermediate &$\tilde{g} \rightarrow g/q\bar{q}+\nino$&400 &100 & 8.6  $\%$ & 0.4 $\%$ & 16.7\% & --     & $7.0 \pm 0.1$ \\
Intermediate &$\tilde{g} \rightarrow g/q\bar{q}+\nino$&600 &100 & 8.9  $\%$ & 6.0  $\%$ & 15.5\% & 28.0\% \\
Intermediate &$\tilde{g} \rightarrow g/q\bar{q}+\nino$&800 &100 & 8.4  $\%$ & 7.4  $\%$ & 15.5\% & 16.0\% \\
Intermediate &$\tilde{g} \rightarrow g/q\bar{q}+\nino$&1000&100 & 7.4 $\%$ & 6.9  $\%$ & 16.1\% & 16.5\% \\
\hline
Regge  &$\tilde{g} \rightarrow g/q\bar{q}+\nino$&400 &100 & 16.7 $\%$ & 0.7  $\%$ & 15.9\% & --     & $5.2 \pm 0.1$ \\
Regge  &$\tilde{g} \rightarrow g/q\bar{q}+\nino$&600 &100 & 19.3 $\%$ & 13.4 $\%$ & 15.4\% & 30.6\% \\
Regge  &$\tilde{g} \rightarrow g/q\bar{q}+\nino$&800 &100 & 19.4 $\%$ & 17.2 $\%$ & 17.8\% & 15.7\% \\
Regge  &$\tilde{g} \rightarrow g/q\bar{q}+\nino$&1000&100 & 19.6 $\%$ & 18.4 $\%$ & 18.8\% & 17.7\% \\
\hline
Generic &$\tilde{t}\rightarrow t + \nino$&300&100 & 11.0 $\%$ & --        &15.6\%  & --    &$10.1 \pm 0.1$ \\
Generic &$\tilde{t}\rightarrow t + \nino$&400&100 & 10.1 $\%$ & --        &16.2\%  & --     \\
Generic &$\tilde{t}\rightarrow t + \nino$&600&100 & 9.6  $\%$ & 4.6  $\%$&16.7\%  & 44.4\% \\
Generic &$\tilde{t}\rightarrow t + \nino$&800&100 & 10.2 $\%$ & 7.7  $\%$&16.5\%  & 17.0\% \\
Generic &$\tilde{t}\rightarrow t + \nino$&400&200 & 10.9 $\%$ & --        &15.5\%  & --     \\
Generic &$\tilde{t}\rightarrow t + \nino$&600&400 & 10.9 $\%$ & --        &15.9\%  & --     \\
Generic &$\tilde{t}\rightarrow t + \nino$&800&600 & 10.7 $\%$ & --        &16.1\%  & --     \\
\hline
Regge   &$\tilde{b}\rightarrow b + \nino$&300&100 & 7.6  $\%$ & --        & 17.8\%& --   & $5.3 \pm 0.1$ \\
Regge   &$\tilde{b}\rightarrow b + \nino$&400&100 & 10.8 $\%$ & --        & 17.3\%& --     \\
Regge   &$\tilde{b}\rightarrow b + \nino$&600&100 & 11.9 $\%$ & 5.8  $\%$& 16.6\%& 59.6\% \\
Regge   &$\tilde{b}\rightarrow b + \nino$&800&100 & 12.5 $\%$ & 10.5 $\%$& 15.8\%& 17.0\% \\
Regge   &$\tilde{b}\rightarrow b + \nino$&400&300 & 3.0  $\%$ & --        & 59.1\%& --     \\
Regge   &$\tilde{b}\rightarrow b + \nino$&600&500 & 3.9  $\%$ & --        & 50.0\%& --     \\
Regge   &$\tilde{b}\rightarrow b + \nino$&800&700 & 3.4  $\%$ & --        & 35.2\%& --     \\
Regge   &$\tilde{b}\rightarrow b + \nino$&300&200 & 1.9  $\%$ & --        & 74.0\%& --     \\
\hline
Regge   &$\tilde{t}\rightarrow t + \nino$&300&100 & 10.5 $\%$ & --        & 16.0\%& --   & $8.1 \pm 0.1$\\
Regge   &$\tilde{t}\rightarrow t + \nino$&400&100 & 10.2 $\%$ & --        & 16.2\%& --     \\
Regge   &$\tilde{t}\rightarrow t + \nino$&600&100 & 10.1 $\%$ & 4.5  $\%$& 17.2\%& 39.8\% \\
Regge   &$\tilde{t}\rightarrow t + \nino$&800&100 & 10.6 $\%$ & 8.0  $\%$& 16.5\%& 19.5\% \\

\hline \hline
\end{tabular}
\end{small}
\end{center}
\end{table*}

The stopping locations are used as input for a second step of {\sc Pythia} where
the decays of the $R$-hadrons are simulated. A uniform random time translation is applied in a 25~ns time window, from $-15$ to $+10$~ns, relative to the bunch-crossing time, since the $R$-hadron would decay at a random time relative to the bunch
structure of the LHC, but would be triggered by the ATLAS detector during the corresponding empty BCID.
These simulated events then proceed through the standard ATLAS digitization simulation~\cite{sim_infra} and event reconstruction (but with cosmic ray muon settings).
The effects of cavern background, a long-lived background component made up of low-energy $\gamma$-rays and x-rays from low-energy neutrons in the cavern, are not included in the simulation directly, but they are accounted for
by measuring the muon activity in the randomly-triggered empty data (see Sec.~\ref{sec:efficiency}). Using the randomly-triggered data, the calorimeter activity due to preceding interactions 
is found to be negligible compared to the jet energy uncertainty and is ignored. 

Different models allow the gluinos to decay via the radiative process, $\tilde{g}\rightarrow g \nino$, 
or via $\tilde{g}\rightarrow q\bar{q}\nino$. The results are
interpreted assuming either a 50\% branching ratio to $g \nino$  and 50\% to $q\bar{q} \nino$, or 100\% to $t\bar{t} \nino$ as would be the case if the top squark was significantly lighter than the other squarks. Reconstruction efficiencies are typically $\approx 20\%$ higher for $q\bar{q} \nino$ compared to $g \nino$ decays.
The stop (sbottom) is assumed to always decay to a top (bottom) quark and a neutralino.
The neutralino mass, $m_{\nino}$, is fixed either to 100~GeV, or such that there is only 100~GeV of free energy left in the decay (a compressed scenario).


\section{Candidate selection}
\label{sec:selection}

First, events are required to pass tight data quality constraints that verify that all parts of the detector were operating normally.  
Events with calorimeter noise bursts are rejected; this has negligible impact on signal efficiency. 
The basic selection criteria, imposed to isolate signal-like events
from background events, demand at least one high energy jet and
no muon segments reconstructed in the muon system passing selections.
Since most of the $R$-hadrons are produced centrally in
$\eta$, the analysis uses only the central barrel of the calorimeter and requires that the leading
jet satisfies $|\eta|<1.2$. In order to reduce the background, the analysis demands the leading jet energy $>$ 50~GeV. 
Up to five additional jets are allowed, more than expected for the signal models considered.

The fractional missing \et\ is the \met\ divided by the leading jet \pt\ and is required to be $>0.5$. This eliminates
background from beam-gas and residual $pp$ events, and has minimal impact on the
signal efficiencies. To remove events with a single, narrow energy spike in the calorimeter, due to noise in the
electronics or data corruption, events are vetoed where the smallest number of cells containing 90$\%$ of the
energy deposit of the leading jet (n90) is fewer than four. 
This n90$>$3 requirement also reduces other background significantly since most large energy deposits from muons in the calorimeter result from hard bremsstrahlung photons, which create short, narrow electromagnetic showers. Large, broad, hadronic showers from deep-inelastic scattering of the muons off nuclei are far rarer.
To further exploit the difference between calorimeter energy deposits from
muons and the expected signal, the jet width is required to be $>$ 0.04, where jet width is the \pt-weighted
$\Delta R$ average of each constituent from the jet axis and $\Delta R=\sqrt{(\Delta\eta)^2+(\Delta\phi)^2}$.
The fraction of the leading jet energy deposited in the TileCal must be $>$ 0.5, to reduce background from beam-halo where the incoming muon cannot be detected due to lack of MS coverage at low radius in the forward region.

\begin{figure*}[htbp]
 \begin{center}
\includegraphics[width=0.48\textwidth]{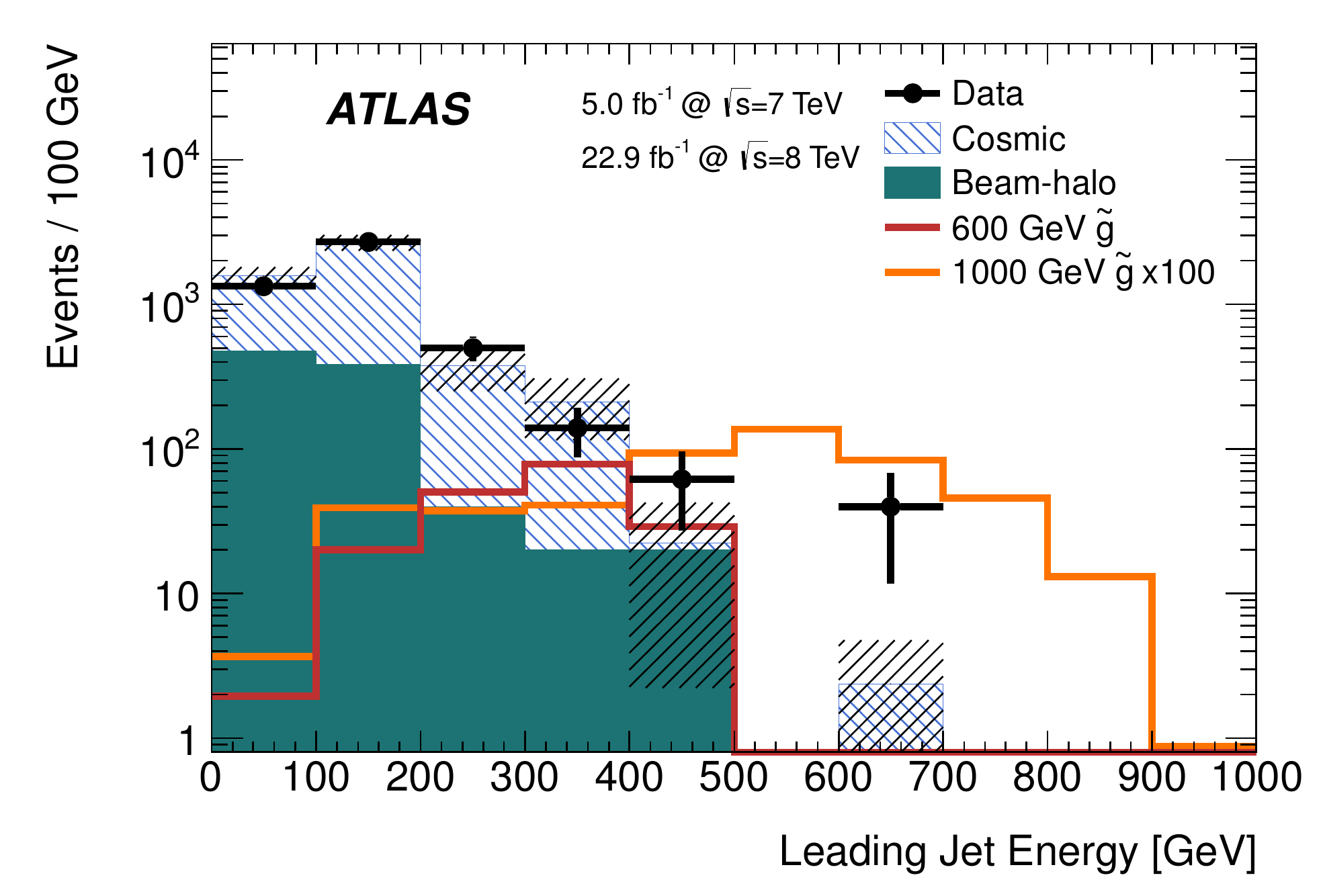} \includegraphics[width=0.48\textwidth]{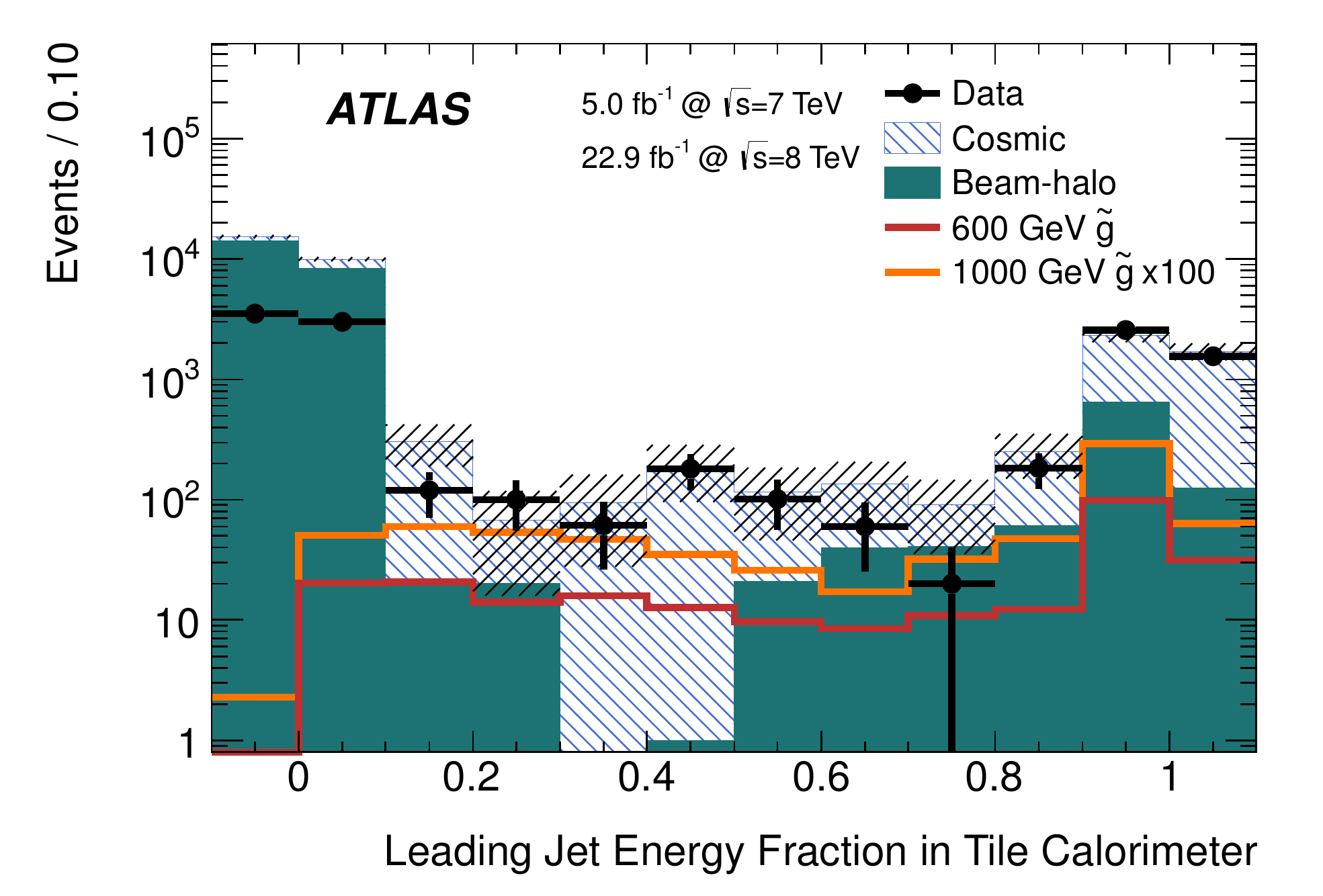}
\includegraphics[width=0.48\textwidth]{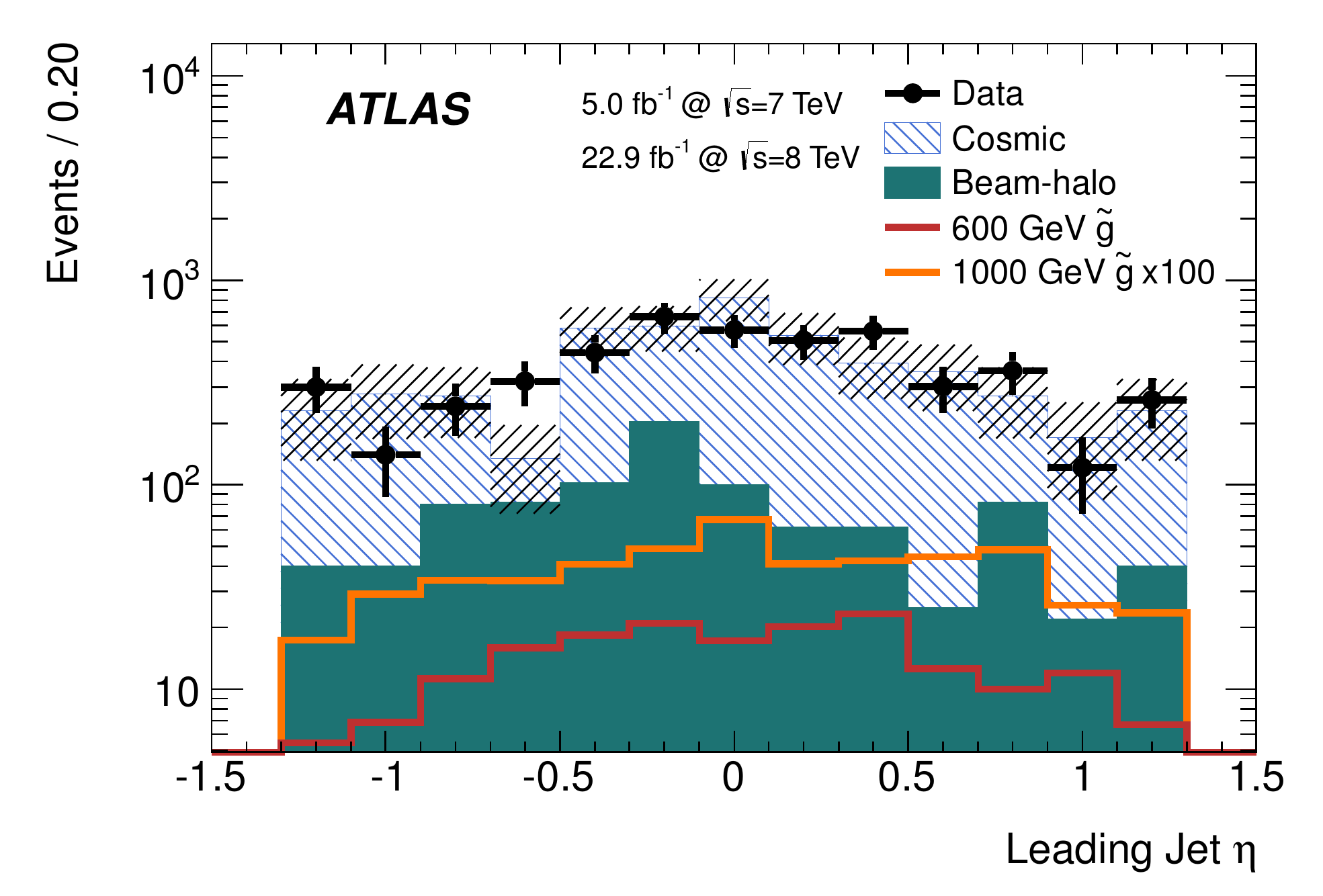} \includegraphics[width=0.48\textwidth]{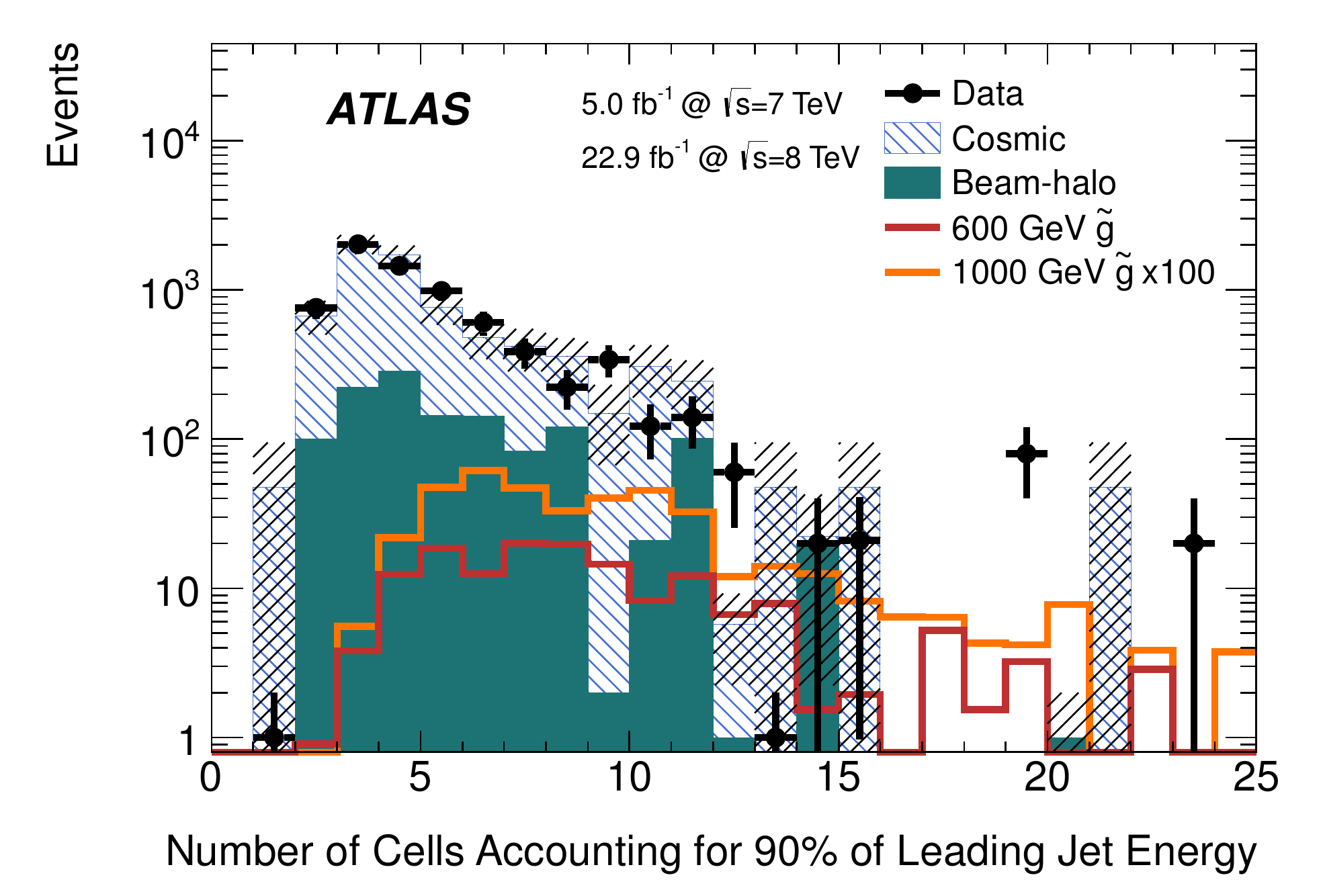}
\includegraphics[width=0.48\textwidth]{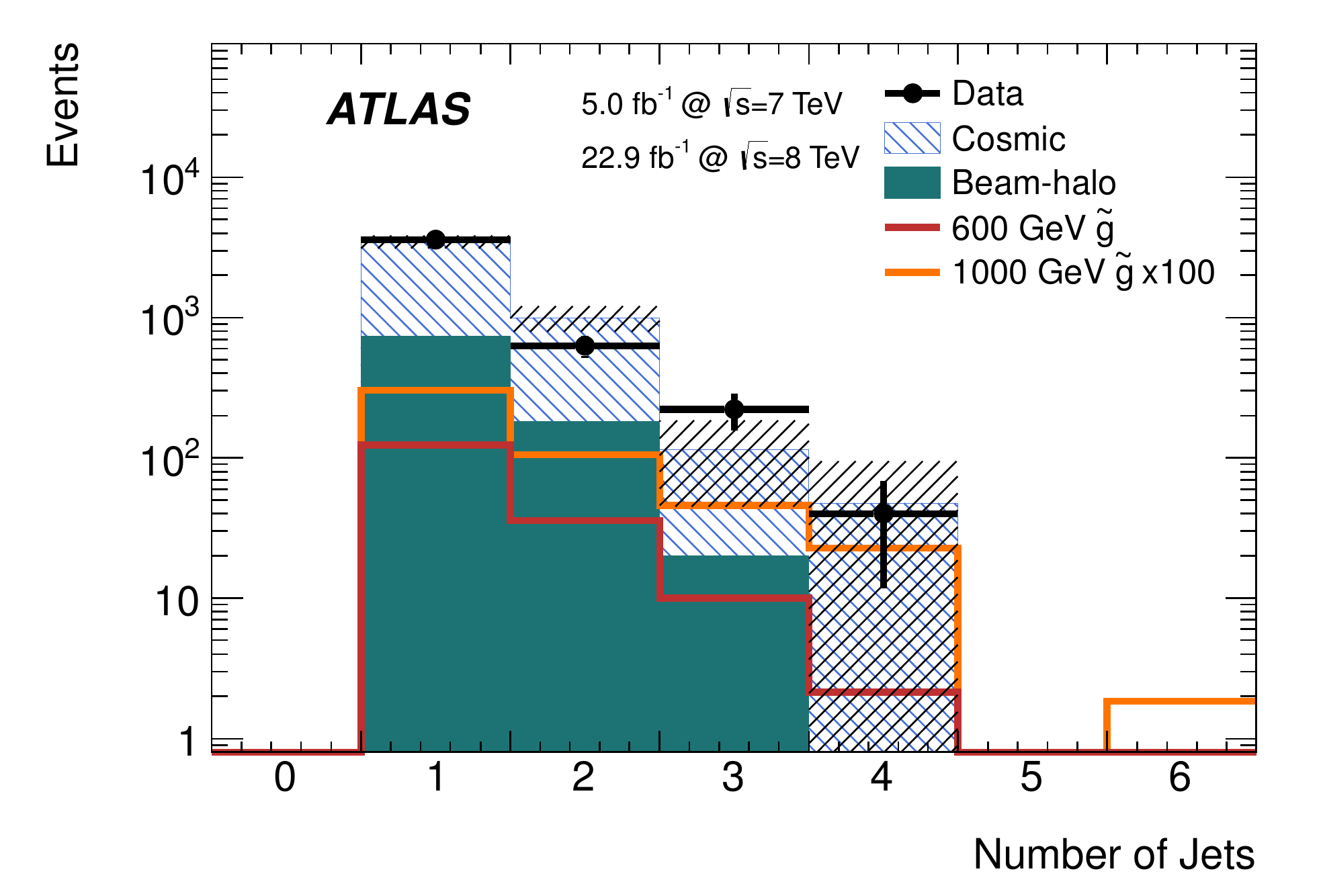} \includegraphics[width=0.48\textwidth]{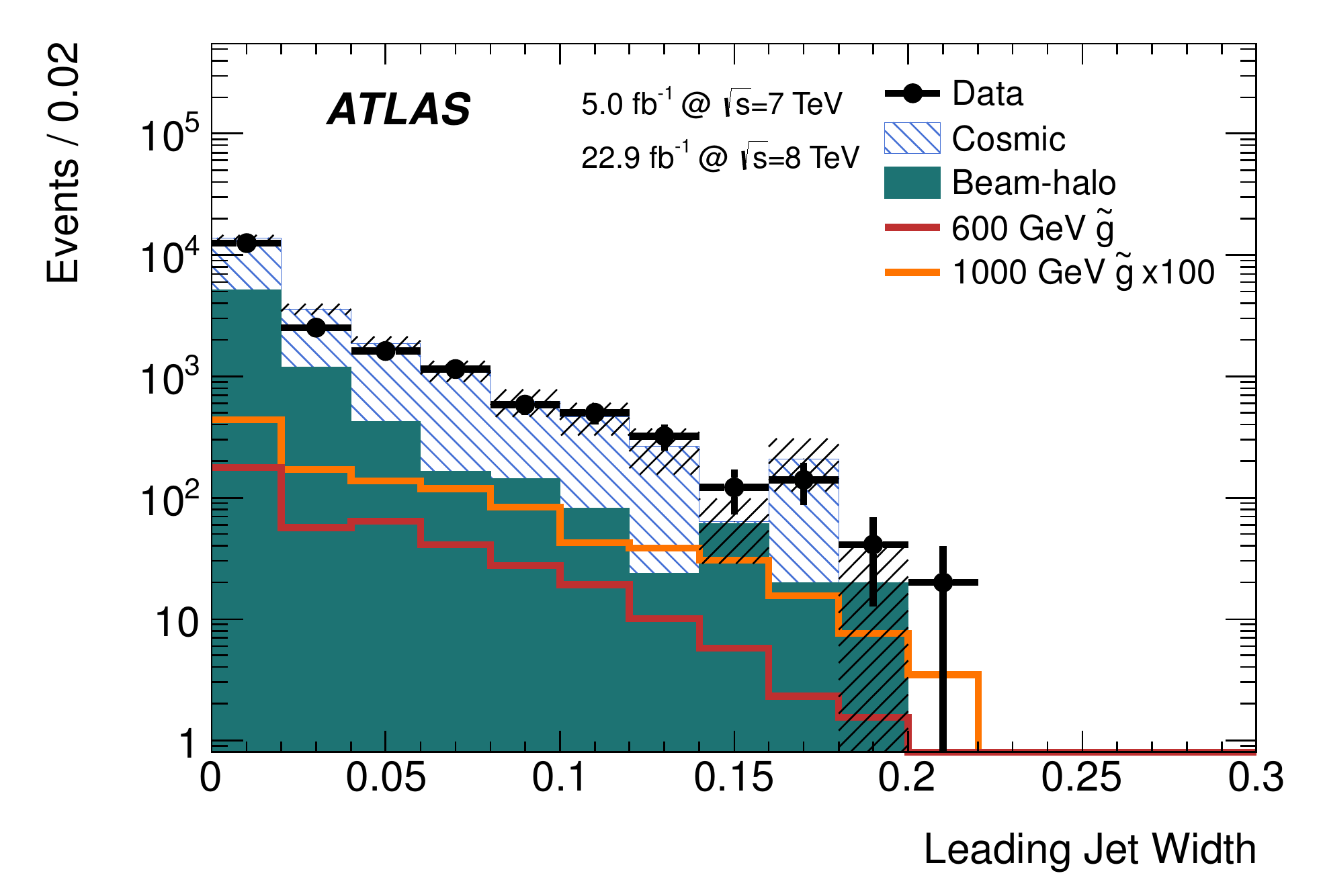}
 \caption{Jet variables for the empty crossing signal triggers. The
    requirements in Table~\ref{tab:cutflow} are applied except for leading jet energy 
    $>100$~GeV and the muon segment veto. For the quantity being plotted, the corresponding selection criterion  
    is not applied. Histograms are normalized to the expected number of events 
    in the search region. Note that only 5\% of data events with more than two muon track segments were kept; they are scaled on these figues by a factor of 20. The top hashed band shows the total statistical uncertainty on the background estimate.
The background estimate does not correctly describe beam-halo with low TileCal energy fraction; this region is not used in the analysis. The TileCal energy fraction can be negative or greater than one due to pileup-subtraction corrections.
 \label{fig:cleaned_muons}}
 \end{center}
\end{figure*}

\begin{figure*}[htbp]
 \begin{center}
 \includegraphics[width=0.48\textwidth]{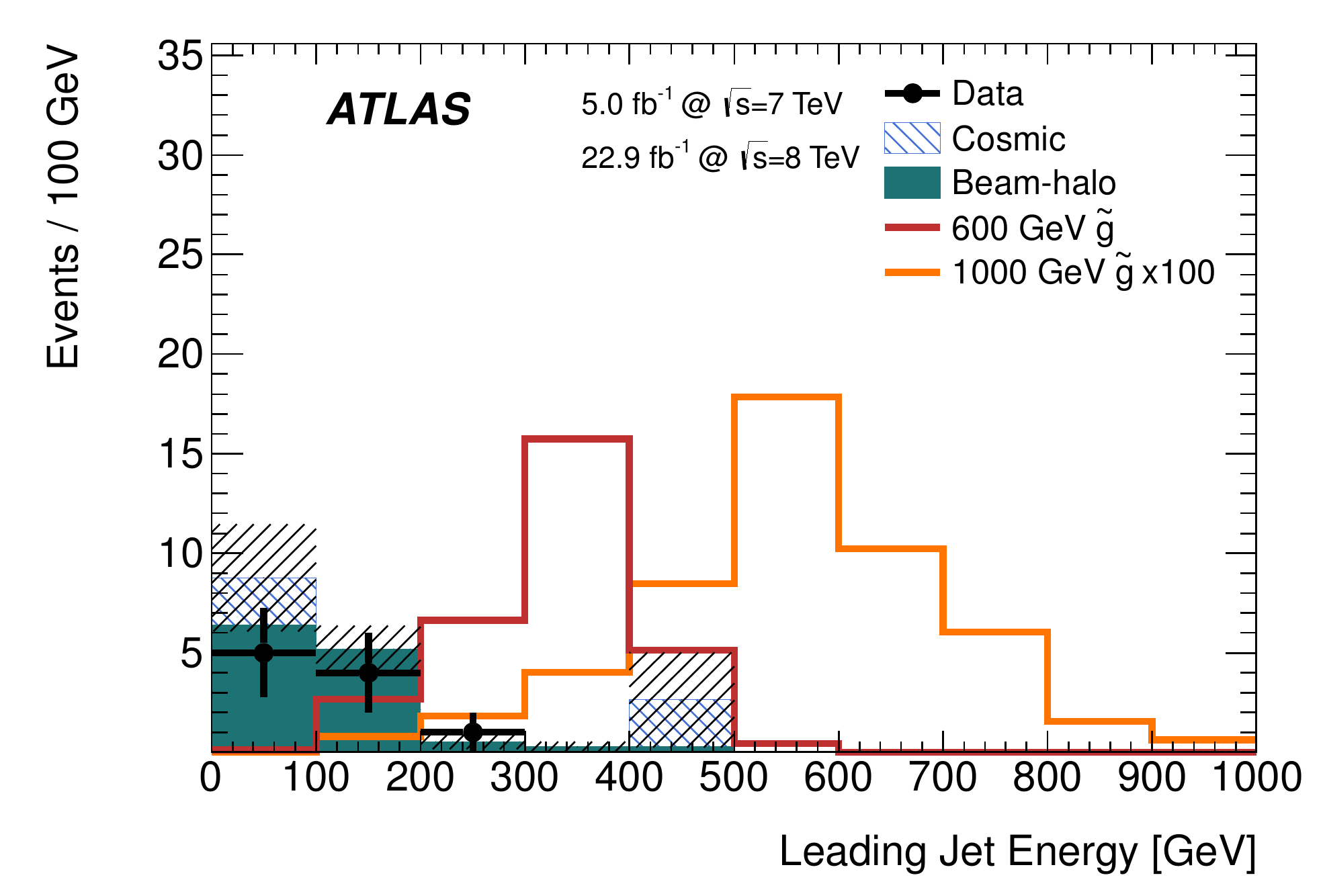} \includegraphics[width=0.48\textwidth]{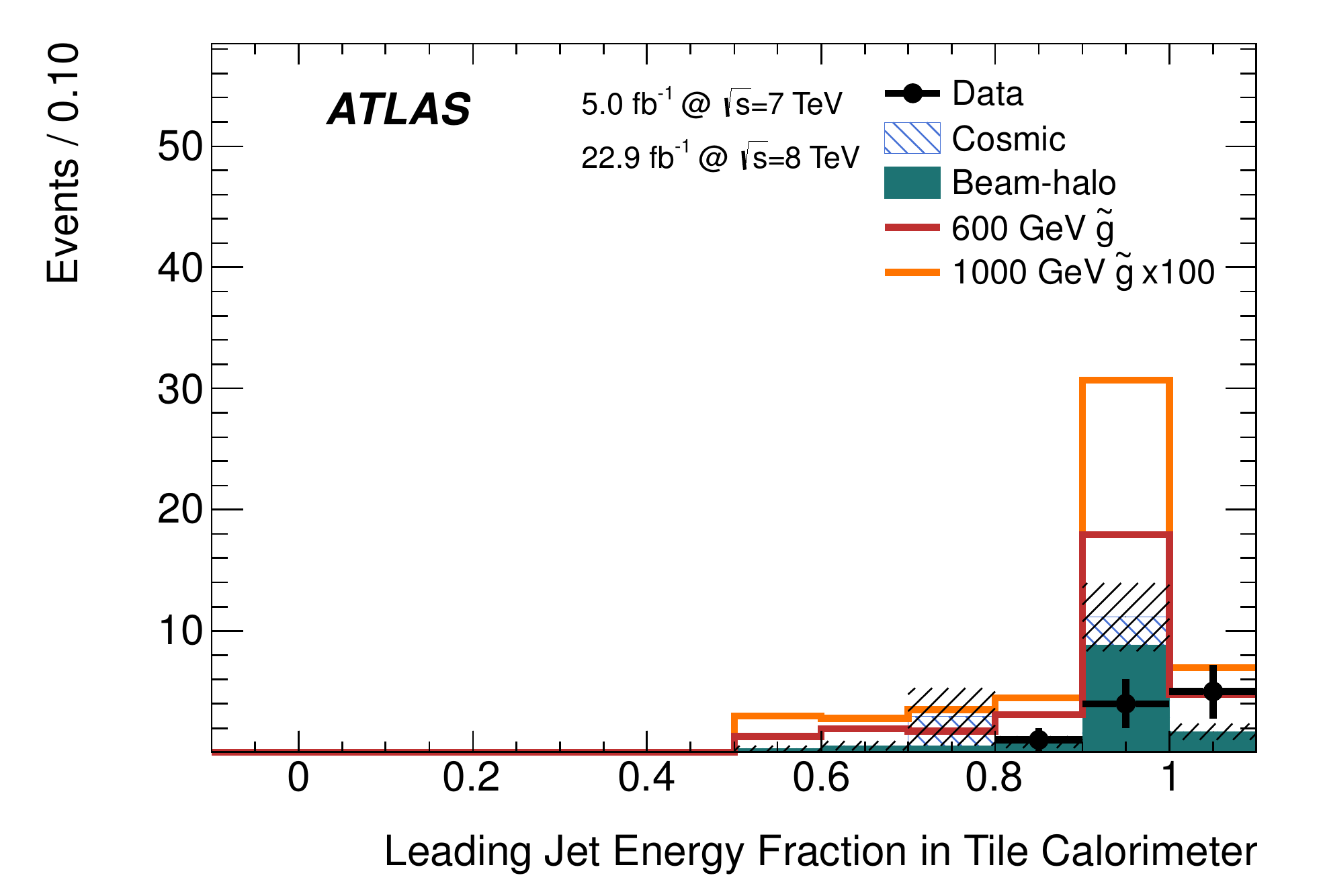}
 \includegraphics[width=0.48\textwidth]{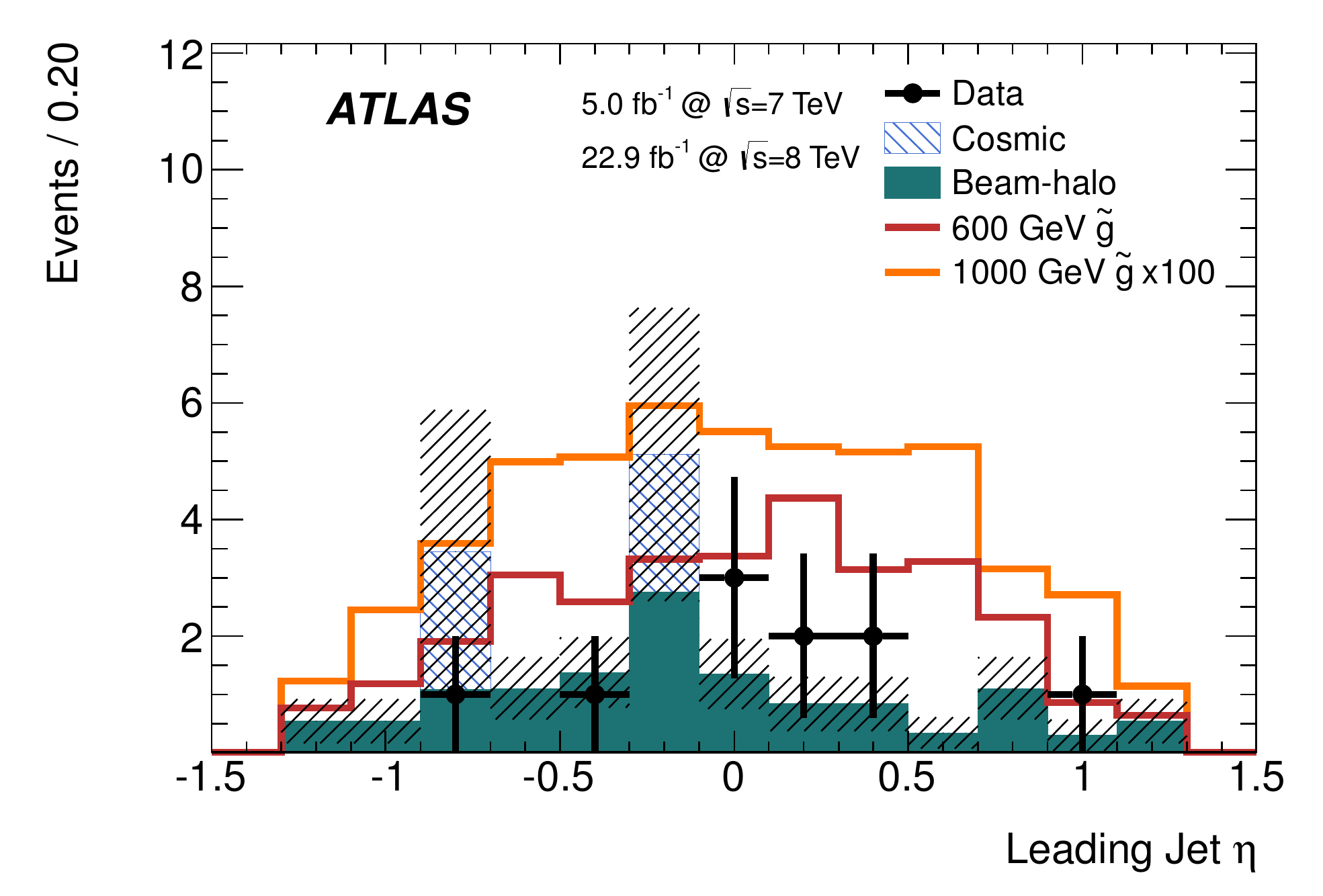} \includegraphics[width=0.48\textwidth]{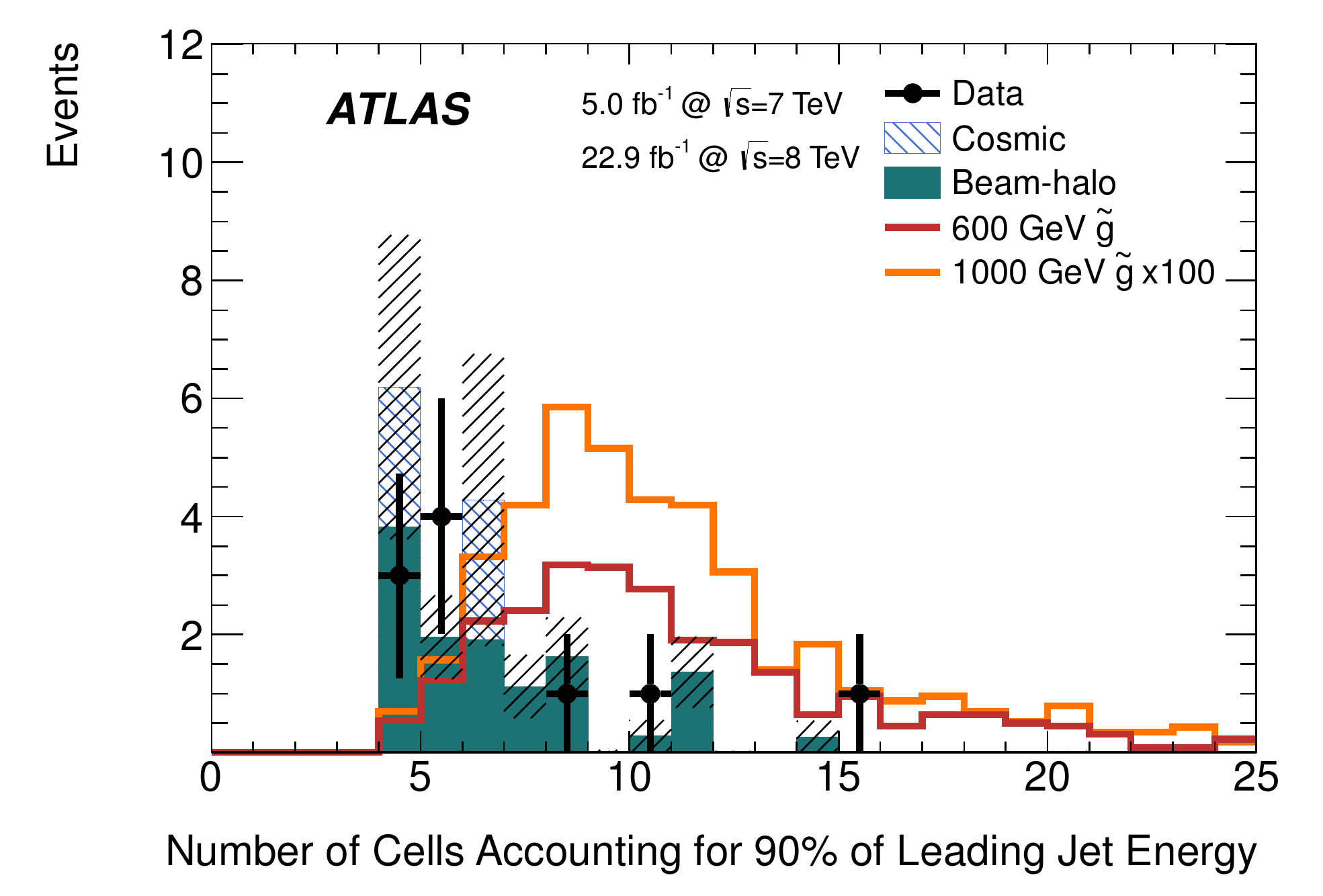}
 \includegraphics[width=0.48\textwidth]{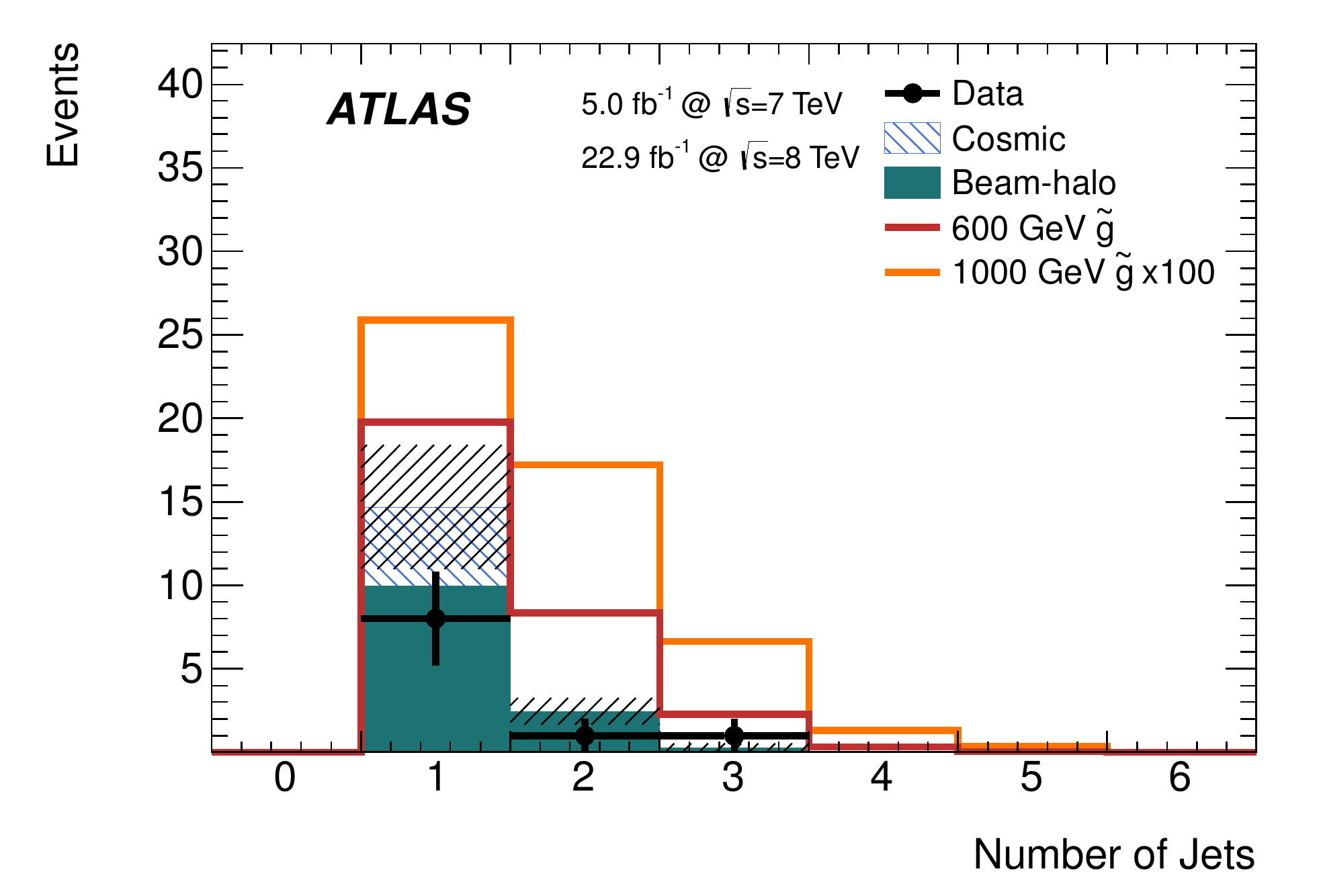} \includegraphics[width=0.48\textwidth]{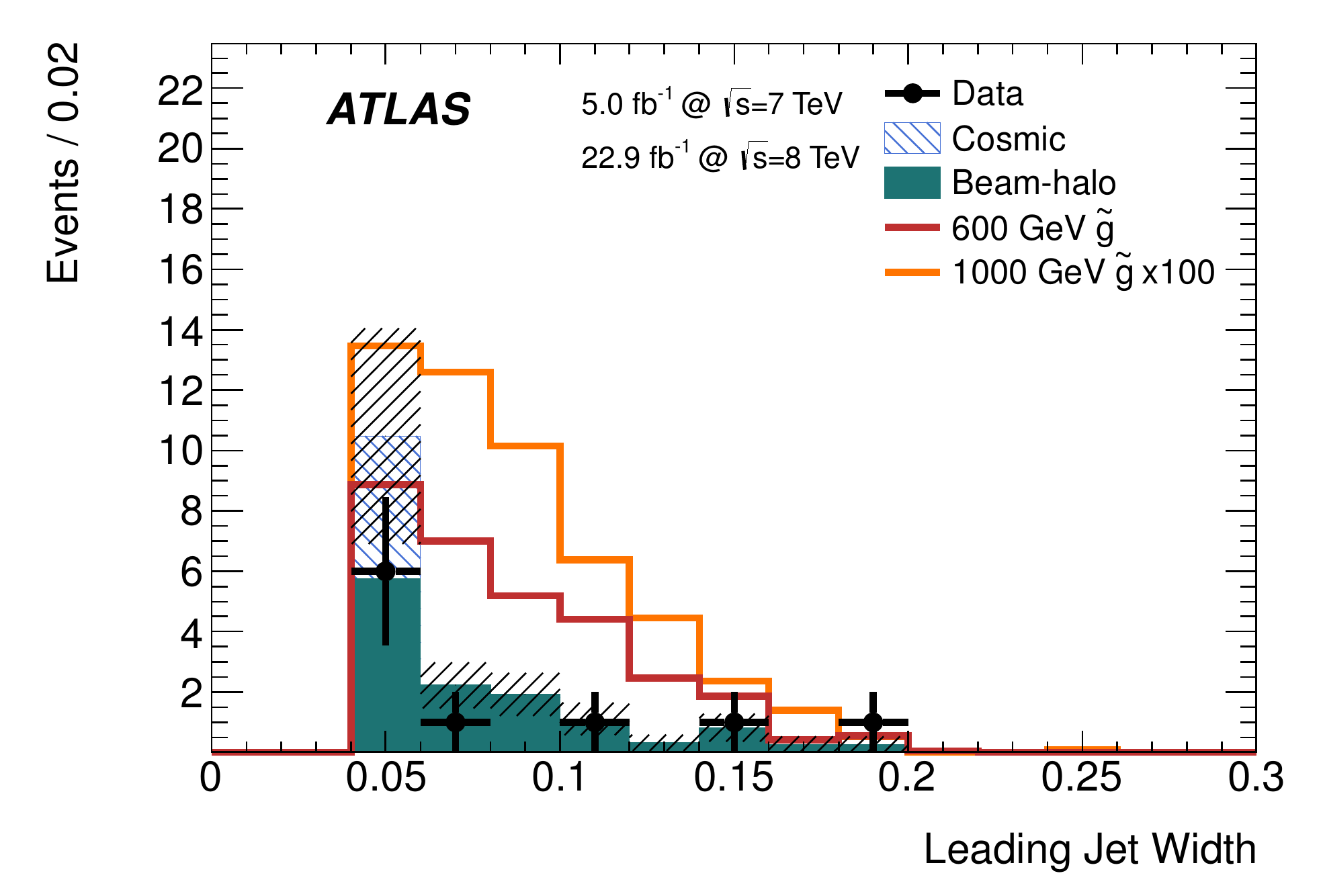}
 \caption{
The event yields in the signal region for candidates with all selection criteria applied (in Table~\ref{tab:cutflow}) including the muon segment veto, but omitting the jet energy $>100$~GeV requirement. 
All samples are scaled to represent their anticipated yields in the search
region. The top hashed band shows the total statistical uncertainty on the background estimate.
 \label{fig:loose_signalRegion1}}
 \end{center}
\end{figure*}

The analysis then requires that no muon segment with more than four associated hits in a muon station be reconstructed in the event. Muon segments with a small number of measurements are often present from cavern background, noise, and pile-up, as studied in the randomly-triggered data.
The events before the muon segment veto, only requiring the leading jet energy $>$ 50~GeV, are studied as a control sample, since the expected signal-to-background ratio is small.
Comparisons of the distributions of several jet variables between backgrounds and data can be seen in Fig.~\ref{fig:cleaned_muons} for events in this control sample.
The backgrounds shown in these figures are estimated using the techniques described in Sec.~\ref{sec:backgroundestimation}.
To remove overlap between the cosmic ray and beam-halo backgrounds in these plots, an event is not considered ``cosmic'' if it 
has a muon segment with more than four hits and an angle within 0.2 radians from parallel to the beamline.
The same distributions are shown for events after the muon segment veto in Fig.~\ref{fig:loose_signalRegion1}.
Finally, a leading jet energy requirement of $>100$ or $>300$~GeV defines two signal regions, sensitive to either a small or large mass difference between 
the $R$-hadron and the neutralino in the signal model, respectively.
Table~\ref{tab:cutflow_signal_summary} shows the efficiencies of these selections on the signal simulations, and 
Table~\ref{tab:cutflow} presents the number of data events surviving each of the imposed selection criteria.

\begin{figure*}[htbp]
 \begin{center}
\includegraphics[width=0.48\textwidth]{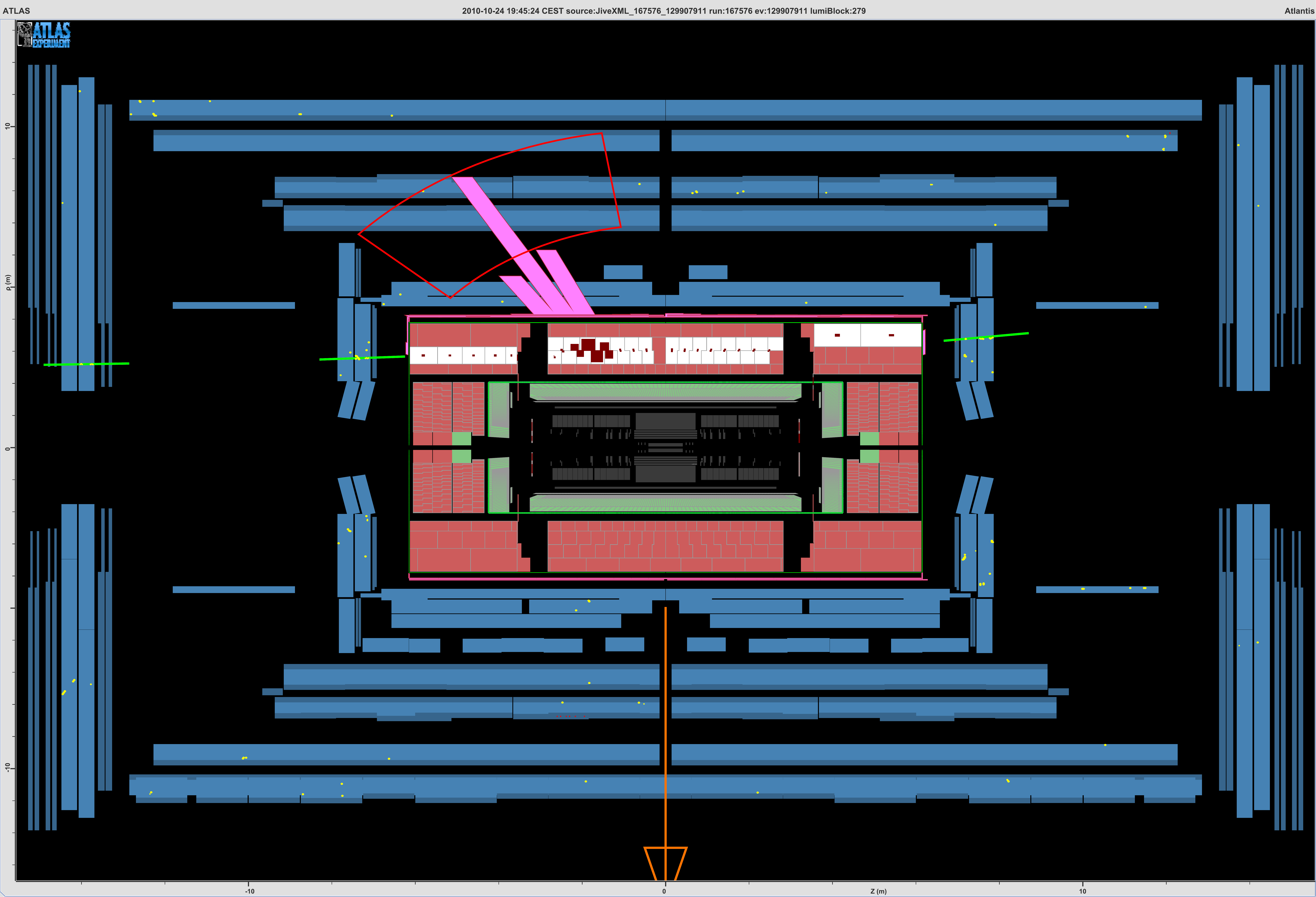}
\includegraphics[width=0.48\textwidth]{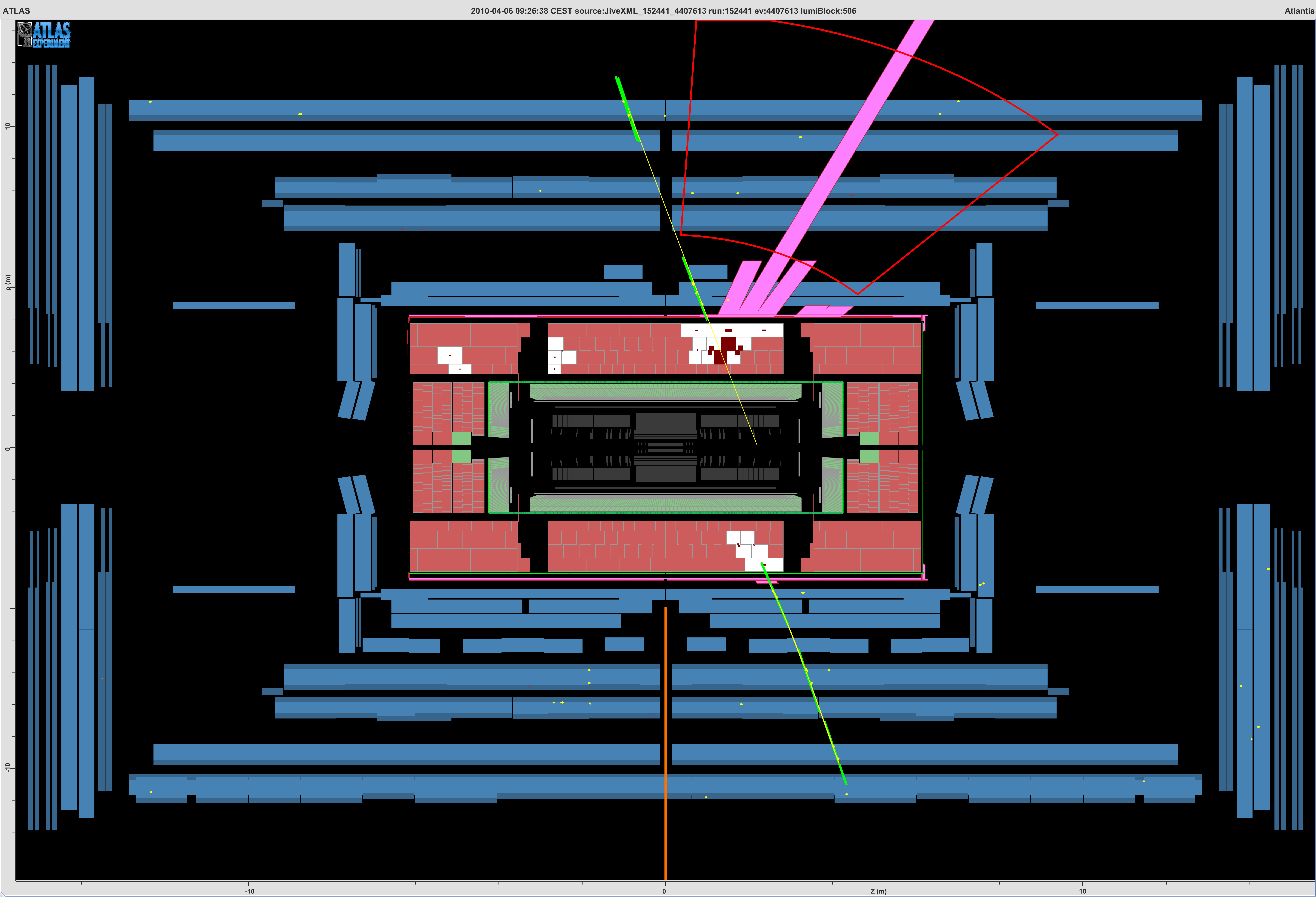}
 \end{center}
 \caption{Left: A beam-halo candidate event during an unpaired crossing in data.
This event passed all the selection criteria except for the muon segment veto.
 Right: A cosmic ray muon candidate event during an empty crossing in data.
    This event passed all the selection criteria except for the muon segment veto.
In both plots, white squares filled with red squares show reconstructed energy deposits in TileCal cells above noise threshold (the fraction of red area indicates the amount of energy in the cell), purple bars show a histogram of total energy in projective TileCal towers, jets are shown by open red trapezoids, muon segments by green line segments in each muon station, and muon tracks by continuous thin yellow lines. \met is shown as an orange arrow.
\label{fig:backgrounddisplays}}
\end{figure*}

\begin{table*}
\begin{center}
    \caption{
Number of events after selections for data in the cosmic background and search regions, defined in
Table~\ref{tab:datasamples}. 
The cosmic background region data are shown before and after scaling to the search region, which
accounts for the different detector live time and accidental muon segment veto efficiency (if applicable)
between the background and search regions. The cumulative efficiency after each cut for a simulated signal of $\tilde{g} \rightarrow g/qq \nino$ decays with a gluino mass of 800 GeV and fixed neutralino mass of 100 GeV is also shown. The uncertainties are statistical only. 
    \label{tab:cutflow}}
\begin{tabular}{l|c|c|c|c}
 \hline\hline
Selection criteria & Cosmic region & Scaled cosmic region & Search region & $\tilde{g} \rightarrow g/qq \nino$ \\
 & & & & $m_{\tilde{g}}=800$ GeV, $m_{\nino}=100$ GeV \\
\hline
Trigger                             &49390 $\pm$ 920  &152800 $\pm$ 2800 &218076  & 71.1$\pm$0.7 \% \\
Leading jet $|\eta|<$ 1.2           &44760 $\pm$ 870  &138500 $\pm$ 2700 &202015   & 60.5$\pm$0.7 \%  \\
Number of jets $<$ 6                        &44690 $\pm$ 870  &138300 $\pm$ 2700 &201628    & 60.4$\pm$0.7 \% \\
Fractional \met\ $>$ 0.5            &44680 $\pm$ 870  &138200 $\pm$ 2700 &201618   & 60.2$\pm$0.7 \%  \\
Leading jet n90 $>$ 3               &12680 $\pm$ 470  &39200 $\pm$ 1500 &85866   & 54.2$\pm$0.7 \%   \\
Leading jet width $>$ 0.04          &4130  $\pm$ 260  &12770 $\pm$ 810  &34445     & 29.9$\pm$0.7 \% \\
Leading jet tile $E$ frac. $>$ 0.5    &1640  $\pm$ 180  &5070  $\pm$ 560  &5396     & 22.3$\pm$0.6 \%  \\
Leading jet energy $>$ 50~GeV       &1640  $\pm$ 180  &5070  $\pm$ 560  &5396    & 22.3$\pm$0.6 \%   \\
Muon segment veto                   &2.0   $\pm$ 1.4  &4.7   $\pm$ 3.4  &10    & 15.5$\pm$0.6 \%     \\
Leading jet energy $>$ 100~GeV      &1.0   $\pm$ 1.0  &2.4   $\pm$ 2.4  &5        & 15.5$\pm$0.6 \%  \\
Leading jet energy $>$ 300~GeV      &1.0   $\pm$ 1.0  &2.4   $\pm$ 2.4  &0       & 13.9$\pm$0.5 \%   \\
\hline\hline
\end{tabular}
\end{center}
\end{table*}


\section{Background Estimation}
\label{sec:backgroundestimation}

\subsection{Beam-Halo Background}
\label{sec:backgroundestimationbeamhalo}

Protons in either beam can interact with residual gas in the beampipe, or with the beampipe itself if they stray off orbit, leading to a hadronic shower. If the interaction takes place several hundred meters from ATLAS, most of the shower is absorbed in shielding or surrounding material before reaching ATLAS. The muons from the shower can survive and enter the detector, traveling parallel to the beamline and in time with the (filled) proton BCIDs~\cite{Azhgirey:786364,Aad:2013zwa}.
The unpaired-crossing data with a jet passing the selection criteria are dominantly beam-halo background.
Figure~\ref{fig:backgrounddisplays} (left) shows an event display of an example beam-halo background candidate event.

To estimate the number of beam-halo events in the empty crossings of the search region, an orthogonal sample of events from the
unpaired crossings is used. The ratio of the number of beam-halo events that pass the jet criteria but fail to have a muon
segment identified to those that do have a muon segment identified is measured. 
This ratio is then multiplied by the number of beam-halo events observed in the signal region that do
have an identified muon segment to give the estimate of the number that do not have a muon segment and thus 
contribute to background in the signal region. Beam-halo events in the unpaired-crossing data are identified by applying a modified version of
the search selection criteria. The muon segment veto is removed, events with leading jet energy $>50$~GeV are used, and the n90$>$3 requirement is not applied. 
Studies show that the muon efficiency is not significantly correlated with the energy or shape of the jet in the calorimeter for beam-halo events.
A muon segment is required to be nearly parallel to the beam pipe, $\theta<0.2$ or $\theta>(\pi-0.2)$, and have more than four muon station measurements. 
Next, beam-halo events that failed to leave a muon segment are counted, allowing the ratio of beam-halo events with no muon segment identified to be calculated.
Then the number of beam-halo muons in the search
region (the empty crossings) that did leave a muon segment is counted. The
same selection criteria as listed in Table~\ref{tab:cutflow} are used, omitting the
100~GeV requirement.  However, instead of a muon segment veto, a
parallel muon segment is required. If no events are present, the uncertainty is taken as $\pm$1 event.
Findings are summarized in Table~\ref{tab:bh_estimation}.

\begin{table*}
\begin{center}
    \caption{Estimate of beam-halo events entering the search region,
            as described in Sec.~\ref{sec:backgroundestimationbeamhalo}.
            The ratio of the number of beam-halo muons that do not leave
            a segment to the number that do leave a segment is calculated from the unpaired data. This ratio is then applied
            to the number of events in the search region where a segment was
            reconstructed to yield the beam-halo estimate.
            The quoted uncertainties are statistical only.
    \label{tab:bh_estimation}}
\begin{tabular}{cc|c|c|c|c}
 \hline\hline
Data     &Leading jet
    &\multicolumn{2}{|c}{Unpaired (all data combined)}  & \multicolumn{2}{|c}{Empty} \\
region  & energy (GeV)
    & Parallel $\mu$& No $\mu$        & Parallel $\mu$ & Predicted No $\mu$\\
\hline
Cosmic & 50 & 1634 & 22 &82    $\pm$ 40   & 1.1  $\pm$ 0.6  \\
Search & 50 & 1634 & 22 &900   $\pm$ 130   & 12  $\pm$ 3   \\
Cosmic & 100& 1634 & 22 &61    $\pm$ 35    & 0.8  $\pm$ 0.5  \\
Search & 100& 1634 & 22 &445   $\pm$ 94    & 6   $\pm$ 2   \\
Cosmic & 300& 1634 & 22 &$0.0^{+1.0}_{-0}$ & $0.000^{+0.01}_{-0}$ \\
Search & 300& 1634 & 22 &40   $\pm$ 28    & 0.5  $\pm$ 0.4  \\
\hline\hline
\end{tabular}
\end{center}
\end{table*}

\subsection{Cosmic Ray Muon Background}
\label{sec:backgroundestimationcosmicmuon}

The background from cosmic ray muons is estimated using the cosmic background region (described in Sec.~\ref{sec:samples}). 
The beam-halo background is estimated for this data sample
as described above, and this estimate is subtracted from the observed events passing all selections. Finally, this number of cosmic ray events in
the cosmic region is scaled by the ratio of the signal-region to cosmic-region live times to estimate the cosmic ray background in the signal region. 
Additionally, the cosmic background estimate is multiplied by the muon-veto efficiency (see Sec.~\ref{sec:efficiency}) to account for the rejection of background caused by the muon veto.
An example cosmic-ray muon background event candidate is shown in Figure~\ref{fig:backgrounddisplays} (right).

\section{Event Yields}
\label{sec:eventyields}

\begin{figure*}[htbp]
 \begin{center}
\includegraphics[width=0.48\textwidth]{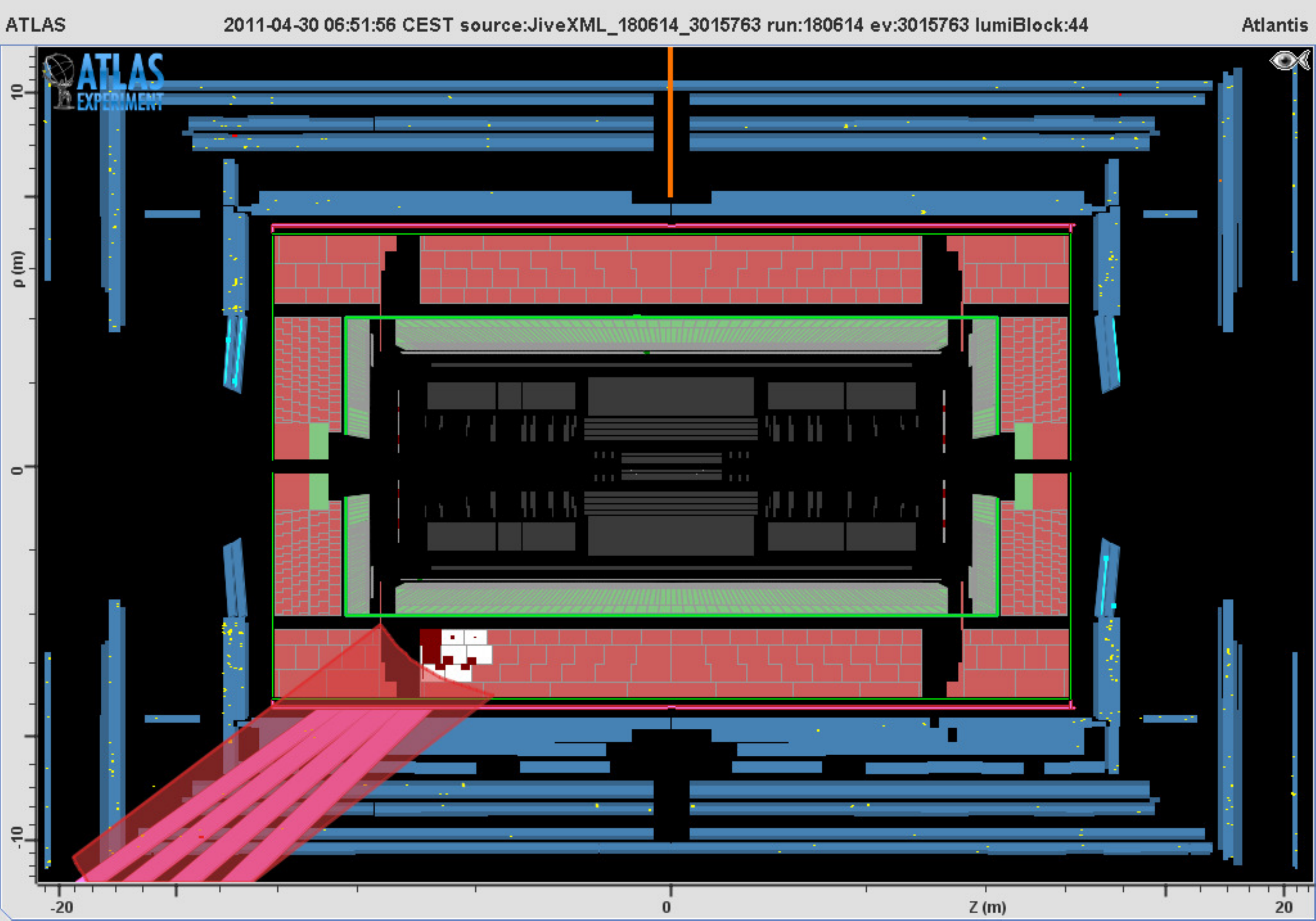} \includegraphics[width=0.48\textwidth]{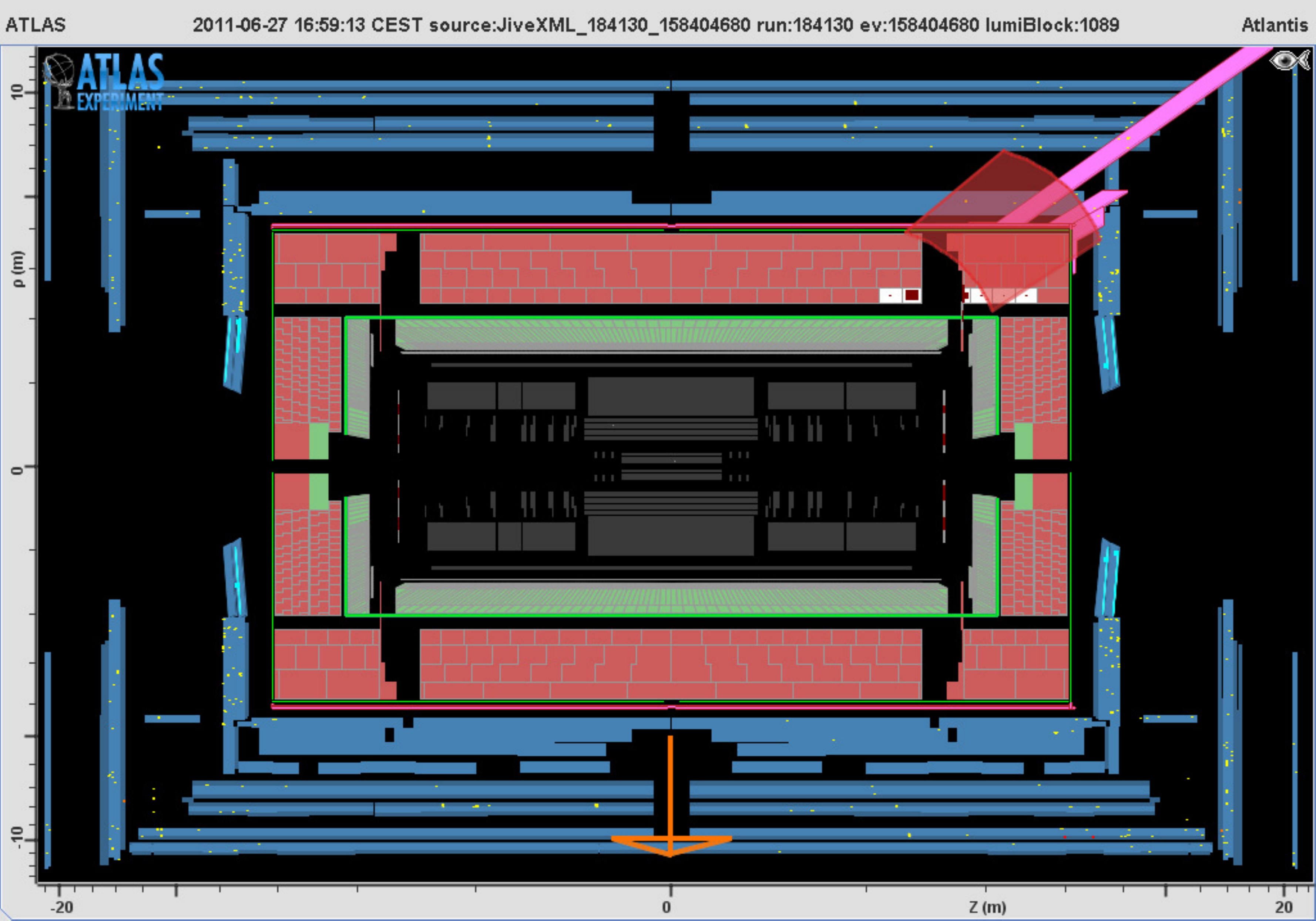}
\includegraphics[width=0.48\textwidth]{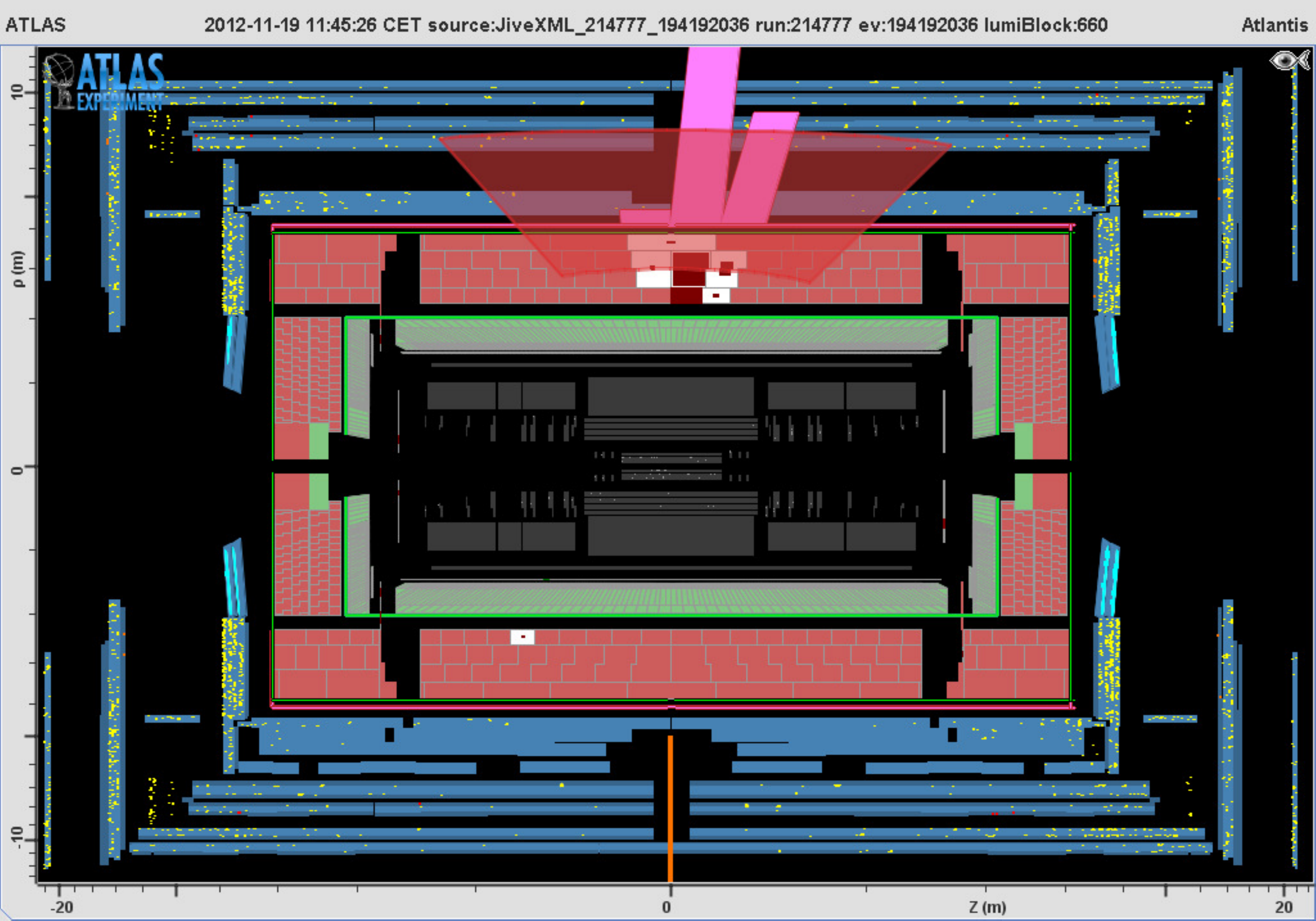} \includegraphics[width=0.48\textwidth]{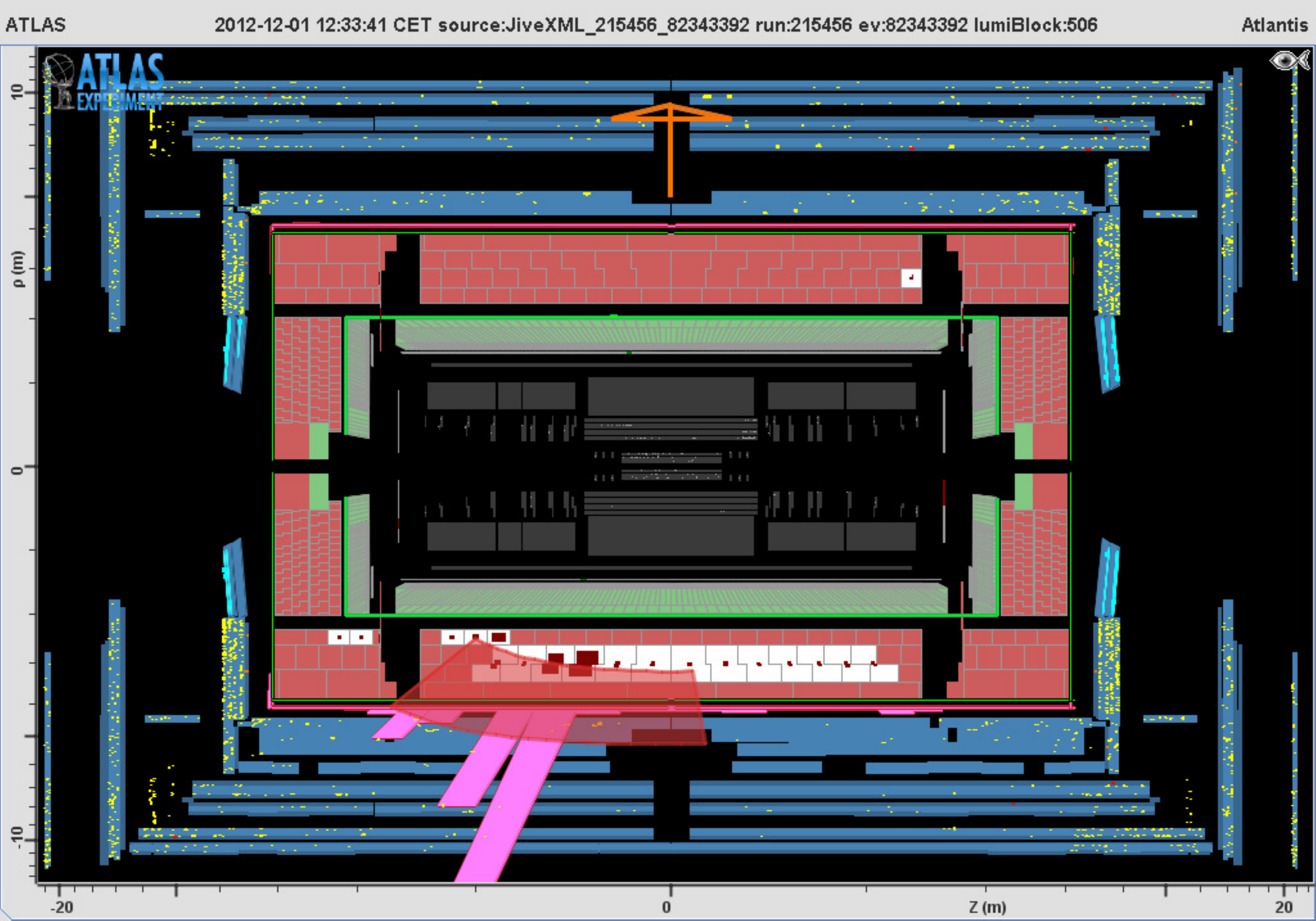} \end{center}
 \caption{Some candidate event displays from 2011 (top) and 2012 (bottom) data passing all selections.
White squares filled with red squares show reconstructed energy deposits in TileCal cells above noise threshold (the fraction of red area indicates the amount of energy in the cell), purple bars show a histogram of total energy in projective TileCal towers, and jets are shown by red semi-transparent trapezoids. Muons segments are drawn but none are reconstructed in these events. \met is shown as an orange arrow.
 \label{fig:candidatedisplays}}
\end{figure*}

\begin{figure*}[htbp]
 \begin{center}
\includegraphics[width=0.48\textwidth]{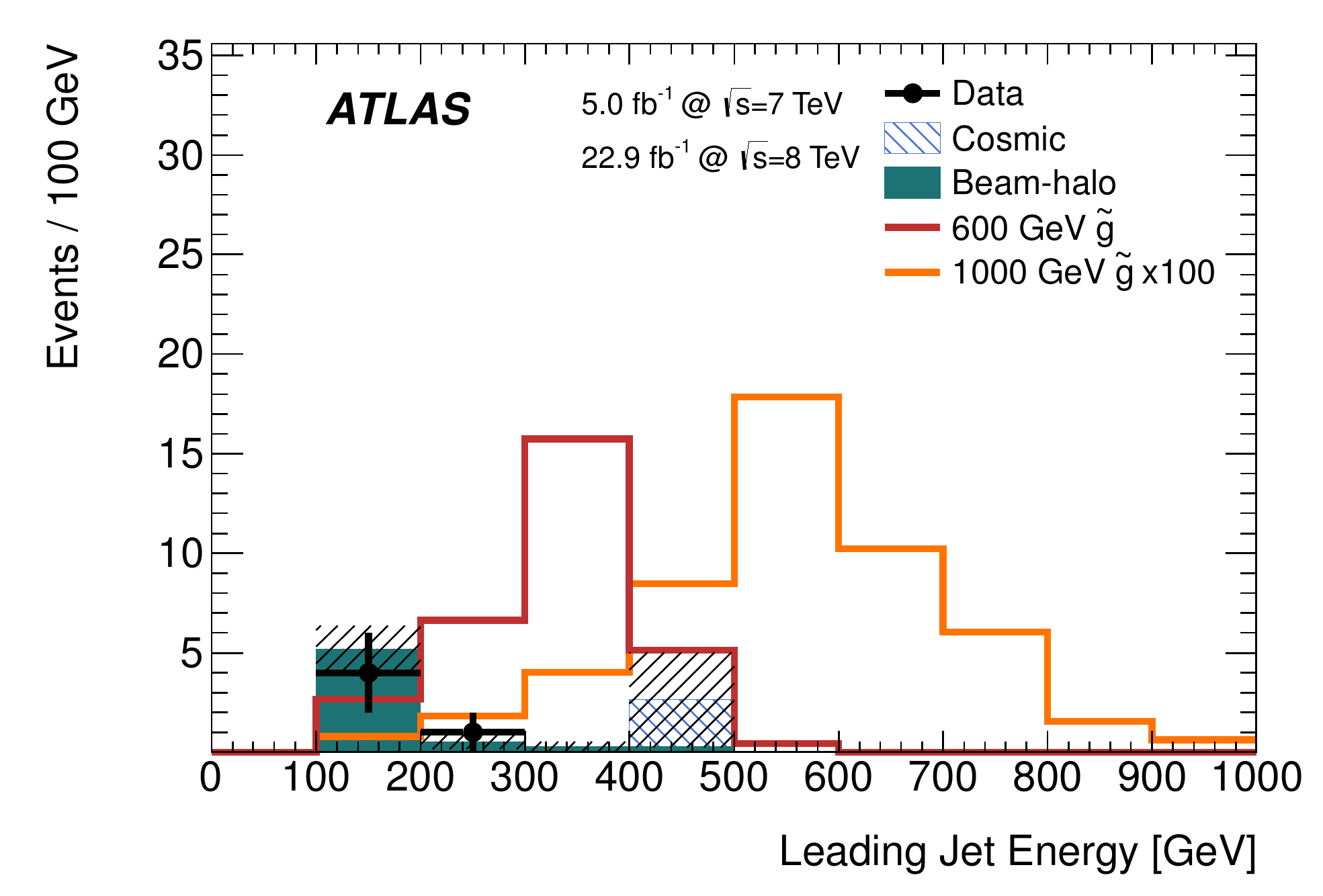} \includegraphics[width=0.48\textwidth]{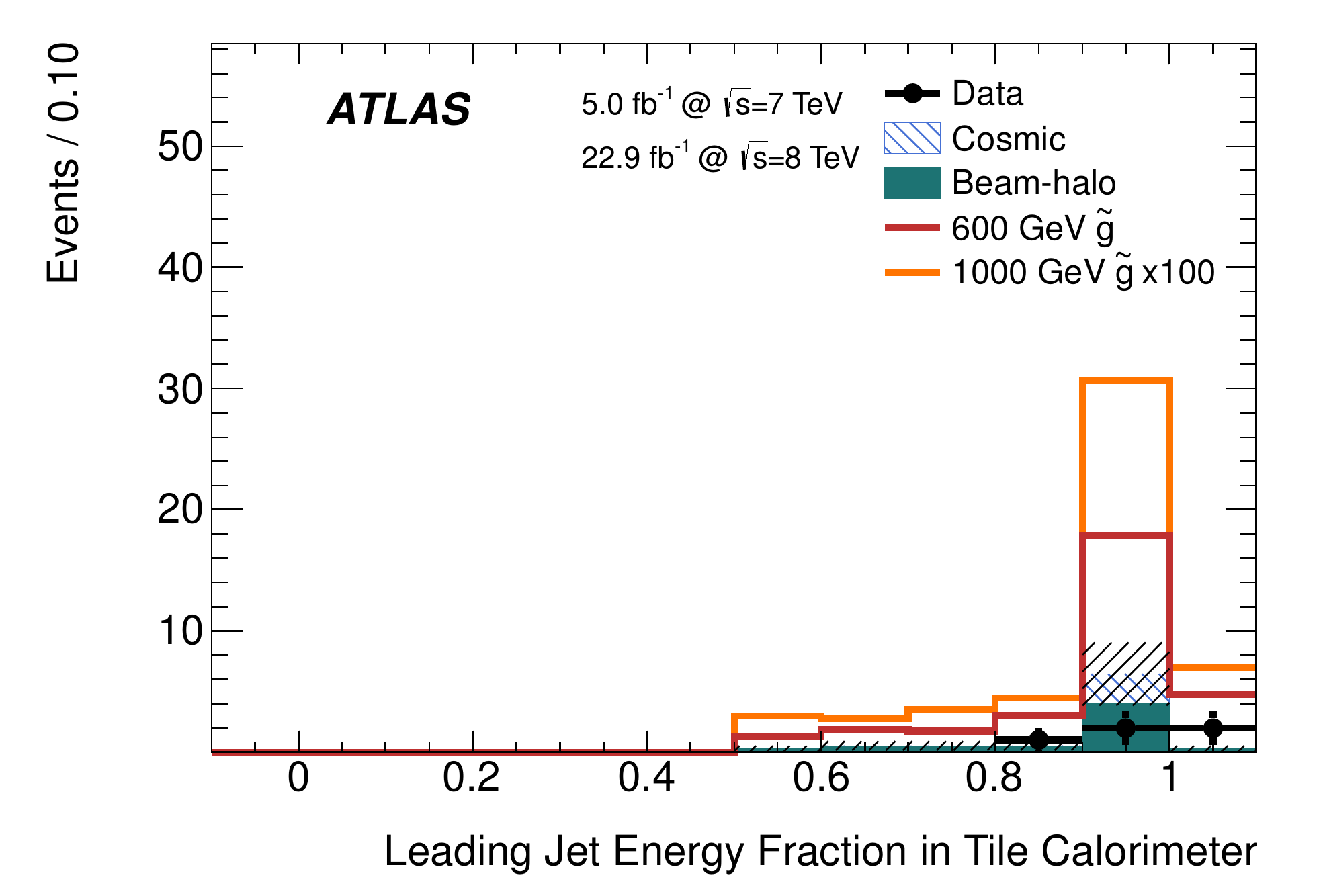}
\includegraphics[width=0.48\textwidth]{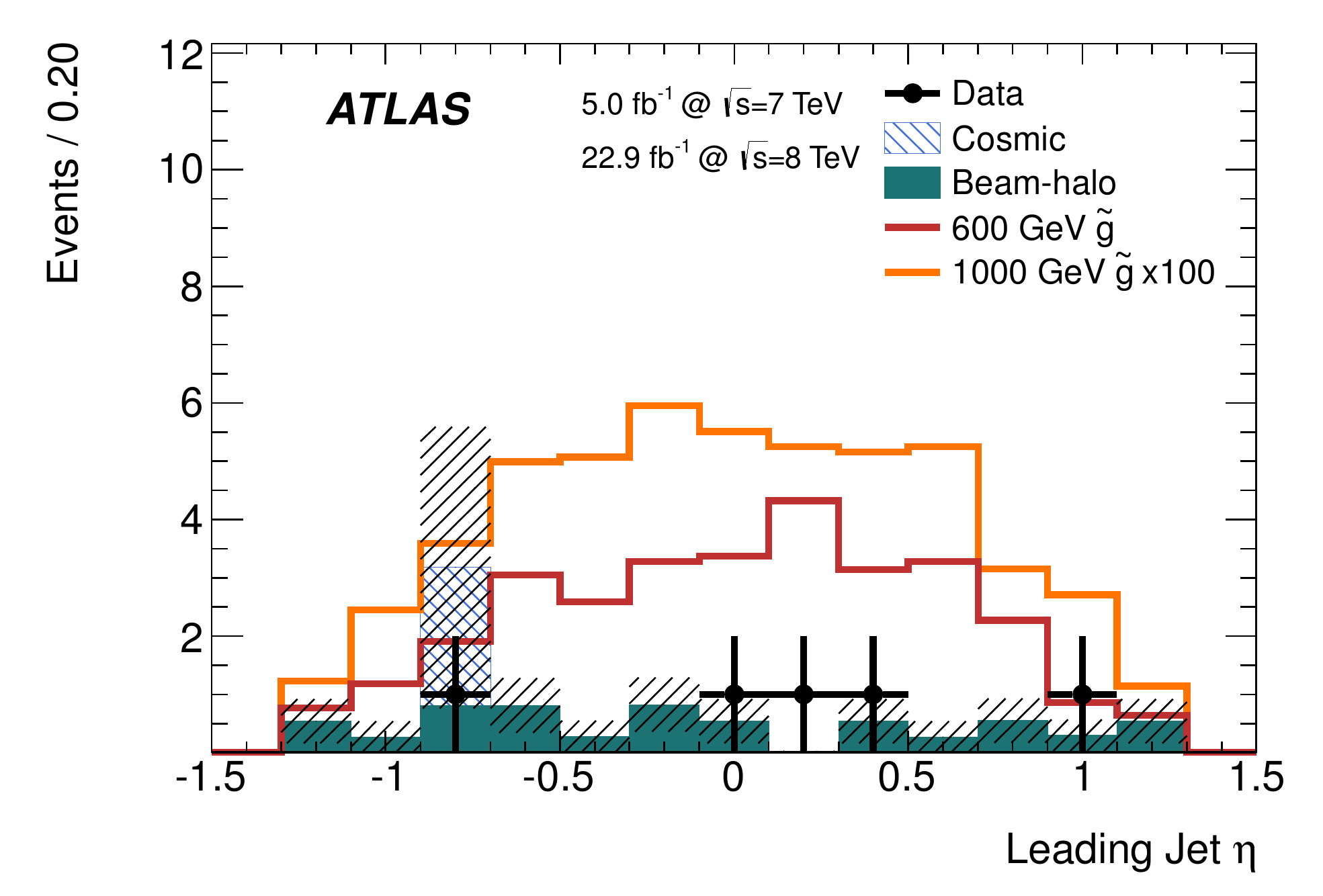} \includegraphics[width=0.48\textwidth]{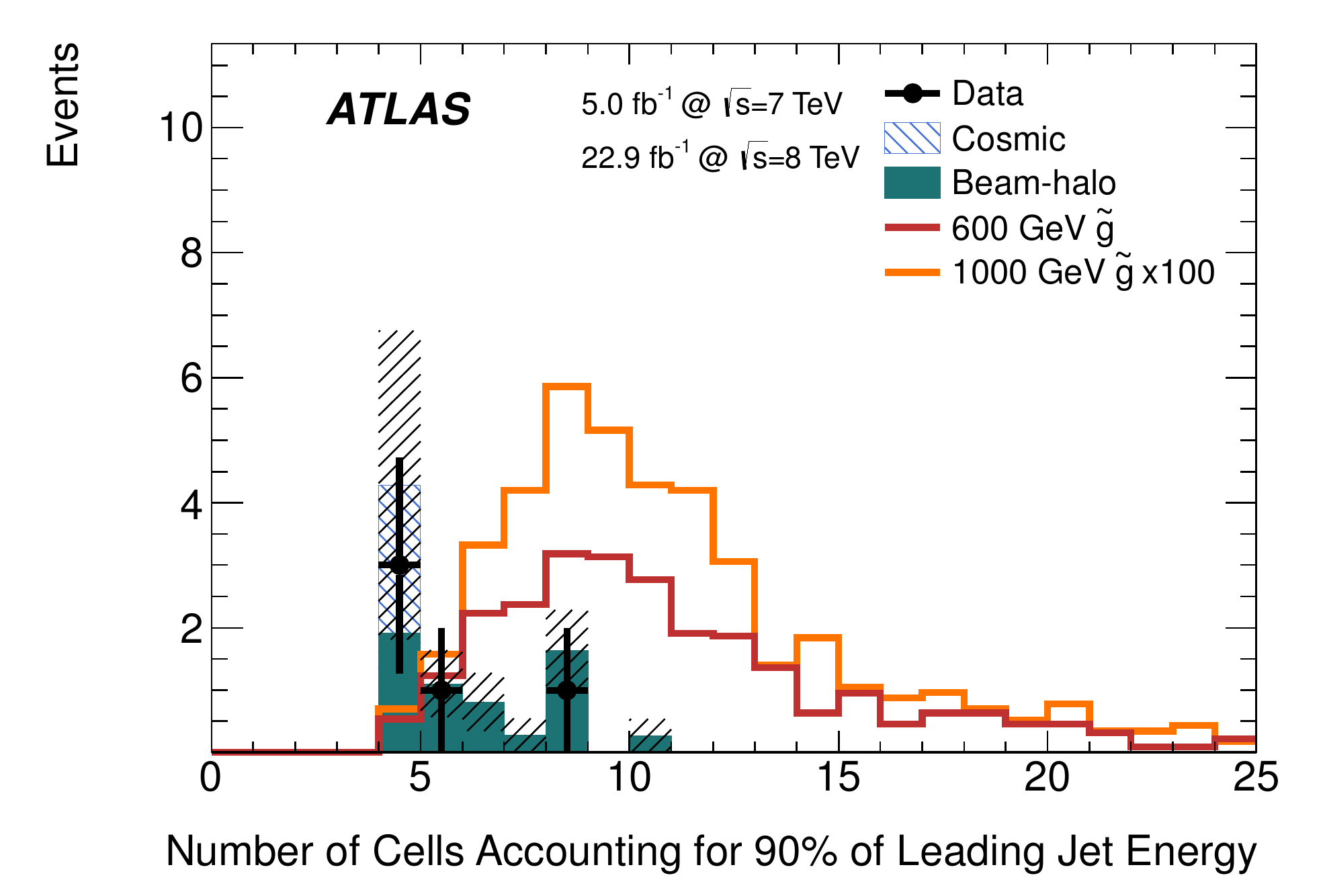}
 \includegraphics[width=0.48\textwidth]{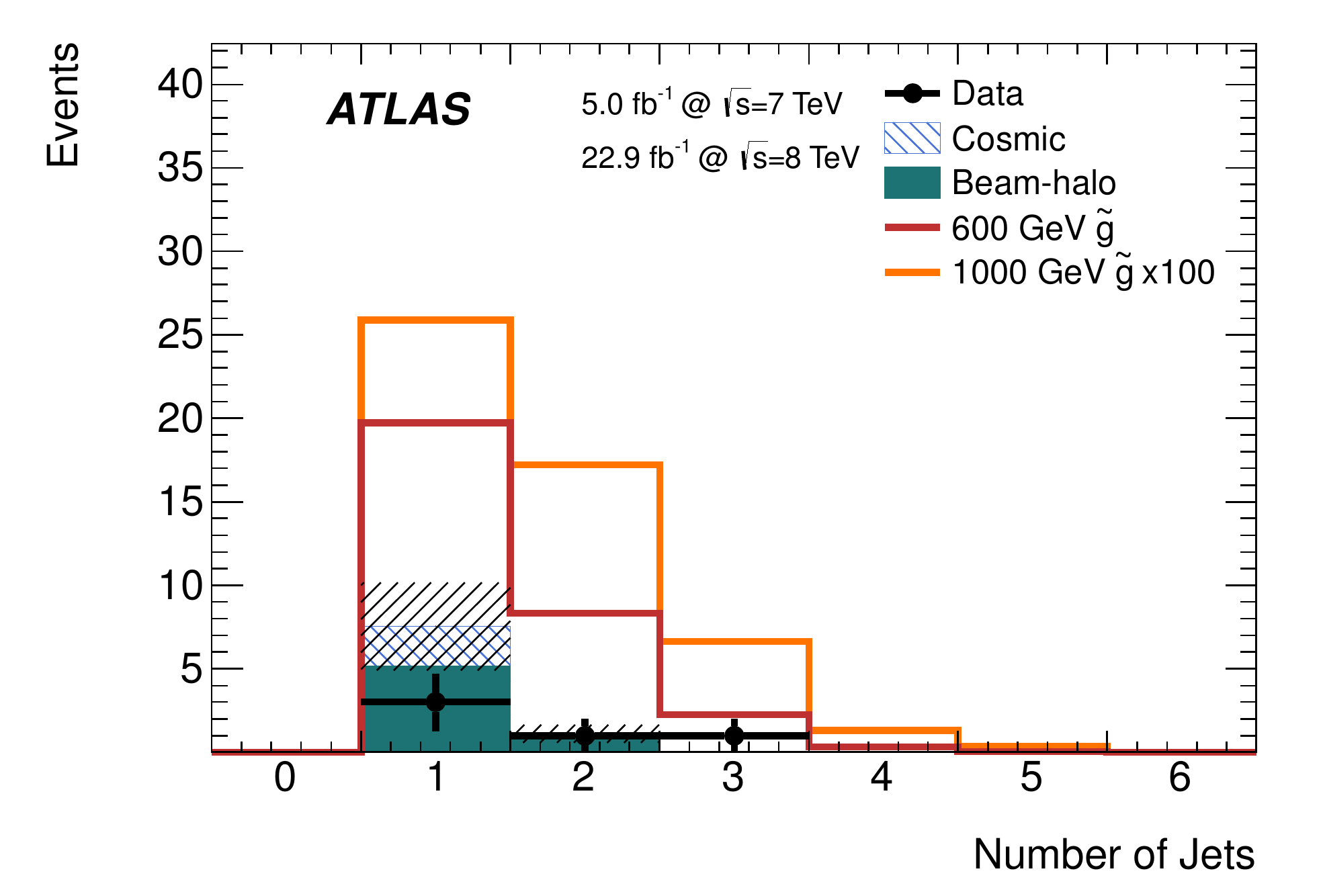} \includegraphics[width=0.48\textwidth]{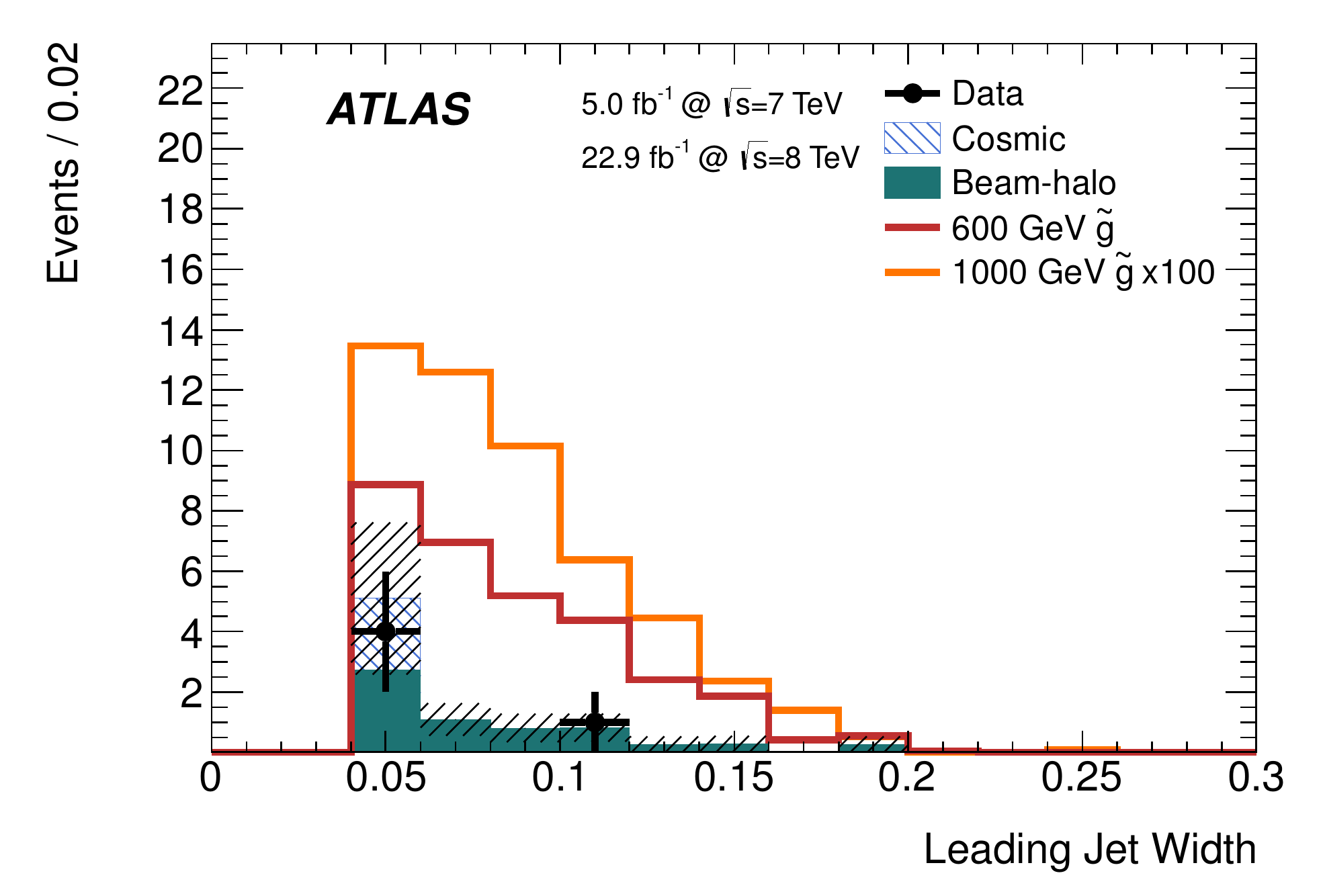}
 \caption{
The event yields in the signal region for candidates with 
all selections (in Table~\ref{tab:cutflow}) except jet energy $>300$~GeV. 
All samples are scaled to represent their anticipated yields in the search
region. The top hashed band shows the total statistical uncertainty on the background estimate.
 \label{fig:signalRegion1}}
 \end{center}
\end{figure*}

Some candidate event displays are shown in Fig.~\ref{fig:candidatedisplays}.
Distributions of jet variables are plotted for events in the jet energy $>100$~GeV signal region
after applying all selection criteria 
and are compared to the estimated backgrounds in Fig.~\ref{fig:signalRegion1}. The
shapes of these distributions and event yields are consistent. Table~\ref{tab:limits} shows the signal region event yields and background estimates. 
There is no evidence of an excess of events over the background estimate.

\section{Contributions to Signal Efficiency}
\label{sec:efficiency}

Quantifying the signal efficiency for the stopped $R$-hadron search
presents several unique challenges due to the non-prompt nature of
their decays. Specifically there are four sources of inefficiency:
stopping fraction (Sec.~\ref{sec:simulation}), reconstruction efficiency (Table~\ref{tab:cutflow_signal_summary}), 
accidental muon veto, and probability to have the decay occur in an empty crossing (timing acceptance).
Since the first two have been discussed above, only the accidental muon veto and timing acceptance are described here.

\subsection{Accidental Muon Veto}
Operating in the empty crossings has the significant advantage
of eliminating collision backgrounds.  However because such a
stringent muon activity veto is employed, a significant number
of events are rejected in the offline analysis 
due to spurious track segments in the muon system, which are not modeled in the signal simulation.  
Both $\beta$-decays from activated nuclei and $\delta$-electrons could produce segments with more than four muon station measurements.
This effect is separate from a signal decay producing a muon segment
that then vetoes the event. To study the rate of muon segments, events from the empty random trigger data in 2011 and 2012 are examined as a function of run number, since the effect can depend strongly on beam conditions. The rate of these events that have a muon segment from noise or other background is calculated. The efficiency per run to pass the muon segment veto is applied on a live-time weighted basis to the cosmic background estimate and varies from 98\% at the start of 2011 to 70\% at the end of 2012.
It is still applied to the cosmic background estimate after the muon veto, since the probability to have the cosmic background event pass the analysis selections and contribute to the signal region events depends on it passing the muon veto. 
The beam-halo background estimate already implicitly accounts for this effect across run periods.
For the signal, this effect is accounted for inside the timing acceptance calculation, on a per-run basis.

\subsection{Timing Acceptance}
The expected signal decay rate does
not scale with instantaneous luminosity. Rather, at any moment in
time, the decay rate is a function of the hypothetical $R$-hadron lifetime and
the entire history of  delivered luminosity. For example, for longer $R$-hadron lifetimes the decay rate
anticipated in today's run is boosted by luminosity
delivered yesterday. To address the complicated
time behavior of the $R$-hadron decays, a timing acceptance is defined
for each $R$-hadron lifetime hypothesis, $\epsilon_T(\tau)$, as the number of
$R$-hadrons decaying in an empty crossing divided by the total
number that stopped. The $\epsilon_T(\tau)$ factor thus accounts for the full history of the delivered luminosity and live time recorded in empty crossings. 
This means the number of $R$-hadrons expected to be reconstructed is 
$L \times \sigma \times \epsilon_{\rm{stopping}} \times \epsilon_{\rm{recon}} 
\times \epsilon_T(\tau)$, where $L$ is the integrated luminosity, $\sigma$ is the $R$-hadron production cross section weighted by integrated luminosity at 7 and 8~TeV, $\epsilon_{\rm{stopping}}$ is the stopping fraction, and $\epsilon_{\rm{recon}}$ is the reconstruction efficiency.

\begin{figure}[htbp]
 \begin{center}
   \includegraphics[width=0.48\textwidth]{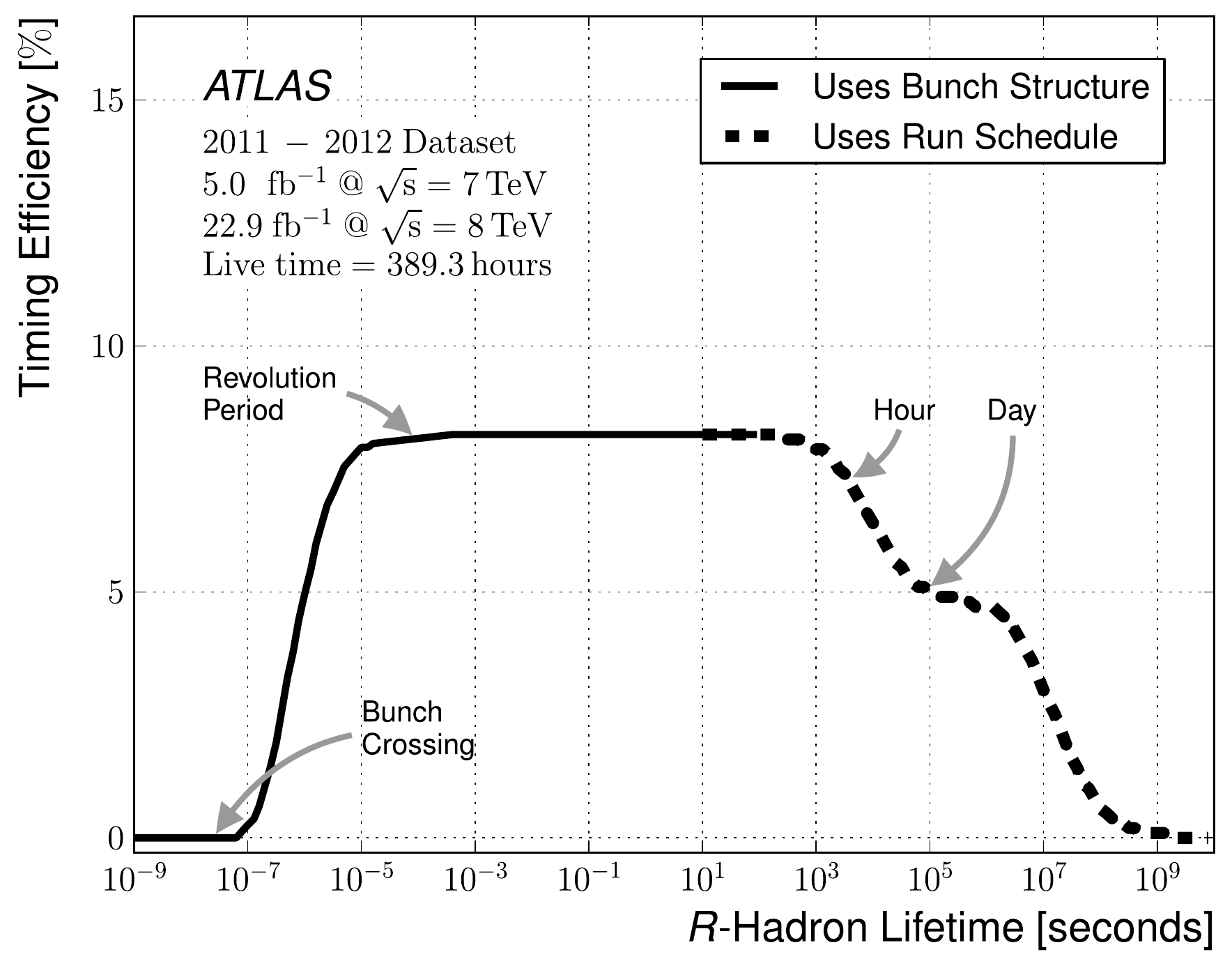}
 \caption{
The timing acceptance for signal as a function of $R$-hadron
lifetime (in seconds). This corresponds to the $\epsilon_T(\tau)$
function described in the text. \label{fig:TimingEfficiency}}
 \end{center}
\end{figure}

To calculate the timing acceptance for the actual 2011 and 2012 LHC and ATLAS run
schedule, measurements are combined of the delivered luminosity in each BCID, the bunch structure of each LHC fill, 
and the live time recorded in empty crossings during each fill, all kept in the ATLAS online conditions database. The efficiency calculation is split into short and
long $R$-hadron lifetimes, to simplify the calculation. For $R$-hadron lifetimes less than 10 seconds, the bunch structure is taken into account, but not the possibility that an $R$-hadron produced in one run could decay in a later one. For longer $R$-hadron lifetimes,
the bunch structure is averaged over, but the chance that stopped $R$-hadrons from one run decay in a later one is considered.
The resulting timing acceptance is presented in Fig.~\ref{fig:TimingEfficiency}.

\begin{table*}
\begin{center}
    \caption{The number of expected and observed events (with statistical uncertainties) corresponding to each of the selection criteria. The number of expected cosmic events in the search region is calculated by subtracting the expected number of beam-halo background events in the cosmic region (see Table~\ref{tab:bh_estimation}) from the number of events observed in the cosmic region (see Table~\ref{tab:cutflow}) and then scaling to the search region. The scaling is simply the ratio of recorded empty live times for the two regions (see Table~\ref{tab:datasamples}) before the muon segment veto is applied. After the muon segment veto is applied, the scaling also accounts for the difference in the muon segment veto efficiency between the cosmic and search regions.
    \label{tab:limits}}
    \begin{tabular}{cc|cc|c|ccc|c}
    \hline\hline
Leading jet      & Muon & \multicolumn{3}{|c}{Cosmic region}  & \multicolumn{4}{|c}{Number of events in search region}       \\
energy (GeV)  & veto  &  Events & Beam-halo bkgd. & Scaling & Cosmic & Beam-halo  & Total background & Observed  \\
    \hline
50 & No     & 1640& 82$\pm$40                     & 3.1     &4820 $\pm$ 570  & 900   $\pm$ 130   & 5720  $\pm$ 590  &5396 \\
50 &Yes     & 2      & 1.1$\pm$0.6                  & 2.4     &2.1  $\pm$ 3.6     & 12   $\pm$ 3         & 14.2  $\pm$ 4.0   & 10 \\
100&Yes    & 1      & 0.8$\pm$0.5                  & 2.4     &0.4  $\pm$ 2.7     & 6   $\pm$ 2          & 6.4   $\pm$ 2.9    & 5 \\
300&Yes    & 1      & $0.000^{+0.01}_{-0}$ & 2.4     &2.4  $\pm$ 2.4     & 0.5  $\pm$ 0.4    & 2.9   $\pm$ 2.4    & 0 \\

\hline\hline
    \end{tabular}
\end{center}
\end{table*}


\section{Systematic uncertainties}
\label{sec:systematics}

Three sources of systematic uncertainty on the signal
efficiency are studied: the $R$-hadron interaction with matter, the out-of-time
decays in the calorimeters, and the effect of the selection
criteria. The total uncertainties, added in quadrature, are shown in Table~\ref{tab:cutflow_signal_summary}.
In addition to these, a $2.6\%$ uncertainty is assigned to the luminosity measurement~\cite{Atlas_lumi}, fully correlated between the 2011 and 2012 data.
To account for occasional dead-time due to high trigger rates,
a 5\% uncertainty is assigned to the timing acceptance; this accounts
for any mismodeling of the accidental muon veto as well.
The gluino, stop, or sbottom pair-production cross-section uncertainty is not included as a systematic uncertainty but is used when extracting limits on their mass by finding the intersection with the cross section $-1\sigma$ of its uncertainty.

\subsection{$R$-hadron--Matter Interactions}
The various simulated signal samples are used to estimate the
systematic uncertainty on the stopping fraction due to the scattering model.
There are two sources of theoretical uncertainty: the spectrum of
$R$-hadrons and nuclear interactions. To estimate the effect from
different allowed $R$-hadron states, three different scattering models
are employed: generic, Regge, and intermediate (see Sec.~\ref{sec:simulation}).  Each allows a
different set of charged states that affect the $R$-hadron's
electromagnetic interaction with the calorimeters. Since these models have large differences for the $R$-hadron stopping fraction, limits are quoted separately for each model, rather than including the differences as a systematic uncertainty on the signal efficiency.
There is also uncertainty from the modeling of nuclear
interactions of the $R$-hadron with the calorimeter since these can 
affect the stopping fraction. The effect is estimated by recalculating the stopping fraction after doubling and halving the nuclear cross
section.  The difference
gives a relative uncertainty of $11\%$, which is used as the systematic
uncertainty in limit setting.

\subsection{Timing in the Calorimeters}

Since the $R$-hadron decay is not synchronized with a BCID it is possible that the calorimeters respond differently to the energy deposits in the simulated signals than in data. 
The simulation considers only a single BCID for each event; 
it does not simulate the trigger in multiple BCIDs and the firing of the trigger for the first BCID that passes the trigger.
In reality, a decay at $-15$~ns relative to a given BCID might fire the trigger for that BCID, or it may fire the trigger for the following BCID.
The reconstructed energy response of the calorimeter can vary between these two cases by up to 10\% since the reconstruction is optimized for in-time energy deposits.
To estimate the systematic uncertainty, the total number of simulated signal events
passing the offline selections is studied when varying the timing offset by $5$ ns in each direction (keeping
the $25$~ns range). This variation conservatively covers the timing difference observed between simulated signal jets and cosmic ray muon showers.
The minimum and maximum efficiency for each mass
point is calculated, and the difference is used as the uncertainty, which is always less than $3\%$ across all mass points.

\subsection{Selection Criteria}

The systematic uncertainty on the signal efficiency due to selection criteria is evaluated by varying each criterion up and down by its known uncertainty.
The uncertainties from each criterion are combined in quadrature
and the results are shown in Table~\ref{tab:cutflow_signal_summary}.
Varying only the jet energy scale produces most of
the total uncertainty from the selection criteria. 
The jet energy scale uncertainty is taken to be $\pm 10\%$ to allow for non-pointing $R$-hadron decays and is significantly larger than is used in standard ATLAS analyses.
Although test-beam studies showed the energy response agrees between data and simulation for hadronic showers to within a few percent~\cite{Dotti2006106}, even for non-projective showers, a larger uncertainty is conservatively assigned to cover possible differences between single pions and full jets, 
and between the test-beam detectors studied and the final ATLAS calorimeter.

\subsection{Systematic Uncertainties on Background Yield}

The systematic uncertainty on the estimated cosmic background arises from the small number of events in the cosmic background region. 
This statistical uncertainty is scaled by the same factor used to propagate the cosmic background region data yield into
expectations of background events in the search regions. 
Similarly, for the beam-halo background, a systematic uncertainty 
is assigned based on the statistical uncertainty of the estimates in the search regions.


\section{Results}
\label{sec:results}

The predicted number of background events agrees well with the observed number of events in the search region, as shown in Table~\ref{tab:limits}. 
Using these yields, upper limits on the number of pair-produced gluino, stop, or sbottom signal events are calculated with a simple event-counting method and then interpreted as a function of their masses for a given range of lifetimes.

\subsection{Limit Setting}

A Bayesian method is used to set 95$\%$ credibility-level upper limits on
the number of signal events that could have been produced. For each limit 
extraction, pseudo-experiments are run. The number of events is sampled from a Poisson distribution, with mean equal to the signal plus background expectation.  The systematic uncertainties are taken into account by varying the Poisson mean according to the effect of variations of the sources of the systematic uncertainties~\cite{2004physics9129H}. The latter variations are assumed to follow a Gaussian distribution, which is convolved with the Poisson function.
A flat prior is used for the signal strength, to be consistent with previous analyses.
A Poisson prior gives less conservative limits that are within 10\% of those obtained with the flat prior.
Since little background
is expected and no pseudo-experiment may produce fewer than zero
observed events, the distribution of upper limits is bounded
from below at $-1.15\sigma$.
The input data for the limit-setting algorithm can be seen in Table~\ref{tab:limits}.
The leading jet energy $>$ 300~GeV region is used to set the limits, except for the compressed models with a small difference between the gluino or squark mass and $m_{\nino}$, where the leading jet energy $>$ 100~GeV signal region is used.

\begin{figure*}[htbp]
 \begin{center}
\includegraphics[width=0.48\textwidth]{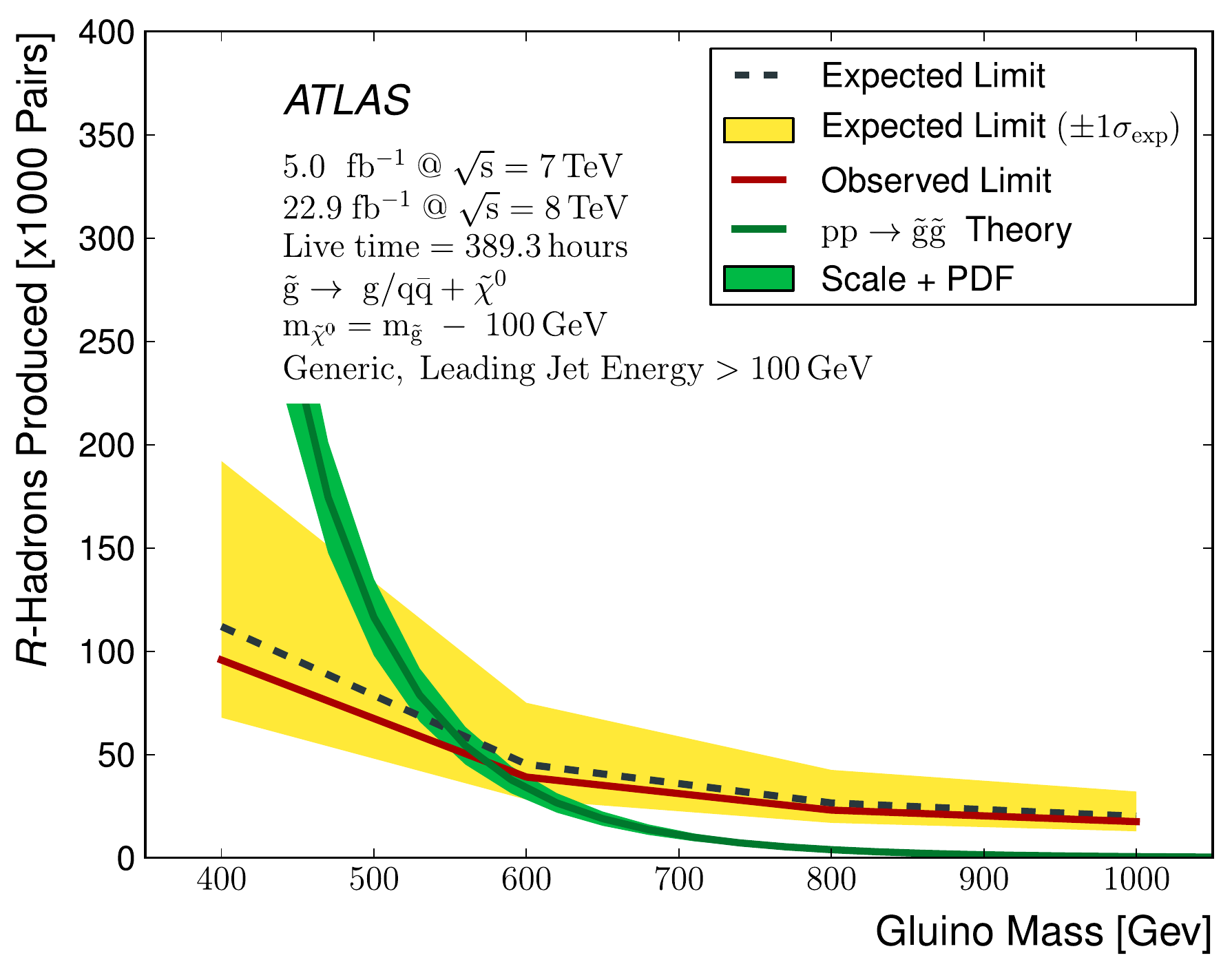} \includegraphics[width=0.48\textwidth]{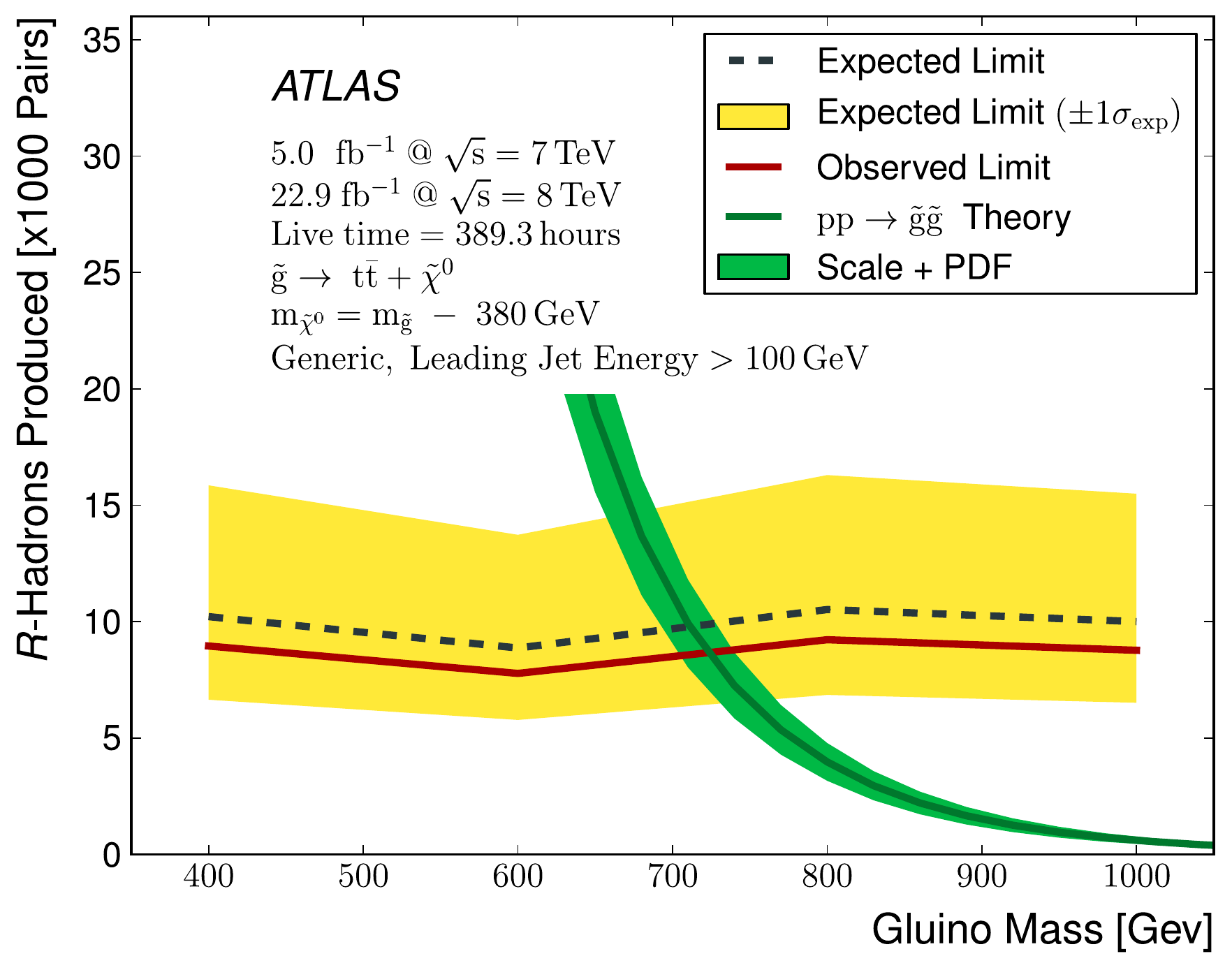}
\includegraphics[width=0.48\textwidth]{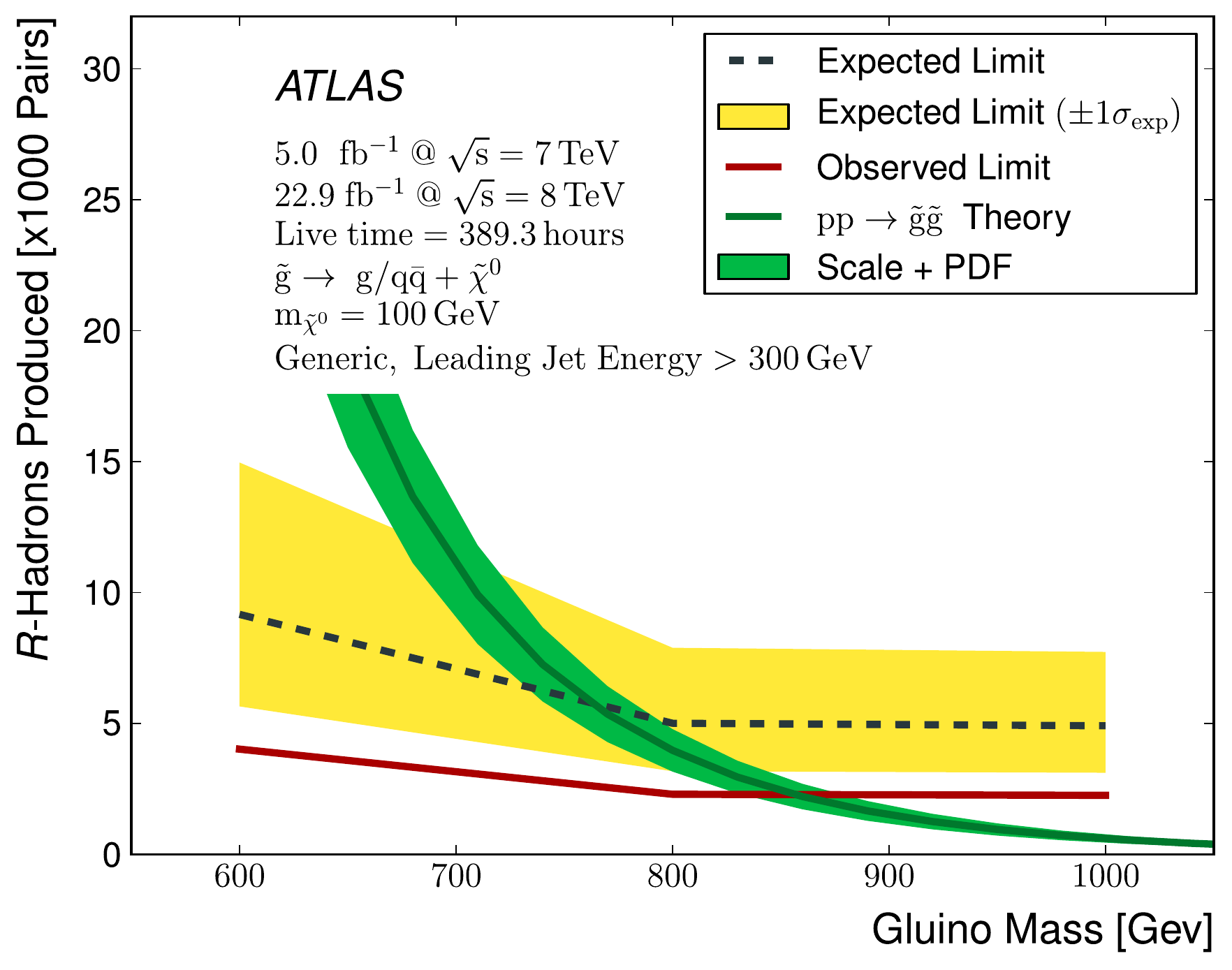} \includegraphics[width=0.48\textwidth]{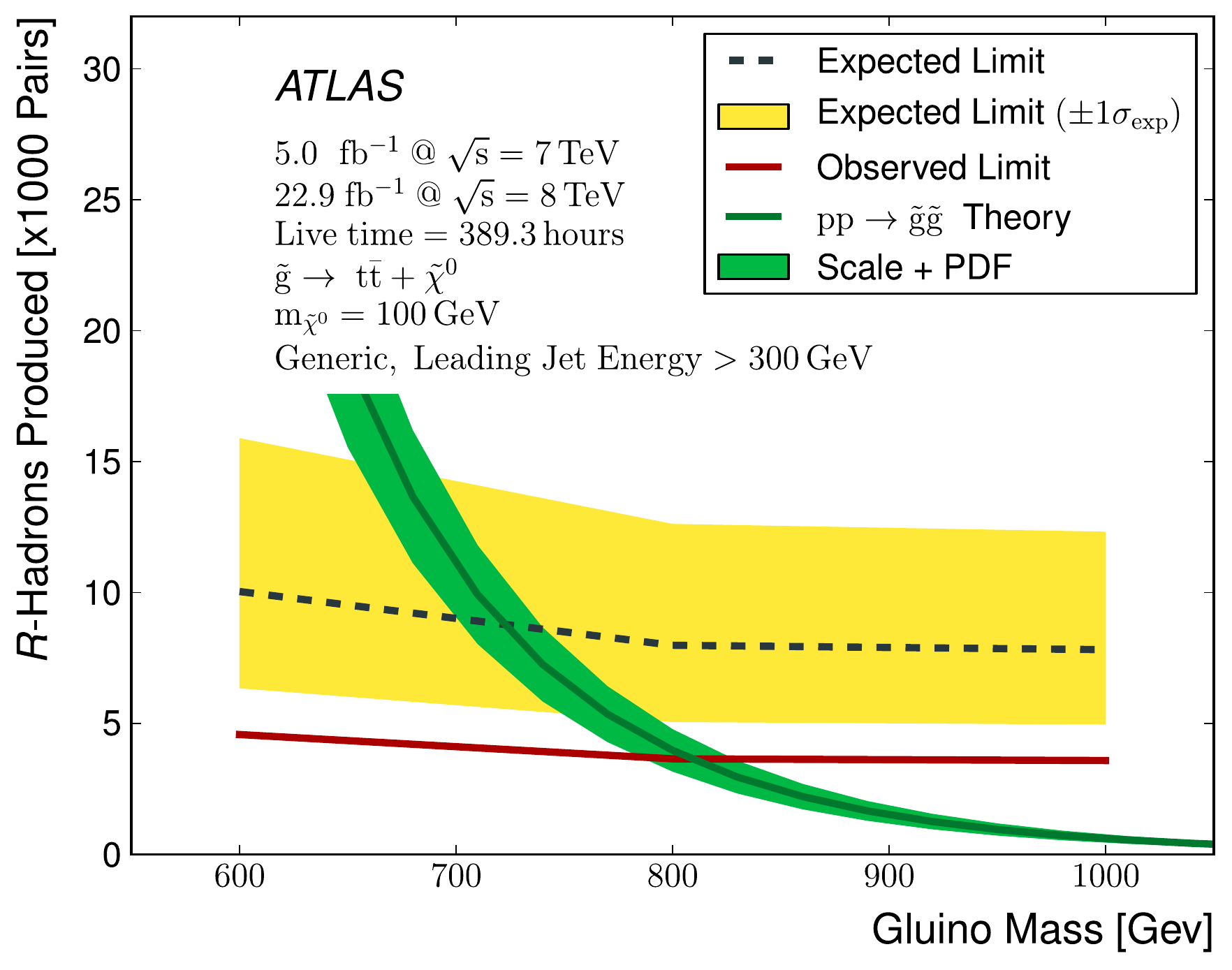}
\includegraphics[width=0.48\textwidth]{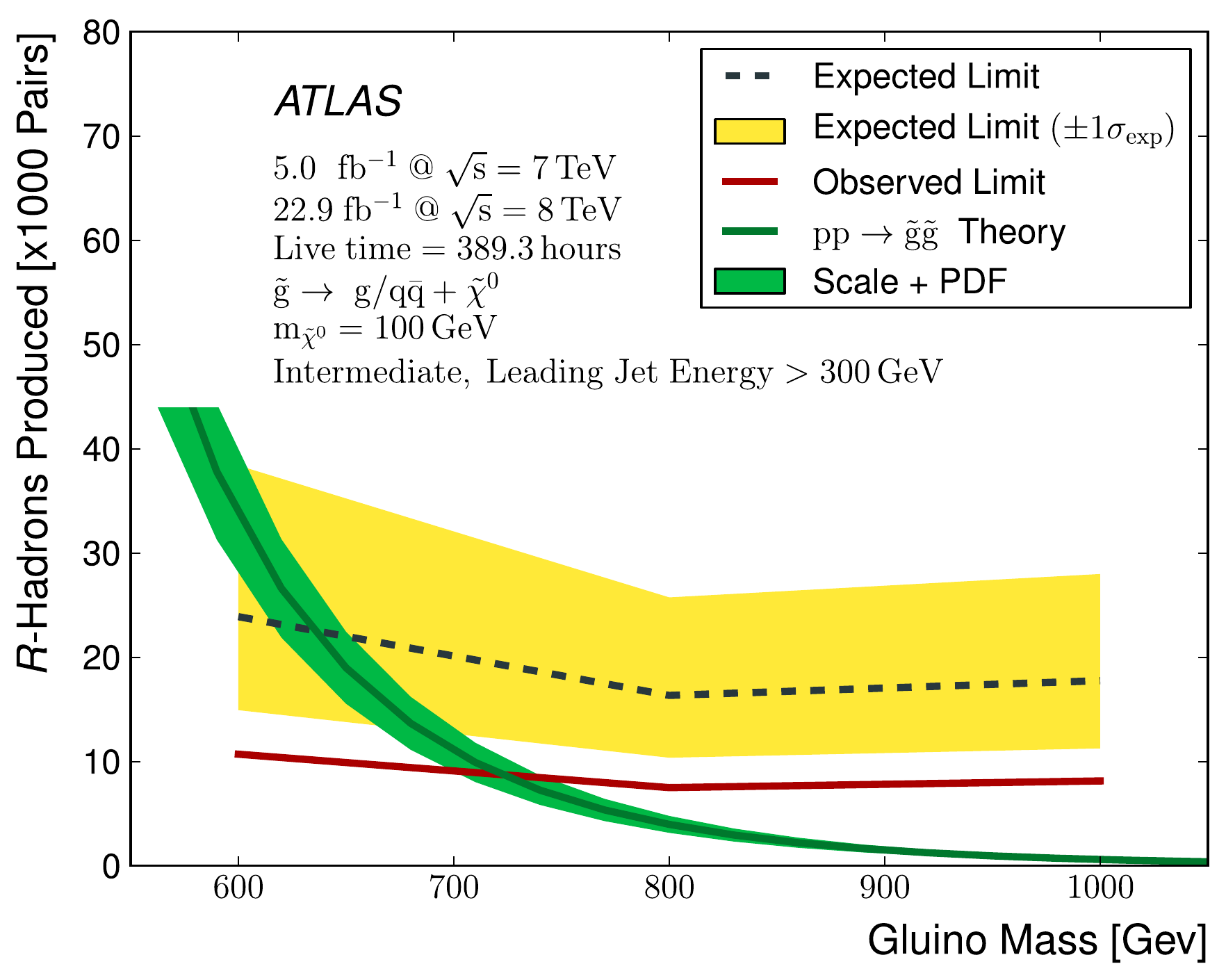} \includegraphics[width=0.48\textwidth]{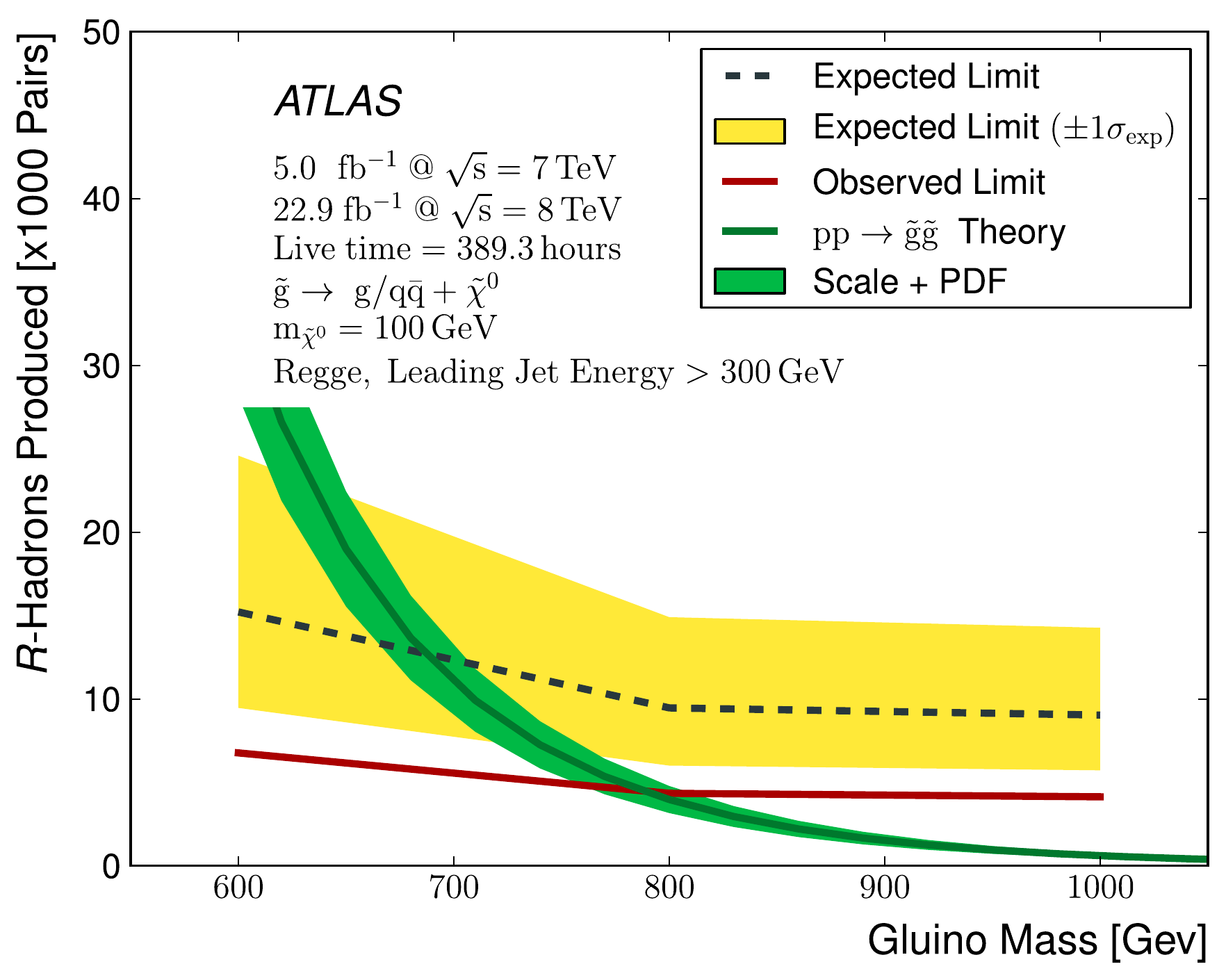}
 \caption{Bayesian upper limits on gluino events produced versus gluino mass for the various signal models considered,
with gluino lifetimes in the plateau acceptance region between $10^{-5}$ and $10^3$ seconds, compared to the theoretical expectations.
 \label{fig:mass_limits}}
 \end{center}
\end{figure*}

\begin{figure*}[htbp]
 \begin{center}
\includegraphics[width=0.48\textwidth]{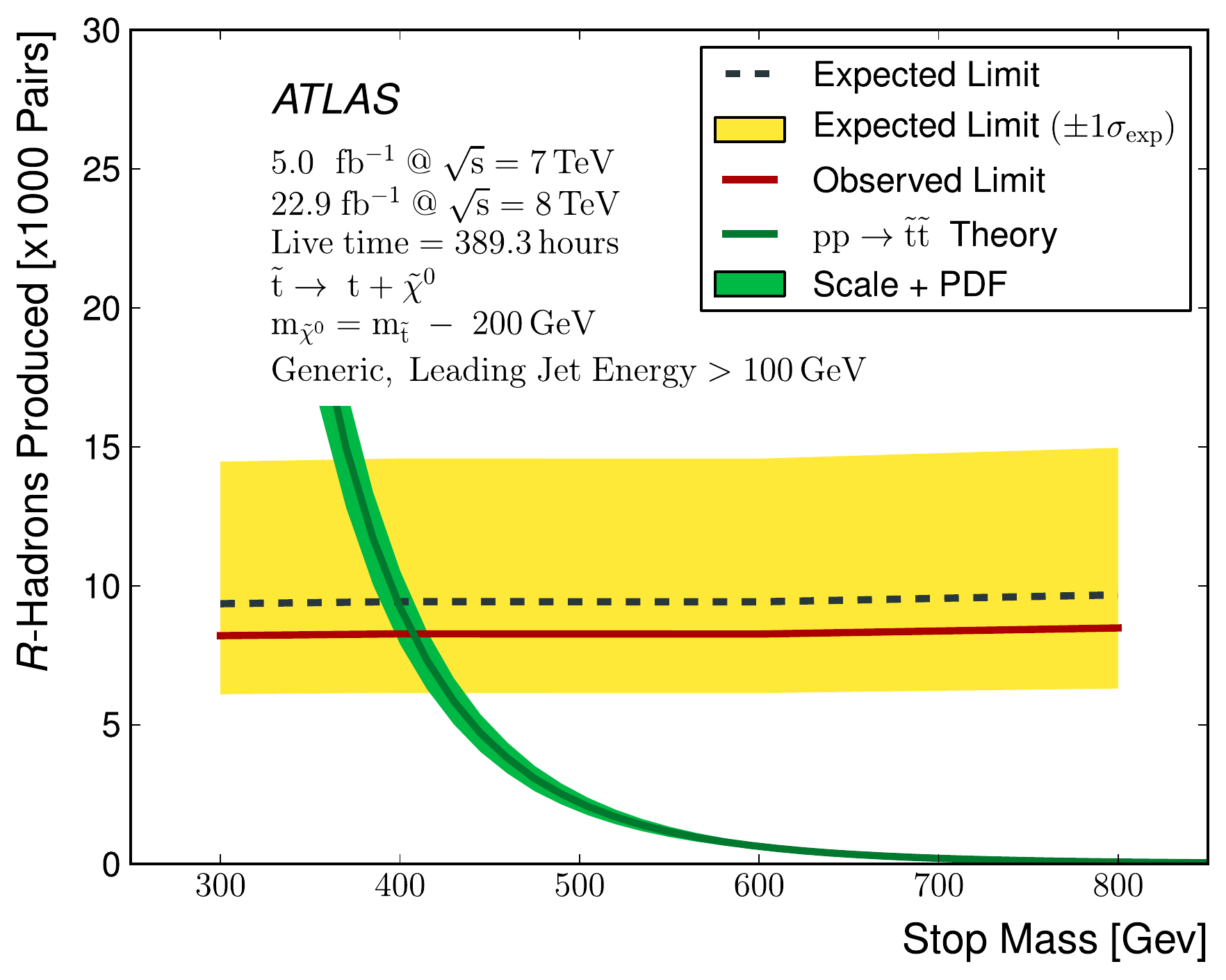}
\includegraphics[width=0.48\textwidth]{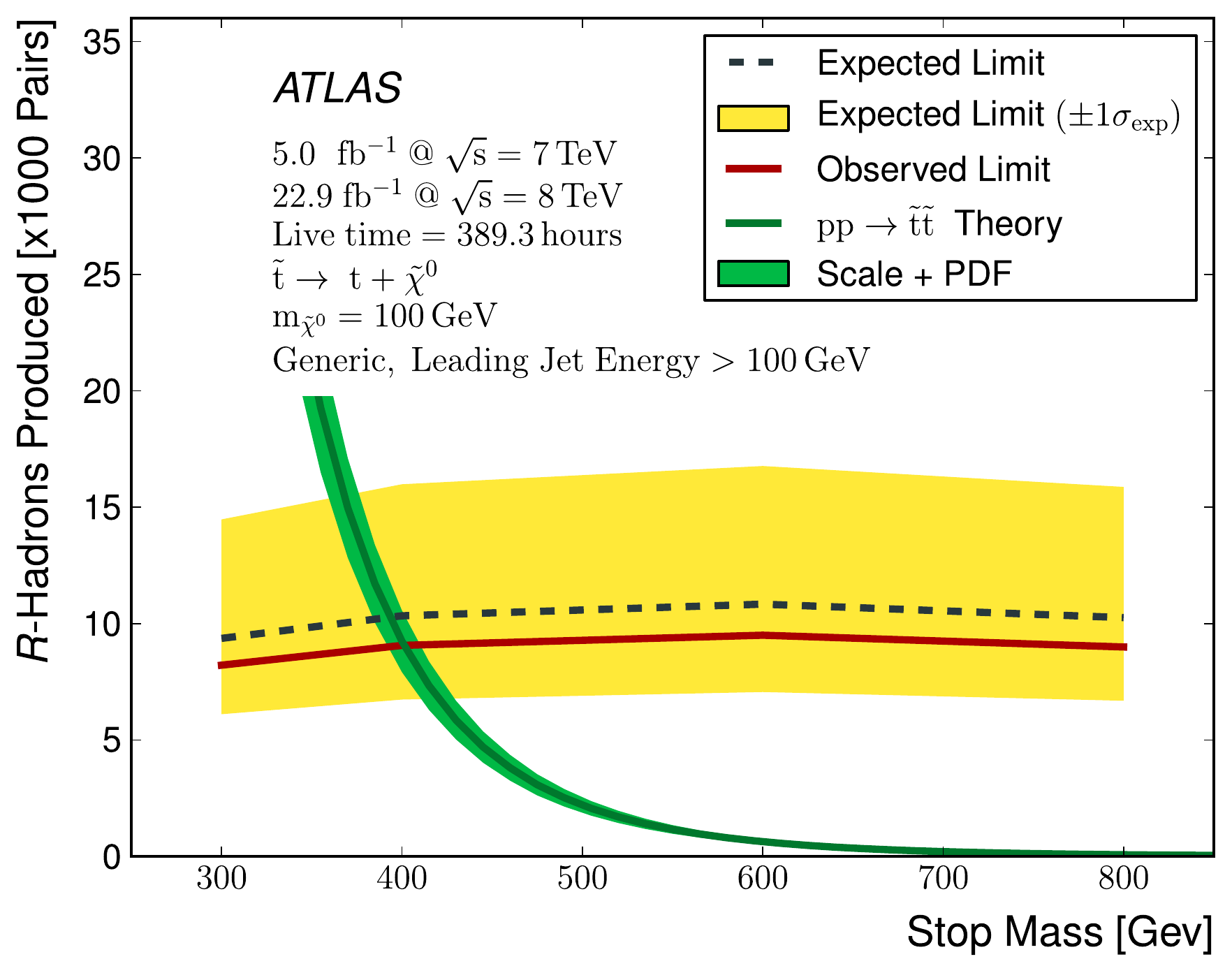}
\includegraphics[width=0.48\textwidth]{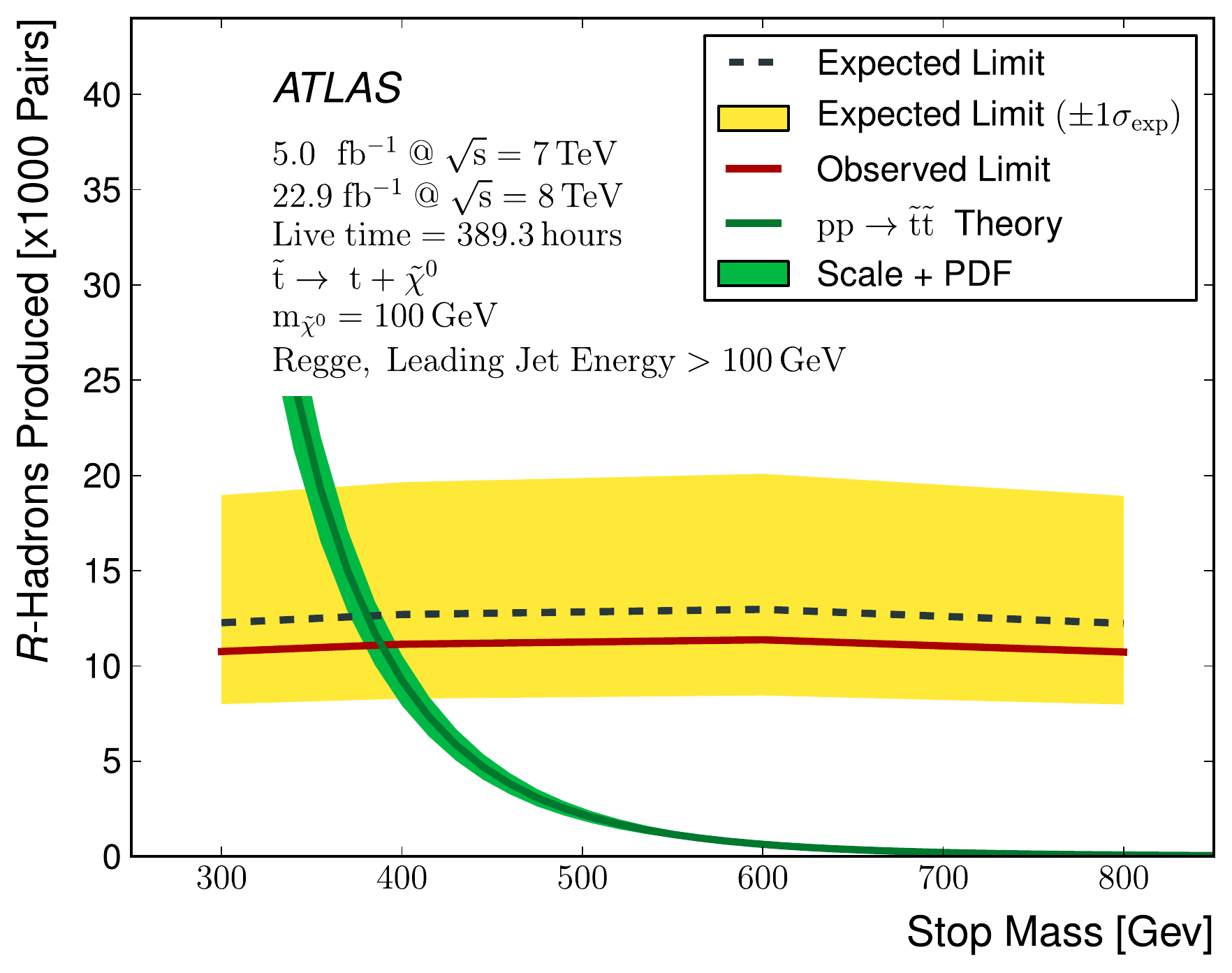}
\includegraphics[width=0.48\textwidth]{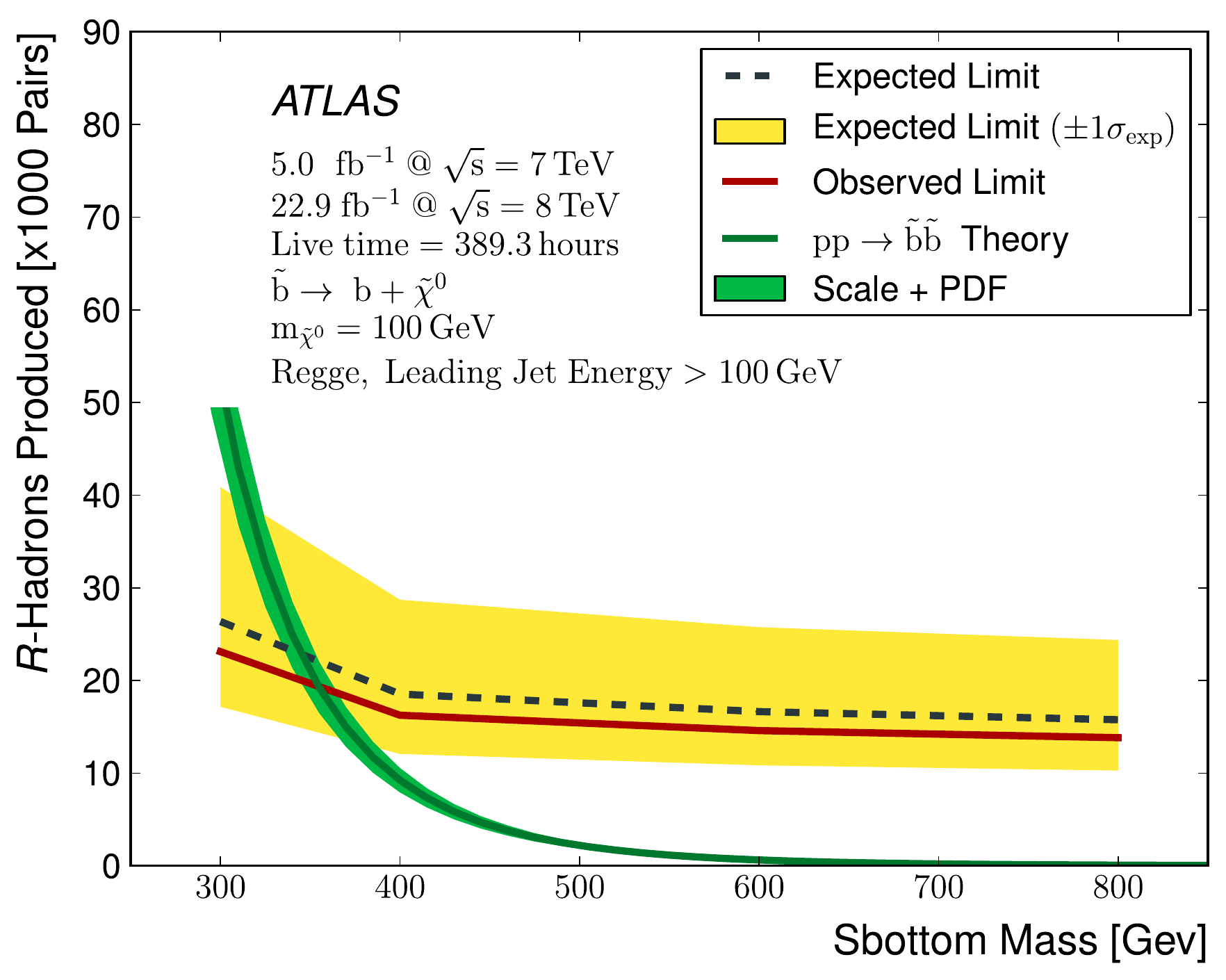}
 \caption{Bayesian upper limits on stop or sbottom events produced versus stop/sbottom mass for the various signal models considered, with stop/sbottom lifetimes in the plateau acceptance region between $10^{-5}$ and $10^3$ seconds, compared to the theoretical expectations.
 \label{fig:mass_limits_squark}}
 \end{center}
\end{figure*}

Signal cross sections are calculated to next-to-leading order in the strong coupling constant, adding the resummation of soft gluon emission at next-to-leading-logarithmic accuracy (NLO+NLL)~\cite{Beenakker:1996ch,Kulesza:2008jb,Kulesza:2009kq,Beenakker:2009ha,Beenakker:2011fu}. The nominal cross section and the uncertainty are taken from an envelope of cross-section predictions using different PDF sets and factorization and renormalization scales, as described in Ref.~\cite{Kramer:2012bx}.
The number of expected signal events is given by the signal cross sections at 7 and 8~TeV, weighted by the integrated luminosities in the 2011 and 2012 data. 
Figures~\ref{fig:mass_limits} and \ref{fig:mass_limits_squark} show the limits on the number of produced signal events 
for the various signal models considered, for $R$-hadron lifetimes in the plateau acceptance region between $10^{-5}$ and $10^3$ seconds.

\begin{figure*}[htbp]
 \begin{center}
\includegraphics[width=0.48\textwidth]{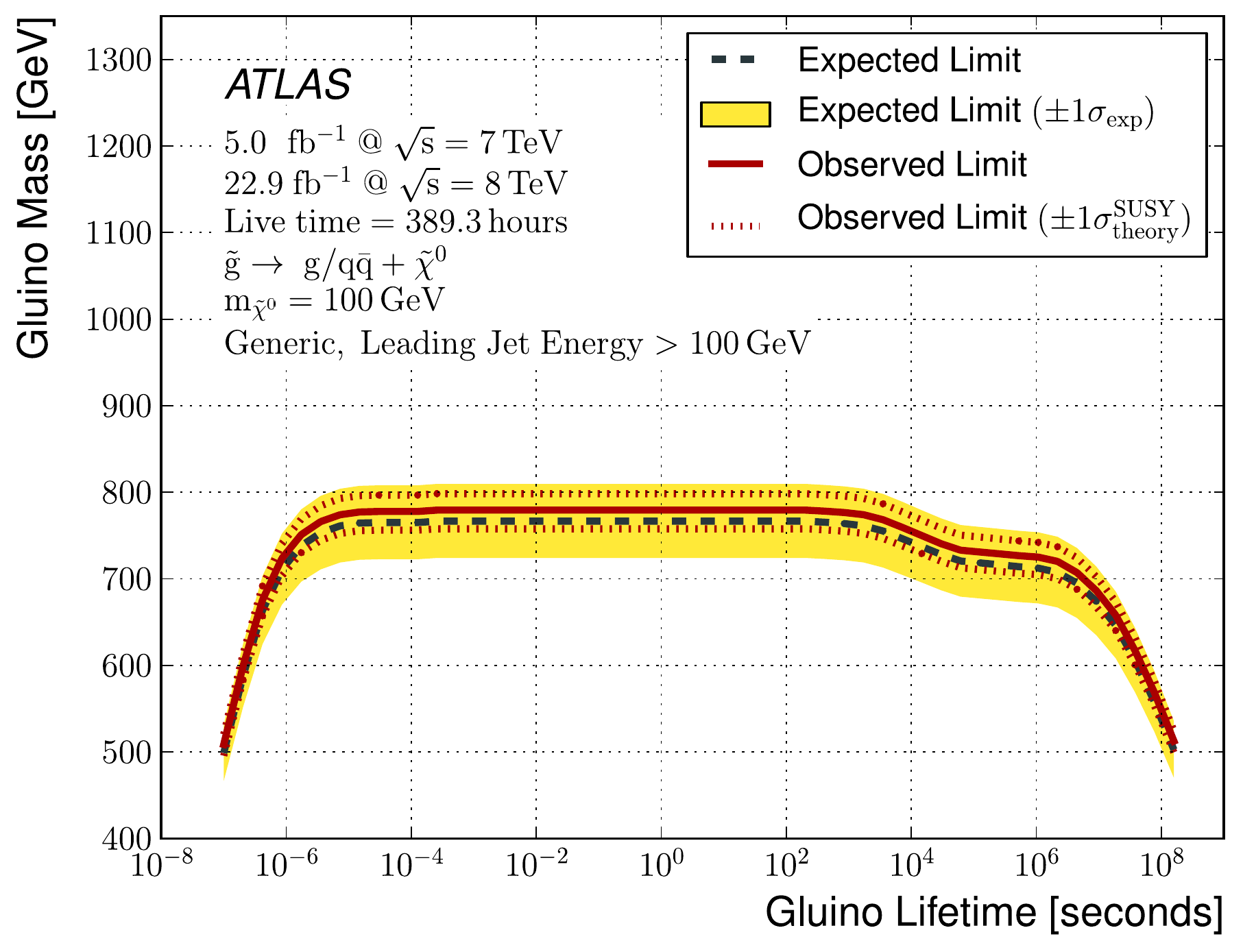} 
\includegraphics[width=0.48\textwidth]{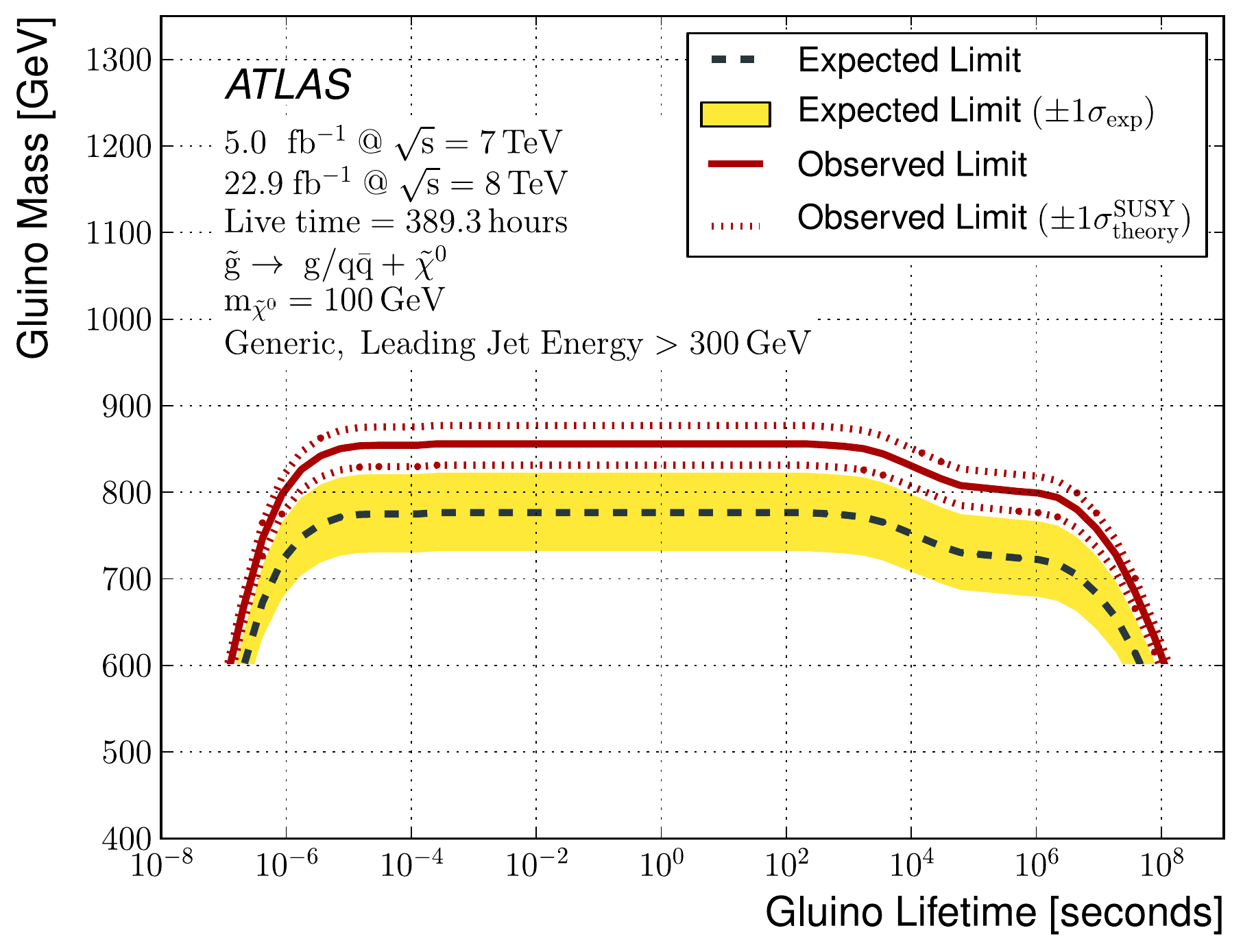}
\includegraphics[width=0.48\textwidth]{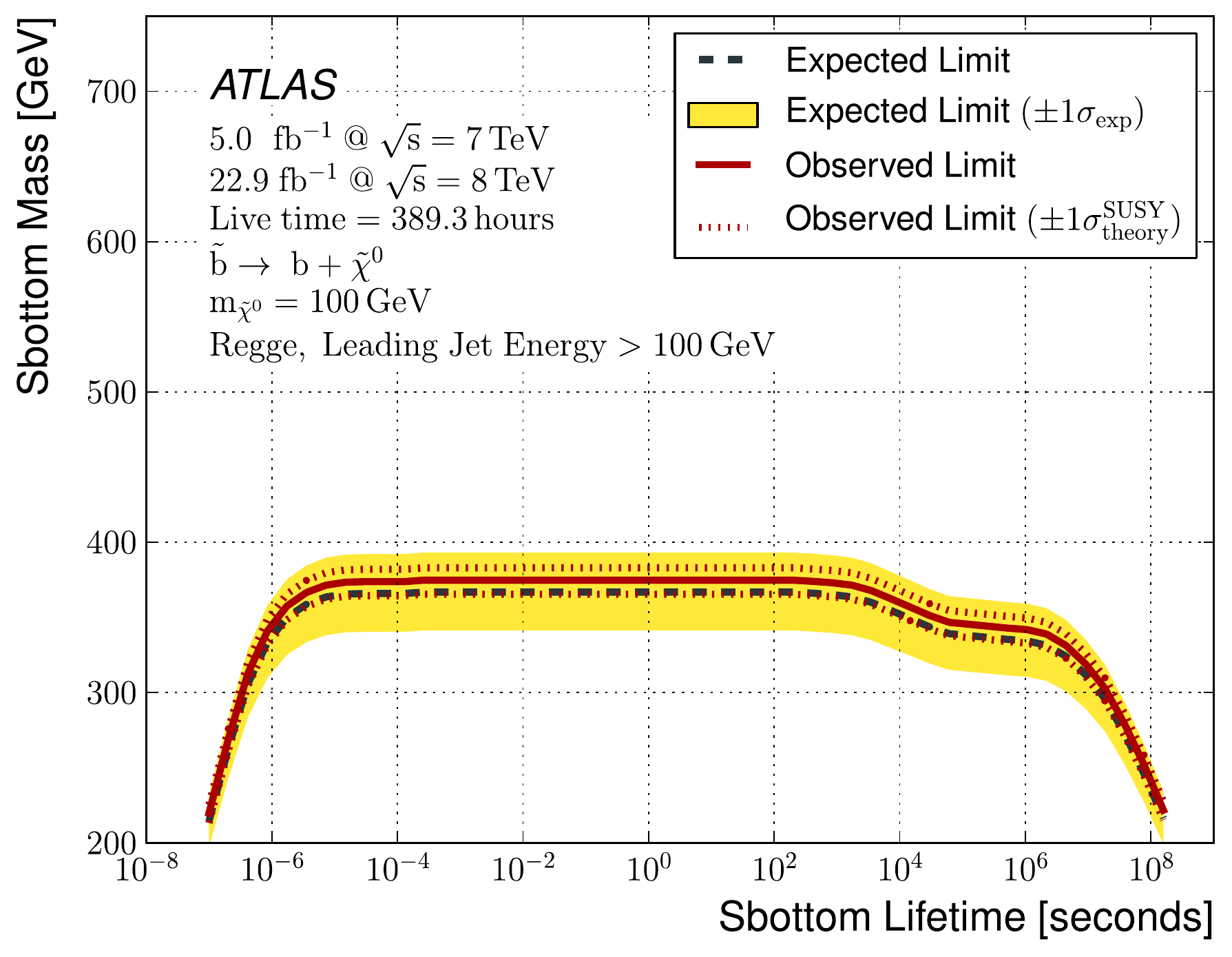}
\includegraphics[width=0.48\textwidth]{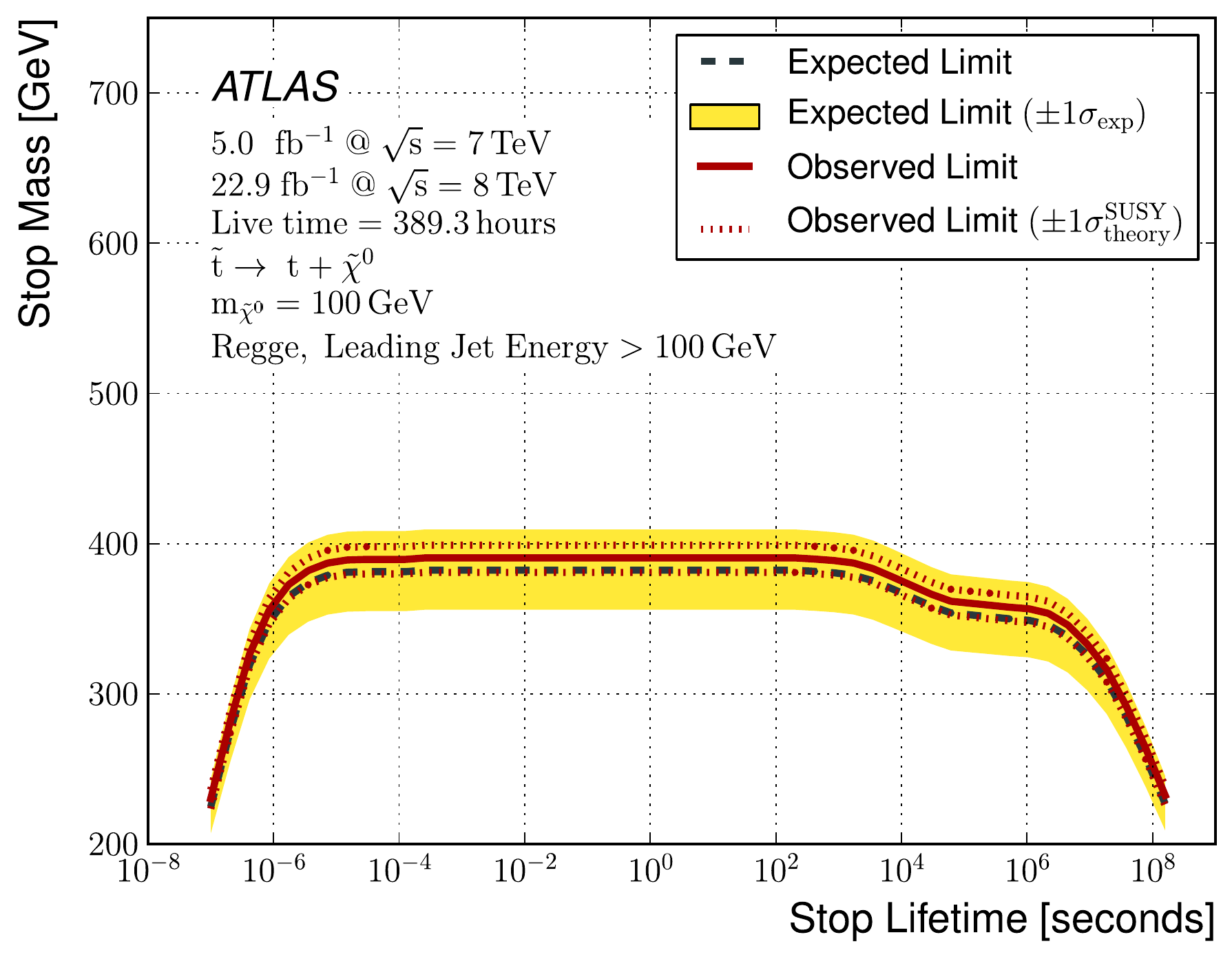}
 \caption{Bayesian lower limits on gluino, stop, or sbottom mass versus its lifetime, for the two signal regions, with $R$-hadron lifetimes in the plateau acceptance region between $10^{-5}$ and $10^3$ seconds. An 800~GeV gluino (stop or sbottom), in the generic (Regge) $R$-hadron model is used as a reference for the stopping fraction and reconstruction efficiency.
 \label{fig:mass_limits_tau}}
 \end{center}
\end{figure*}

\begin{figure*}[htbp]
 \begin{center}
\includegraphics[width=0.48\textwidth]{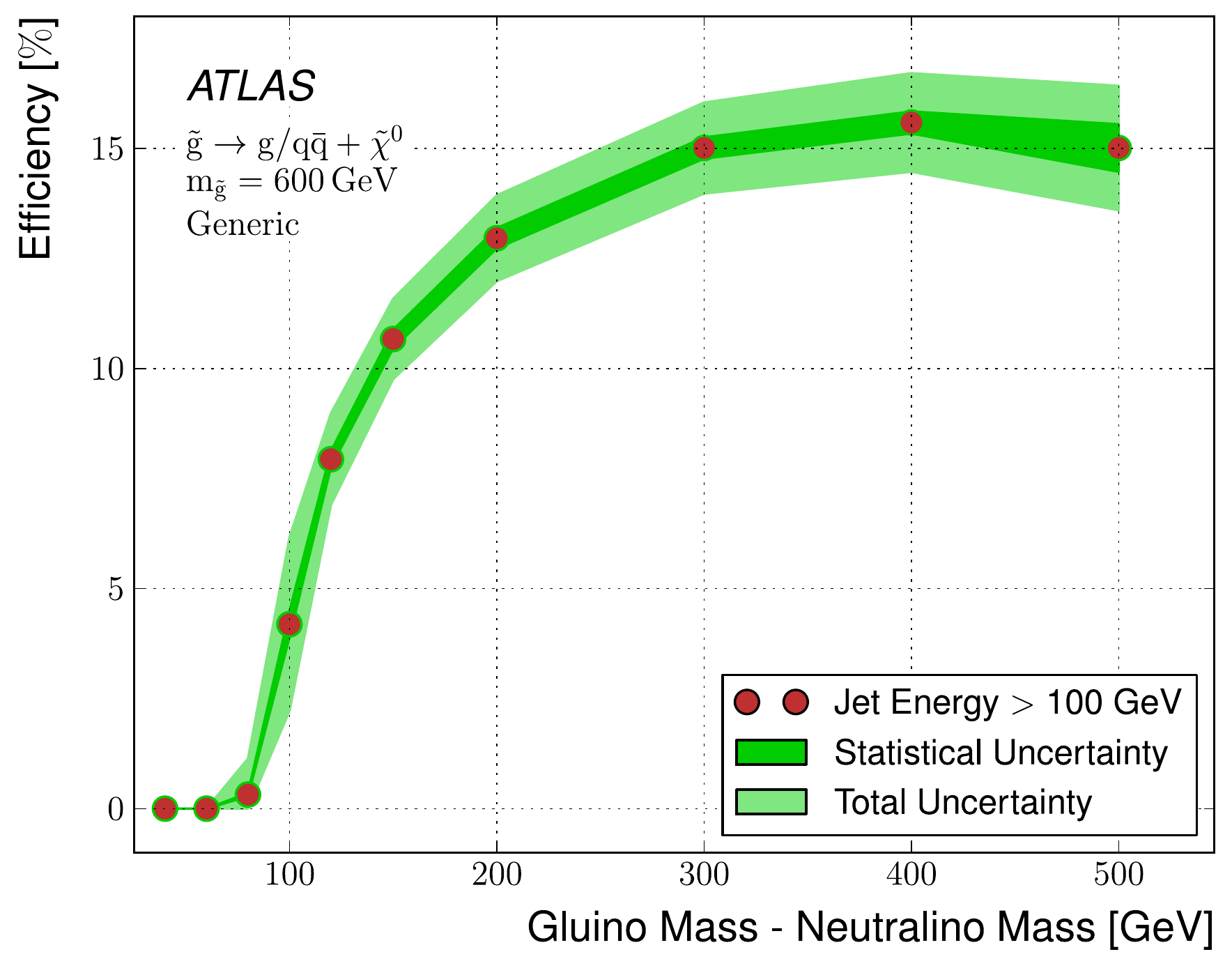} 
\includegraphics[width=0.48\textwidth]{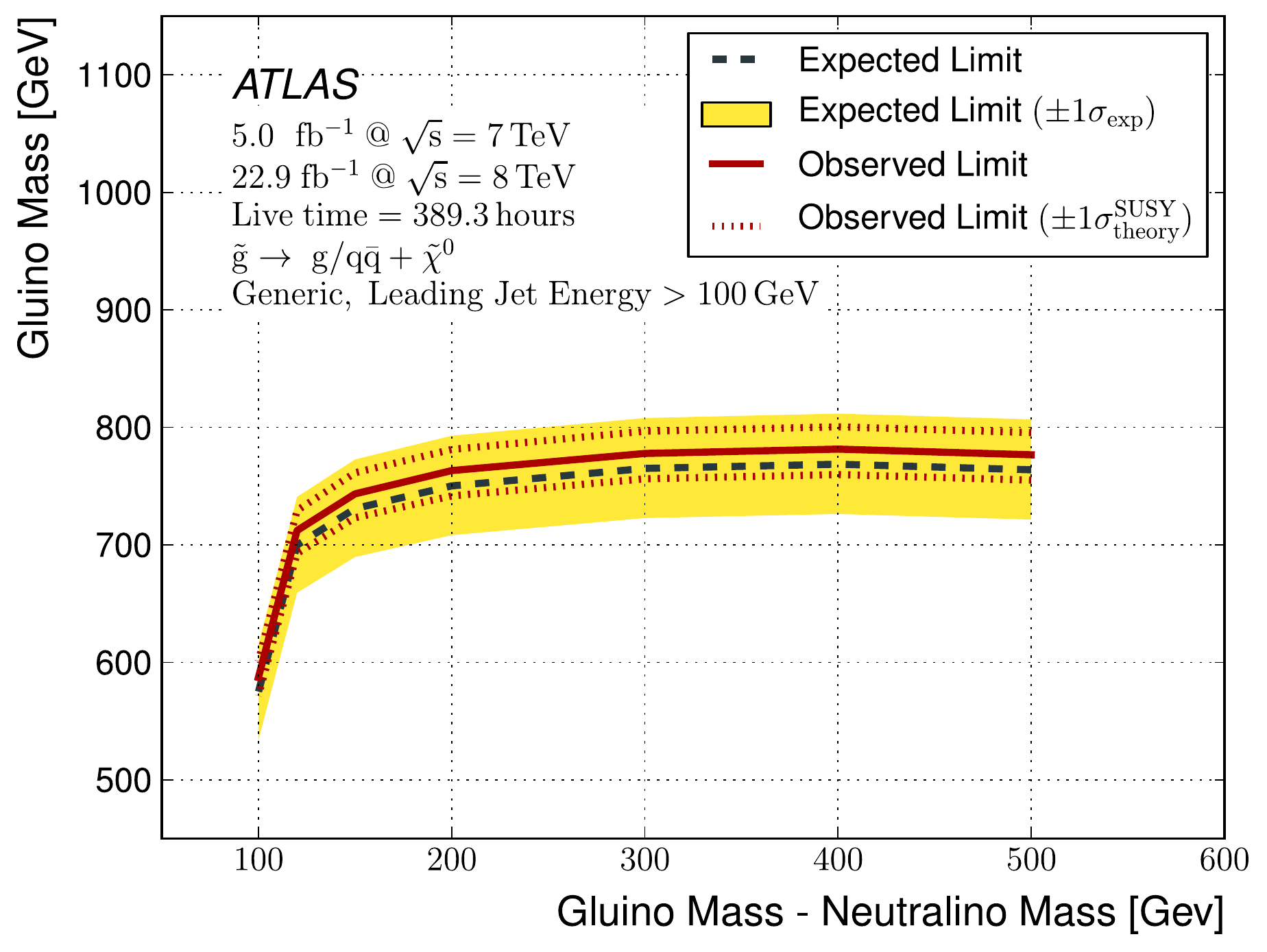}
 \caption{Total reconstruction efficiency and Bayesian lower limits on gluino mass, as a function of the mass splitting between the gluino and the neutralino, in the generic $R$-hadron model with gluino lifetimes in the plateau acceptance region between $10^{-5}$ and $10^3$ seconds. A 600~GeV gluino is used as a reference for the reconstruction efficiencies.
 \label{fig:splitting}}
 \end{center}
\end{figure*}

To provide limits in terms of the gluino, stop, or sbottom masses, the mass is found where the theoretically predicted number of signal events, using the signal cross sections at $-1\sigma$ of their uncertainties, intersects the experimental limit on the number of signal events produced.
The gluino, stop, and sbottom mass limits for each of the signal models, for lifetimes in the plateau acceptance region between $10^{-5}$ and $10^3$ seconds, can be seen in Table~\ref{tab:crosssection_limit}.
Figure~\ref{fig:mass_limits_tau} shows the mass limits for various signal models as a function of lifetime.
Limits are also studied as a function of the mass splitting between the gluino or squark and the neutralino. Figure~\ref{fig:splitting} shows the total reconstruction efficiency for a gluino mass of 600~GeV and mass limits for gluino $R$-hadrons, in the generic $R$-hadron model with lifetimes in the plateau acceptance region between $10^{-5}$ and $10^3$ seconds, as a function of this mass splitting between the gluino and the neutralino. A similar dependence exists for stop and sbottom efficiencies and mass limits.

\begin{table*}
\begin{center}
    \caption{Bayesian lower limits on gluino, stop, and sbottom masses for the various signal models considered,
with lifetimes in the plateau acceptance region between $10^{-5}$ and $10^3$ seconds.
        \label{tab:crosssection_limit}}
    \begin{tabular}{cccc|cc}
    \hline \hline
Leading jet  &  $R$-hadron & Gluino/squark  & Neutralino &\multicolumn{2}{|c}{Gluino/squark mass limit (GeV)} \\
energy (GeV)&  model    & decay & mass (GeV) & Expected & Observed       \\
    \hline
100  &  Generic  &  $\tilde{g} \rightarrow \ g/q\bar{q} + \tilde{\chi}^{0}$ &
$m_{\tilde{g}} \ - \ 100$ & 526 & 545 \\
100  &  Generic  &  $\tilde{g} \rightarrow \ t\bar{t} + \tilde{\chi}^{0}$ &
$m_{\tilde{g}} \ - \ 380$ & 694 & 705 \\
300  &  Generic  &  $\tilde{g} \rightarrow \ g/q\bar{q} + \tilde{\chi}^{0}$ &
$100$ & 731 & 832 \\
300  &  Generic  &  $\tilde{g} \rightarrow \ t\bar{t} + \tilde{\chi}^{0}$ &
$100$ & 700 & 784 \\
300  &  Intermediate  &  $\tilde{g} \rightarrow \ g/q\bar{q} + \tilde{\chi}^{0}$ &
$100$ & 615 & 699 \\
300  &  Regge  &  $\tilde{g} \rightarrow \ g/q\bar{q} + \tilde{\chi}^{0}$ &
$100$ & 664 & 758 \\
\hline
100  &  Generic  &  $\tilde{t} \rightarrow \ t + \tilde{\chi}^{0}$ &
$m_{\tilde{t}} \ - \ 200$ & 389 & 397 \\
100  &  Generic  &  $\tilde{t} \rightarrow \ t + \tilde{\chi}^{0}$ &
$100$ & 384 & 392 \\
100  &  Regge  &  $\tilde{t} \rightarrow \ t + \tilde{\chi}^{0}$ &
$100$ & 371 & 379 \\
100  &  Regge  &  $\tilde{b} \rightarrow \ b + \tilde{\chi}^{0}$ &
$100$ & 334 & 344 \\
 \hline \hline
\end{tabular}
\end{center}
\end{table*}

\section{Summary}
\label{sec:summary}

An updated search is presented for
stopped long-lived gluino, stop, and sbottom $R$-hadrons decaying in the calorimeter, using a jet trigger operating in empty crossings of the LHC. Data from the ATLAS experiment recorded in 2011 and 2012 are used, from 5.0 and 22.9~fb$^{-1}$ of $pp$ collisions at 7 and 8~TeV, respectively.
The remaining events after all selections are
compatible with the expected rate from backgrounds, predominantly
cosmic ray and beam-halo muons where no muon track segment is identified.
Limits are set on the gluino, stop, and sbottom masses, for different decays, lifetimes, and neutralino masses. 
With a neutralino of mass 100~GeV, the analysis excludes
$m_{\tilde{g}} < \masslimitgeneric$~GeV (with an expected lower limit of $\masslimitexpgeneric$~GeV), for a gluino lifetime between 10~$\upmu$s and 1000~s in the generic $R$-hadron model with equal branching ratios for decays to $q\bar{q}\nino$ and $g\nino$. For the same $m_{\nino}$ and squark lifetime assumptions, stop and sbottom are excluded with masses below $\masslimitstop$ and $\masslimitsbottom$~GeV, respectively, in the Regge $R$-hadron model.

\section*{Acknowledgments}





We thank CERN for the very successful operation of the LHC, as well as the
support staff from our institutions without whom ATLAS could not be
operated efficiently.

We acknowledge the support of ANPCyT, Argentina; YerPhI, Armenia; ARC,
Australia; BMWF and FWF, Austria; ANAS, Azerbaijan; SSTC, Belarus; CNPq and FAPESP,
Brazil; NSERC, NRC and CFI, Canada; CERN; CONICYT, Chile; CAS, MOST and NSFC,
China; COLCIENCIAS, Colombia; MSMT CR, MPO CR and VSC CR, Czech Republic;
DNRF, DNSRC and Lundbeck Foundation, Denmark; EPLANET, ERC and NSRF, European Union;
IN2P3-CNRS, CEA-DSM/IRFU, France; GNSF, Georgia; BMBF, DFG, HGF, MPG and AvH
Foundation, Germany; GSRT and NSRF, Greece; ISF, MINERVA, GIF, DIP and Benoziyo Center,
Israel; INFN, Italy; MEXT and JSPS, Japan; CNRST, Morocco; FOM and NWO,
Netherlands; BRF and RCN, Norway; MNiSW, Poland; GRICES and FCT, Portugal; MNE/IFA, Romania; MES of Russia and ROSATOM, Russian Federation; JINR; MSTD,
Serbia; MSSR, Slovakia; ARRS and MIZ\v{S}, Slovenia; DST/NRF, South Africa;
MINECO, Spain; SRC and Wallenberg Foundation, Sweden; SER, SNSF and Cantons of
Bern and Geneva, Switzerland; NSC, Taiwan; TAEK, Turkey; STFC, the Royal
Society and Leverhulme Trust, United Kingdom; DOE and NSF, United States of
America.

The crucial computing support from all WLCG partners is acknowledged
gratefully, in particular from CERN and the ATLAS Tier-1 facilities at
TRIUMF (Canada), NDGF (Denmark, Norway, Sweden), CC-IN2P3 (France),
KIT/GridKA (Germany), INFN-CNAF (Italy), NL-T1 (Netherlands), PIC (Spain),
ASGC (Taiwan), RAL (UK) and BNL (USA) and in the Tier-2 facilities
worldwide.


\clearpage
\bibliographystyle{atlasstylem}
\bibliography{PRDpaper}

\onecolumngrid
\clearpage
\begin{flushleft}
{\Large The ATLAS Collaboration}

\bigskip

G.~Aad$^{\rm 48}$,
T.~Abajyan$^{\rm 21}$,
B.~Abbott$^{\rm 112}$,
J.~Abdallah$^{\rm 12}$,
S.~Abdel~Khalek$^{\rm 116}$,
O.~Abdinov$^{\rm 11}$,
R.~Aben$^{\rm 106}$,
B.~Abi$^{\rm 113}$,
M.~Abolins$^{\rm 89}$,
O.S.~AbouZeid$^{\rm 159}$
% ...
\bigskip
\\
% ...
$^{11}$ Institute of Physics, Azerbaijan Academy of Sciences, Baku, Azerbaijan\\
$^{12}$ Institut de F{\'\i}sica d'Altes Energies and Departament de F{\'\i}sica de la Universitat Aut{\`o}noma de Barcelona, Barcelona, Spain\\
% ...
$^{21}$ Physikalisches Institut, University of Bonn, Bonn, Germany\\
% ...
$^{48}$ Fakult{\"a}t f{\"u}r Mathematik und Physik, Albert-Ludwigs-Universit{\"a}t, Freiburg, Germany\\
% ...
$^{89}$ Department of Physics and Astronomy, Michigan State University, East Lansing MI, United States of America\\
% ...
$^{106}$ Nikhef National Institute for Subatomic Physics and University of Amsterdam, Amsterdam, Netherlands\\
% ...
$^{112}$ Homer L. Dodge Department of Physics and Astronomy, University of Oklahoma, Norman OK, United States of America\\
$^{113}$ Department of Physics, Oklahoma State University, Stillwater OK, United States of America\\
% ...
$^{116}$ LAL, Universit{\'e} Paris-Sud and CNRS/IN2P3, Orsay, France\\
% ...
$^{159}$ Department of Physics, University of Toronto, Toronto ON, Canada\\
% ...
\end{flushleft}



\end{document}